# Real GDP per capita: global redistribution of economic power

Ivan O. Kitov, Institute of Geosphere Dynamics, RAS
Oleg I. Kitov

**Abstract**

Growth rate of real GDP per capita, GDPpc, is represented as a sum of two components – a monotonically decreasing economic trend and fluctuations related to population change. The economic trend is modelled by an inverse function of GDPpc with a constant numerator which varies for the largest developed economies. In 2006, a statistical analysis conducted for 19 selected OECD countries for the period between 1950 and 2003 showed a very weak linear trend in the annual GDPpc increment for the largest economies: the USA, Japan, France, Italy, and Spain. The UK, Australia, and Canada showed a slightly steeper positive linear trend. The 2012 revision showed that the positive trends became much lower and some of them fell below zero due to the Great Recession. The fluctuations around the trend values are characterized by a quasi-normal distribution with heavy and asymmetric tails. This research revises the previous estimates and extends the set of studied countries by economies in East Europe, Latin America, BRICS, Africa, and Asia including several positive outliers with extremely fast growth. The change in GDP definitions and measuring procedures with time and economic source is discussed in relation to the statistical significance of the trend estimates and data quality requirements for a consistent economic model. The relative performance of all counties since 1960 is compared according to the predicted total GDPpc growth as a function of the initial value. The performance in the 21$^{st}$ century is analyzed separately as revealing potential and actual shifts in the global economic powers.

## 1. Introduction

Real economic growth has been studied numerically since Kuznets' works on accounting of national income and aggregate factor inputs. Hodrick and Prescott [1980] introduced a concept of two-component economic growth – an economic trend and a deviation or business cycle component. The trend component is responsible for the long-term growth and defines economic efficiency. In the long run, the deviation component of economic growth has to have a zero mean value. Many researchers proposed a variety of endogenous and exogenous shocks as the force driving fluctuations in the real GDP growth rate.

Kitov [2006] proposed a model with the long-term GDP growth rate dependent on the attained level of real GDP per capita. In developed countries, real GDP per capita has to grow with time along a straight line, if no large non-economic force (war, pandemic, etc.) is observed. The relative growth rate of the real GDP per capita, GDPpc, has to be an inverse function of the attained level of GDPpc with a potentially constant numerator for developed economies. This study is devoted to re-validation of the model using the GDPpc and population data for selected developed countries since 2011. We added several economies to demonstrate the difference between the most developed and other countries as well as country-specific characteristics of the transition to a stable and stationary economic growth.

## 2. Model

The model assumptions should be supported by data. Figure 1 displays the real GDP per capita trajectories for several developed economics as reported by the Maddison Project Database,

University of Groningen [MPD, 2020]. We discussed the transition from the low-rate development in the second half of the 19th century and first quarter of the 20th century in [Kitov, Kitov, 2012]. There is one common feature for all trajectories observed in the first and second period in Figure 1 – they are well approximated by linear function of time. We excluded the transition period between 1930 (the Great Depression) and 1960 (WWII and its aftermaths) in order to highlight the periods less affected by internal and external non-economic forces. The WWI had much lower, but not negligible, influence on the economies - the trade between the fighting sides was not stopped.

The trajectories in Figure 1 illustrate the assumption that the real economic growth in developed countries was likely linear since 1960. In other words, the annual increment of the real GDPpc is constant on average. The study of fine features of this linear behaviour is the main objective of this paper. The set of developed countries with linear growth is extended by several economics from Eastern Europe, BRICS, Latin America, Asia and the countries with the exceptional economic growth.

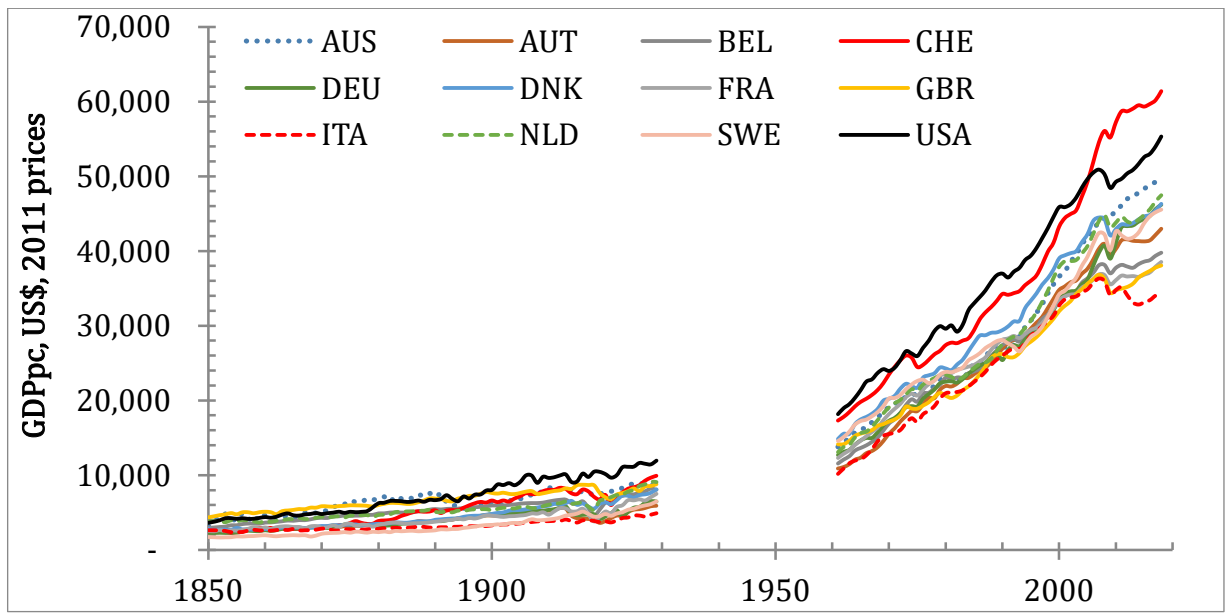

Figure 1. The evolution of real GDP per capita reported by the MPD since 1850. The transition period between 1930 and 1960 is excluded for the clarity of trend change.

The principal model claim is simple – the growth rate, $g(t)$, of real GDP per capita, $G(t)$, is driven by the attained level of real GDP per capita and the change in a country dependent specific age population, $N_s$. The growth rate of the real GDP per capita in developed countries is characterized by a constant annual increment, $A$. All fluctuations around this constant increment can be explained by the change in the number of people of the country-specific age:

$$g(t) = dlnG(t)/dt = A/G(t) + 0.5dlnN_s(t)/dt \qquad (1)$$

Equation (1) is a quantitative model that has been constructed empirically and proved statistically by cointegration tests [Kitov, Kitov, 2012]. With the new data between 2010 and 2019, the model also passes the cointegration tests.

In economic statistics, usually the relative growth rate is published, as represented by $dG(t)/G(t)=dlnG(t)$. For the sake of simplicity, we assume that the second term in (1) is zero. Accordingly, the economic system under study is in a stationary or inertial growth, i.e. $A/G(t)$ is "the inertial growth" or economic trend. The adults between 16 and 64, the working age population, can be also considered as living in a stationary regime since no dramatic organic and functional changes happen to their life process out of the margins of natural variations. One can assume that the stationary regime of economic growth is related to the stationary regime of the human being life.

For the inertial growth, the real GDP per capita grows as a linear function of time:

$$g(t) = dlnG(t)/dt (given \; dN_s(t)=0) \; = A/G(t)$$

$$G(t) = At + C \qquad (2)$$

where $G(t)$ is completely equivalent to the inertial growth, $G_i(t)$, i.e. the first component of the overall growth as defined by (1). Relationship (2) defines the linear trajectory of the GDP per capita, where $C=G_i(t_0)=G(t_0)$ and $t_0$ is the initial time. In the regime of inertial growth, the real GDP per capita increases by the constant value $A$ per time unit. Relationship (3) is equivalent to (2), but holds for the inertial part of the total growth:

$$G_i(t) = G_i(t_0) + At \qquad (3)$$

The relative rate of growth along the inertial linear growth trend, $g_i(t)$, is the reciprocal function of $G_i$ or, equivalently, $G$:

$$g_i(t) = dlnG_i/dt \; = A/G_i = A/G(t) \qquad (4)$$

Relationship (4) implies that the rate of GDP growth will be asymptotically approaching zero, but the annual increment $A$ will be constant. Moreover, the absolute rate of the GDP per capita growth is constant and is equal to $A$ [$/y].

In physics, inertia is the most fundamental property. In economics, it might also be a fundamental property, taking into account the difference between ideal theoretical equilibrium of space/time and the stationary real behaviour of the society. Mechanical inertia implies that no change in motion occurs in the absence of net external force and without change in internal motion energy. In real world, the net force is zero for constant speed, but one should apply extra forces in order to overcome the net traction force and to keep the body moving at a constant speed. For a society, the net force applied by all economic agents is not zero but counteracts all dissipation/discounting processes and creates goods and services in excess of the previous level. The economy does grow with time and its "internal motion energy" as expressed in monetary units does increase at a constant speed.

## 3. Data Quality

The history of physics has many examples when the measured values of fundamental physical constants were corrected according to new measurements using more sensitive instruments and methods. A good example is the change in the measured value of the gravitational constant. In economics, multiple revisions to a once measured value are common practice. As a result, many economic time series are accompanied by disclaimers that the measurements in subsequent years are incompatible due to change in measurement methodology and/or definition of parameters. This is true even for such a simple process as population counting: the population estimates are corrected many times back in the past.

The main objective of this study is an extended analysis of the linear growth in the real GDP per capita (GDPpc). In 2006 [Kitov, 2006] and 2012 [Kitov, Kitov, 2012], the original analysis was based on the GDPpc time series from the Total Economy Database published by the Conference Board [TED, 2012]. At this point, we return to Section 2 and estimate the changes in the MPD time series relative to the 2013 TED. This is a mandatory methodological step in any study based on alternating data, *e.g.* the difference between the 2010 and 2013 TEDs was large for some countries. This issue is described in more details in [Kitov, Kitov, 2021]. For this study, a more detailed investigation of the real GDP per capita data quality is needed as the data consistency from various sources is a big problem for economic modeling.

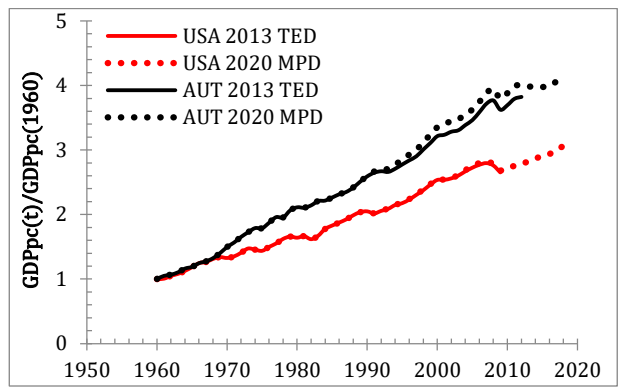

Figure 2. Comparison of the MPD and TED (2013) GDPpc time series for Austria (AUT) and USA. The GDPpc evolution in Austria has important updates in the MPD, which may result in not negligible differences in the growth rate estimation.

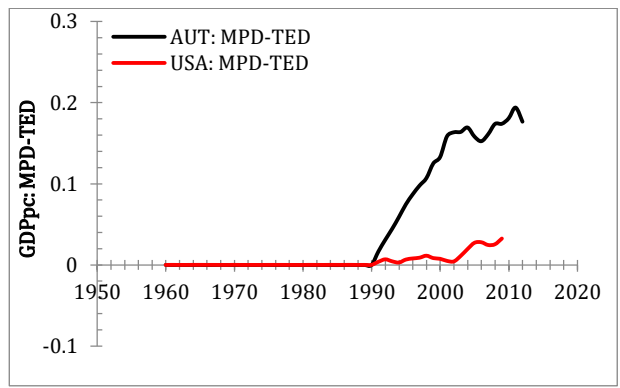

Figure 3. The difference between the MPD 2020 and TED 2013 presented in Figure 2. For Austria, the difference in 2012 was 0.18 units of GDPpc total growth. This is a significant change in the time series.

We start with the change in the estimates provided by the TED and MPD. In the past, the Groningen Growth and Development Centre at the University of Groningen (MPD) was in tight cooperation with the Conference Board (TED). Figure 2 presents the evolution of the real GDPpc in the USA and Austria and compares the 2013 TED and 2020 MPD estimates. In order to compare several time series of the real GDPpc with different reference years, and thus different units of measurements, we select the start of the studied interval as the reference year. All 4 time series in Figure 2 are normalized their respective levels in 1960. Both time series for the USA are practically identical with low-amplitude deviations since 1990 (total growth since 1960 is 2.65 for the TED 2013 and 2.68 for the MPD, i.e. (2.65-2.68)/2.68 =0.9988 or -1.1%), as Figure 3 shows, where the difference between the normalized MPD 2020 and TED 2013 are depicted.

For Austria, the difference since 1990 has larger amplitude and the overall growth in the GDPpc since 1960 is 3.82 in the TED 2013 and 4.00 for the MPD, i.e. 4.4%. In absolute values, the difference in the total growth is 0.03 units in the USA and 0.18 units in Austria. Therefore, the GDPpc evolution in Austria has important updates in the MPD, which may result in non-negligible differences in the growth rate estimation. Therefore, the GDPpc estimates are corrected in many revisions within one source [Kitov, Kitov, 2012] and between the sources. The revision dynamics may affect the results of statistical assessment for the previous model versions.

In order to obtain a deeper understanding of the problems related to data quality and consistency we extend the set of data sources by the OECD. This source has a limited set of countries, but includes all developed economies. In Figure 4, the difference between the MPD, OECD, and TED estimates for Austria is illustrated. All three time series were retrieved from corresponding sources in December 2020. The start point is 1970 as defined by the availability of the OECD data for Austria. (For the OECD, the start time varies between countries.) The TED provides the GDPpc estimates from 1950. Since the main goal of the MPD is to provide the longest possible GDPpc estimates, selected continuous time series start in the first half on the 19th century. In the upper panel of Figure 4, the original GDP per capita estimates are presented, with clear difference in the reference year. Three time series in the middle panel are normalized to their levels in 1970. The OECD and TED are practically identical, and the MPD reports a higher total growth between 1970 and 2018. The relative difference between the MPD and OECD is depicted in the lower panel of Figure 4. There is a linear segment between 1990 and 2012. Then the difference has an approximately 4-year-long shelf, and falls since 2016. The linearity in the difference is not a problem for our model since it can be easily compensated by the linear regression coefficient without loss in statistical confidence.

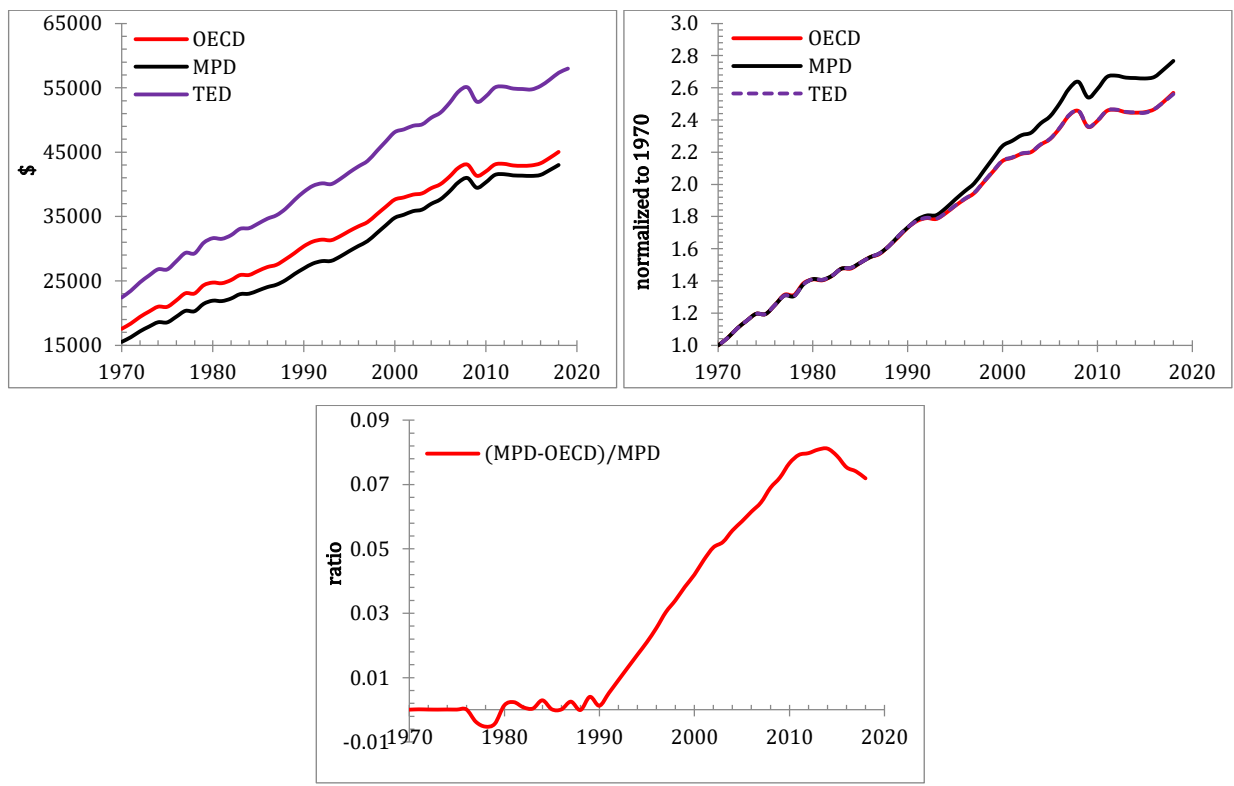

Figure 4. The difference between the MPD, OECD, and TED estimates for Austria. All estimates were obtained on December 21, 2020. The start point is 1970 as defined by the availability of the OECD data. Left panel: The original GDP per capita estimates with clear difference in the reference year. Right panel: Three time series normalized to their levels in 1970. The OECD and TED are practically identical, and the MPD reports much higher total growth between 1970 and 2018. Lower panel: Relative difference between the MPD and OECD. There is a linear segment between 1990 and 2012. Then the difference stalls for four years and falls since 2016.

These features reveal several problems. The closeness of the curves between 1970 and 1990 illustrates the closeness of definitions and raw data used by the three sources for this period. The MPD and OECD start to deviate along a linear time function since 1990 and one can suggest that there was just a constant coefficient between the corresponding GDP deflator definitions. The dynamic change in the relative difference since 2011 is likely explained by the activity of economists redefining major economic variables in order to re-assess the causes of the Great Recession. Seemingly, this type of activity is not finished and the data consistency between the OECD, TED, and MPD estimates will suffer further deviations.

The difference between the MPD and OECD/TED in Austria is not an extreme example. In Figure 5, we present the case of Japan. The TED estimates of the real GDPpc start to deviate from the other two sources in 1977 and the difference reaches 25% in 2016 as the relative growth, (TED-MPD)/TED, shows. Moreover, the relative difference is better approximated by a polynomial function rather than by a linear one. In other words, one cannot replace the MPD by the TED in economic models without loss in statistical agreement. The polynomial approximation cannot be used since it has no simple economic explanation.

In Figure 6, we present the UK. There are 4 sources with the ONS (Office of National Statistics) nominating real GDP in British pounds. There should be no difference between measurements in different currencies in case they have a fixed proportion. The ONS estimates provide the highest total growth (by a factor of 3.49) in the real GDPpc in the UK since 1955 (the OECD time series starts in 1955), but they are close to those reported by the OECD (3.46). The MPD has the lowermost total growth (3.03) and the TED reports the integral growth of 3.31. The pair-wise ratios of the time series demonstrate various types of behavior from identical to essentially nonlinear (*e.g.*, quadratic).

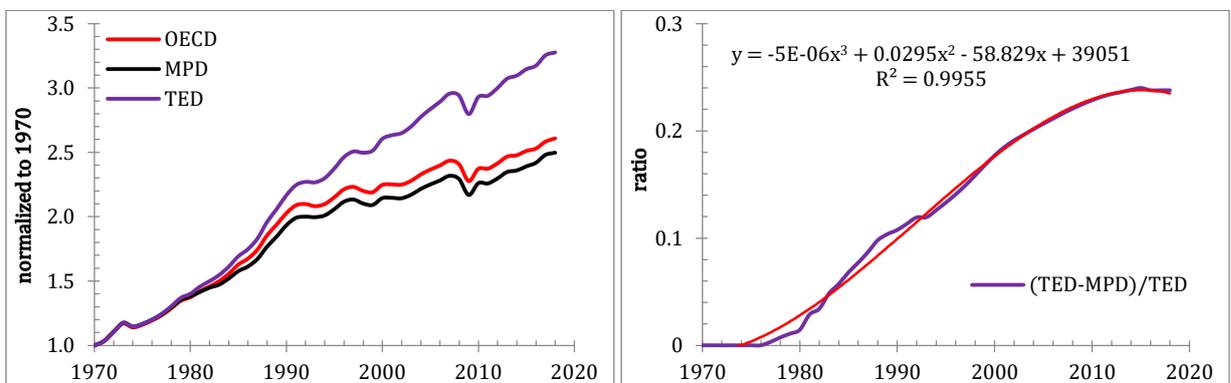

Figure 5. Same as in Figure 4 for Japan

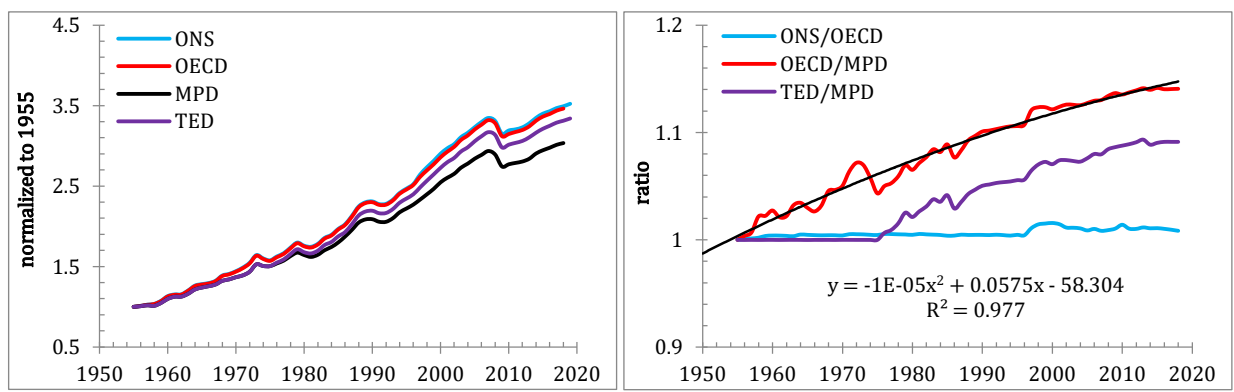

Figure 6. Same as in Figure 4 for the UK.

The difference in the real GDP per capita estimates reported by different economic sources should be considered as a serious issue in economic/econometric modeling. Figure 7 depicts the total GDPpc growth in 17 developed countries as defined by the MPD, OECD, and TED. Two periods are presented: between 1970 and 2018 (all involved economies have GDPpc estimates since 1970) and between 2000 and 2018, *i.e.* in the most recent period with many definitional revisions. The relative position of a given agency in a given country does not reveal many changes (*e.g.* the MPD has the highest total growth in the first 7 countries in the both studied periods), but the relative amplitude may change. Table 1 lists the MPD GDPpc estimate for selected year between 1960 and 2018. It also provides the average annual increment and its standard deviation for each country. The overall average is $533 with standard deviation $570.

Interestingly, the TED estimates are the highest for the USA and Japan. The MPD is most generous to the north European countries (the Netherlands, Denmark, Sweden), Austria, Germany, Italy, Switzerland, Australia, Canada, New Zealand. The OECD likes the UK, France, and Spain. One cannot reject the hypothesis that the economic sources may have some positive bias in favor of the countries they depend on.

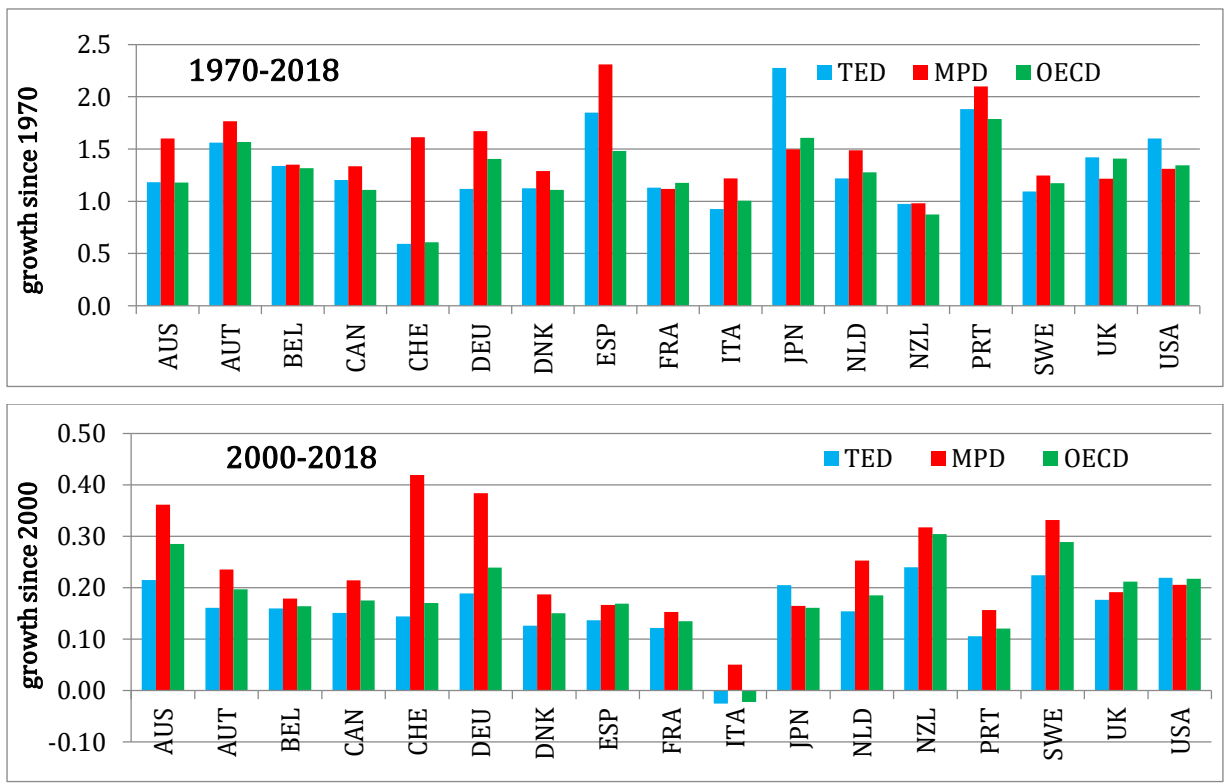

Figure 7. Total growth in real GDP per capita as estimated by three different sources (TED, MPD, and OECD) for two different periods. Strong variations between sources and time intervals are observed. Notice the negative growth in Italy reported by the TED and OECD in the 21st century.

In [Kitov, 2006], some statistical properties of the residual growth, i.e. fluctuations, were estimated. Frequency distributions in $200-wide bins were constructed using the annual GDPpc increments for 19 countries. In this study, we use the same bins and 17 economies. At the same time, the data set is extended by 15 new readings for each country and three different sources of the real GDP per capita estimates are available. The statistical analysis may reveal important differences between sources.

Figure 8 depicts the frequency distribution of the annual GDPpc increments as obtained from the TED. Two versions are presented: from 1960 to 2018 and from 1960 to 2007 in order to highlight the effect of the Great Recession, which induced numerous large amplitude negative and positive (recovery growth or rebound) fluctuations. Near the peak, the 1960-2018 distribution is very close to a normal one with the mean value $820 and standard deviation $460. The 1960-2007 distribution (right panel in Figure 8) is approximated by a normal distribution with the mean value $850 and standard deviation $475. Considering the heavy tails effect on the estimation procedure, there is no significant difference in the parameters of these two approximating normal distributions. The difference between the two real distributions is most striking in the range beyond the standard deviation limits, and the 1960-2018 set shows more large-amplitude deviations, both negative and positive. The tails of the real distributions are above the predicted values of the respective normal distribution. This effect is often observed in natural sciences and is associated with inaccurate measurements, limited amount of readings, and sometimes with action of some real factors. Heavy tails are well known in the time series related to the stock returns.

Table 1. Real GDP per capita in developed countries for selected years; mean annual increment; standard deviation of the annual increment. For the whole set of countries: Mean=$533, StDev=$570.

|  | **2018** | **2010** | **2000** | **1990** | **1980** | **1970** | **1960** | **Mean** | **StDev** |
|---|---|---|---|---|---|---|---|---|---|
| **AUS** | 49831 | 45400 | 36603 | 27373 | 22972 | 19166 | 14013 | 618 | 448 |
| **AUT** | 42988 | 40288 | 34796 | 26930 | 21932 | 15537 | 10391 | 562 | 472 |
| **BEL** | 39756 | 37739 | 33720 | 27412 | 23060 | 16914 | 11081 | 494 | 439 |
| **CAN** | 44869 | 41209 | 36943 | 30082 | 25784 | 19207 | 13952 | 533 | 571 |
| **CHE** | 61373 | 57219 | 43251 | 34250 | 27406 | 23479 | 16358 | **776** | **729** |
| **DEU** | 46178 | 41110 | 33367 | 25391 | 22497 | 17277 | 12282 | 584 | 615 |
| **DNK** | 46312 | 42932 | 39021 | 29412 | 24272 | 20221 | 14046 | 556 | 610 |
| **ESP** | 31497 | 31786 | 26995 | 19215 | 14008 | 9511 | 5037 | 456 | 567 |
| **FRA** | 38,516 | 36,087 | 33,410 | 28,129 | 23,537 | 18,187 | 11,792 | 461 | *368* |
| **ITA** | 34364 | 34766 | 32717 | 26003 | 20959 | 15492 | 9430 | 430 | 546 |
| **JPN** | 38674 | 35011 | 33211 | 29949 | 21404 | 15484 | 6354 | 557 | 570 |
| **NLD** | 47474 | 43812 | 37900 | 27515 | 23438 | 19075 | 13209 | 591 | 582 |
| **NZL** | 35336 | 31586 | 26823 | 21817 | 19681 | 17835 | 15087 | *349* | 523 |
| **PRT** | 27036 | 25463 | 23372 | 17256 | 12822 | 8724 | 4712 | 385 | 434 |
| **SWE** | 45542 | 42635 | 34203 | 28068 | 23809 | 20269 | 13849 | 546 | 697 |
| **UK** | 38058 | 34754 | 31946 | 26189 | 20612 | 17162 | 13780 | 419 | 486 |
| **USA** | 55335 | 49267 | 45886 | 36982 | 29611 | 23958 | 18057 | 643 | 650 |

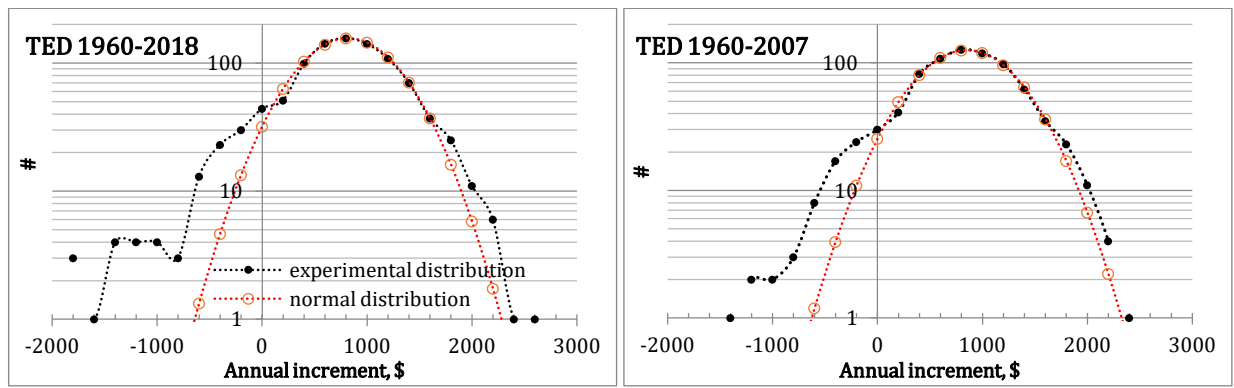

Figure 8. Frequency distribution of the annual increments of real GDP per capita in 17 developed countries (Table 1) as estimated by the TED. Left: Between 1960 and 2018. Right: Between 1960 and 2007. Both distributions are compared with normal distributions fitting the observations near the peak. One can observe the heavy tails, with predominance of negative values. The 1960 to 2007 distribution has less readings in the heavy tails because the Great Recession and the following intensive rebound are excluded.

The results for the other two agencies – MPD and OECD – are presented in Figures 9 and 10. The MPD data demonstrate similar features with slight differences in the tail distributions, *e.g.*, more negative readings with higher fluctuation amplitudes. The features of the approximating normal distributions are similar for the TED and MPD. The OECD demonstrates significant deviations from the approximating normal distribution near the peak, a much lower number of positive large-amplitude fluctuations (annual increments) and heavy negative tail. The differences between the three sources of economic data under study are likely related to their methodologies since the raw data should be the same as obtained by national agencies.

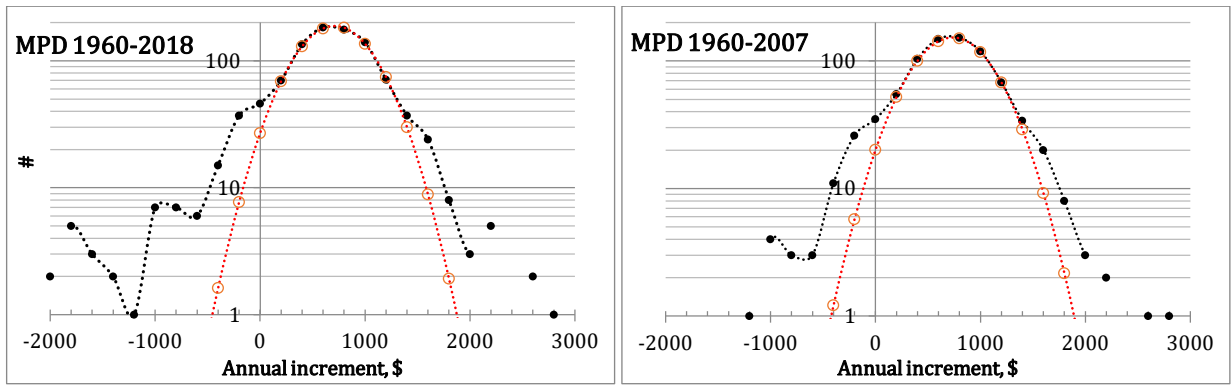

Figure 9. Same as in Figure 8 for the MPD. More positive and negative outliers are observed.

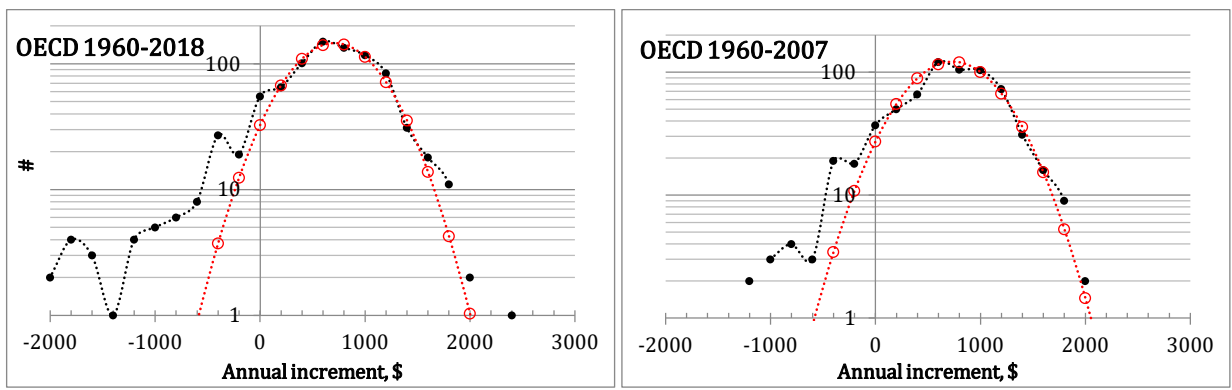

Figure 10. Same as in Figure 8 for the OECD. More negative and less positive outliers are observed.

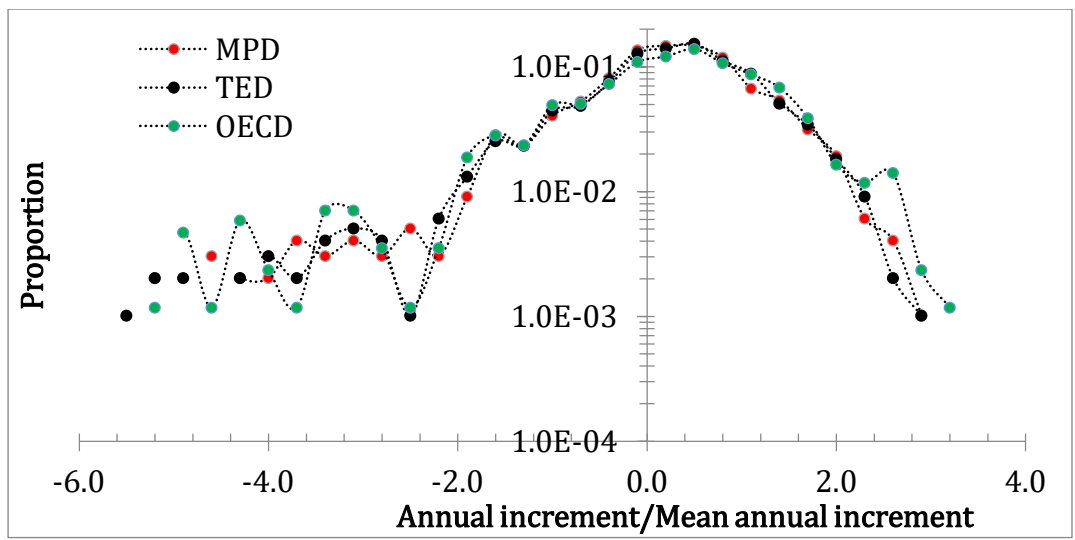

Figure 11. Probability density distribution of the ratio of annual increments and the mean increment for each country separately, i.e. the individual increments are measured in the mean increment values matching the normal distribution requirements. This approach makes all three sources compatible.

Finally, we reduced all three cases to the same unit of measurement, i.e. normalized the annual increments for each country to the corresponding country mean increment. This procedure allows measuring each increment in dimensionless units in a country independent manner. Figure 11 illustrates the difference between TED/MPD and OECD distributions. In this representation, the OECD has heavier positive tail than in the absolute value version. The MPD data are used in further analysis. The other two agencies may likely give slightly different results. However, the OECD does not provide long enough time series for all involved countries.

## 4. Real GDP per capita data revisited

We start with a revision of the period after 1950 used in the previous studies. Originally, the evolution of annual increments before 2003 was estimated. Then data for the period between 2004 and 2007 were added and the model was re-estimated. In 2012, 4 new readings were added for all involved countries. In 2006, we made a model based assumption that all large deviations from the linear trend in the annual GDP per capita should fade away in the near future. With time, this assumption becomes more and more successful statistically – the average annual increment line is the same as the regression line of the annual increment. For exponential economic growth, the best regression line has to be exponential. For illustration, we revisit two countries – Austria and the USA.

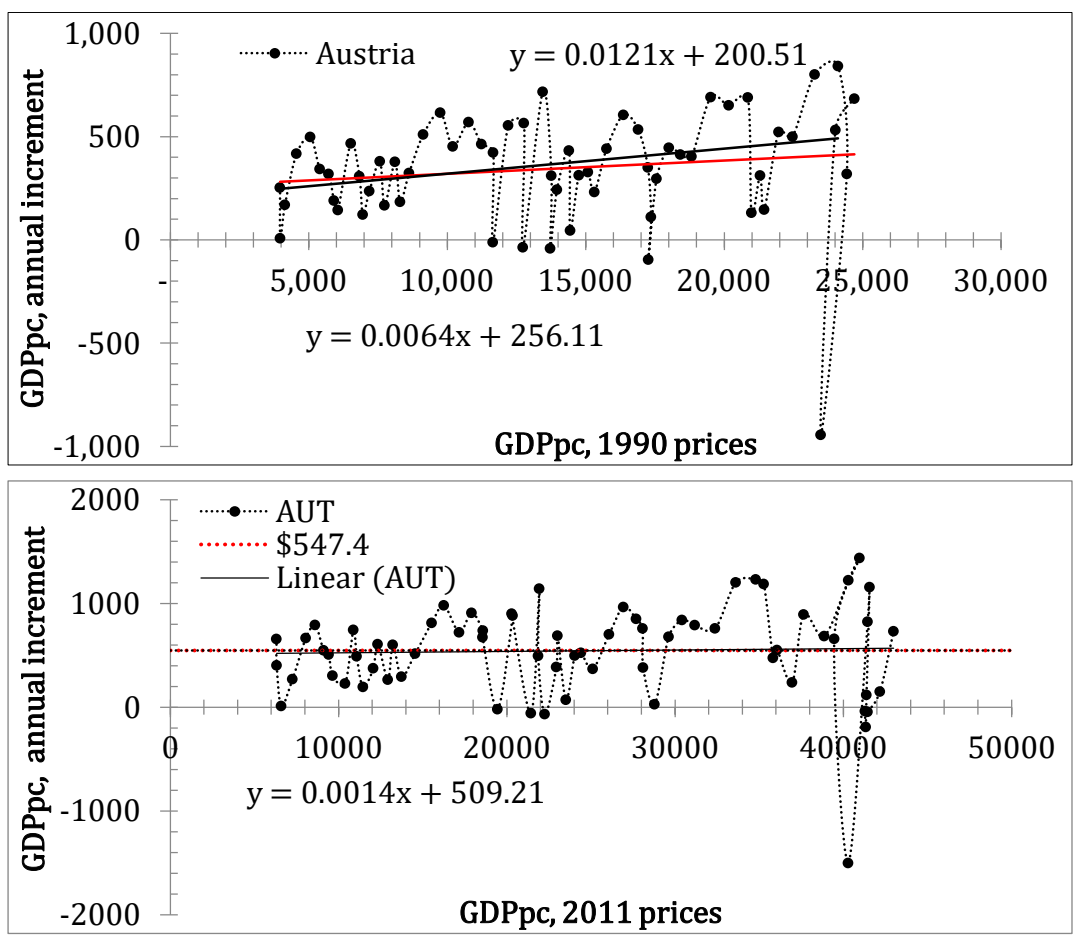

Figure 12. Upper panel: annual increment of real GDP per capita (in 1990 US dollars) as a function of real GDP per capita in Austria for the period between 1951 and 2011. The regression (red) line slope is $0.0064 per dollar. For the period between 1951 and 2007, the regression (black) line has a larger slope of $0.012 per dollar. Lower panel: Same as in the upper panel for the period between 1951 and 2018. The slope fell to $0.014 per $ (2011 prices). The mean GDPpc annual increment value is $547.4.

The upper panel in Figure 12 is borrowed from the 2012 paper and presents the comparison of GDP per capita in Austria for the period between 1950 and 2011. The positive slope reported in 2006 (black line, slope=0.012 $/$) decreased between 2007 and 2011 (red line, slope=0.0064 $/$). Since 2007, the period of poor economic performance with an extended recession resulted in a further decline in the speed of economic growth, and the lower panel in Figure 12 shows that the slope (the MPD data are used for this analysis) fell to 0.0014 $/$ (US$ also became cheaper between 1991 and 2011). In 2018, the linear regression (black) line is not much different from the mean value (red dotted) line. This behavior validates the assumption of long term inertial growth with constant annual increment in the real GDP per capita and fluctuations defined by a random process with zero average.

Figure 13 presents a similar analysis made for the USA where the slope for the period between 1951 and 2018 is close to that the period between 1950 and 2011 because of the major fall in economic growth between 2008 and 2010. The mean value between 1951 and 2018 is $569, i.e. slightly higher than that in Austria. The 2020-2021 COVID-19 pandemic has to reduce the trend slope and even make the slope negative. Obviously, this force does not have inherently economic character, but in real life there are numerous non-economic forces influencing the rate of economic growth in developed countries.

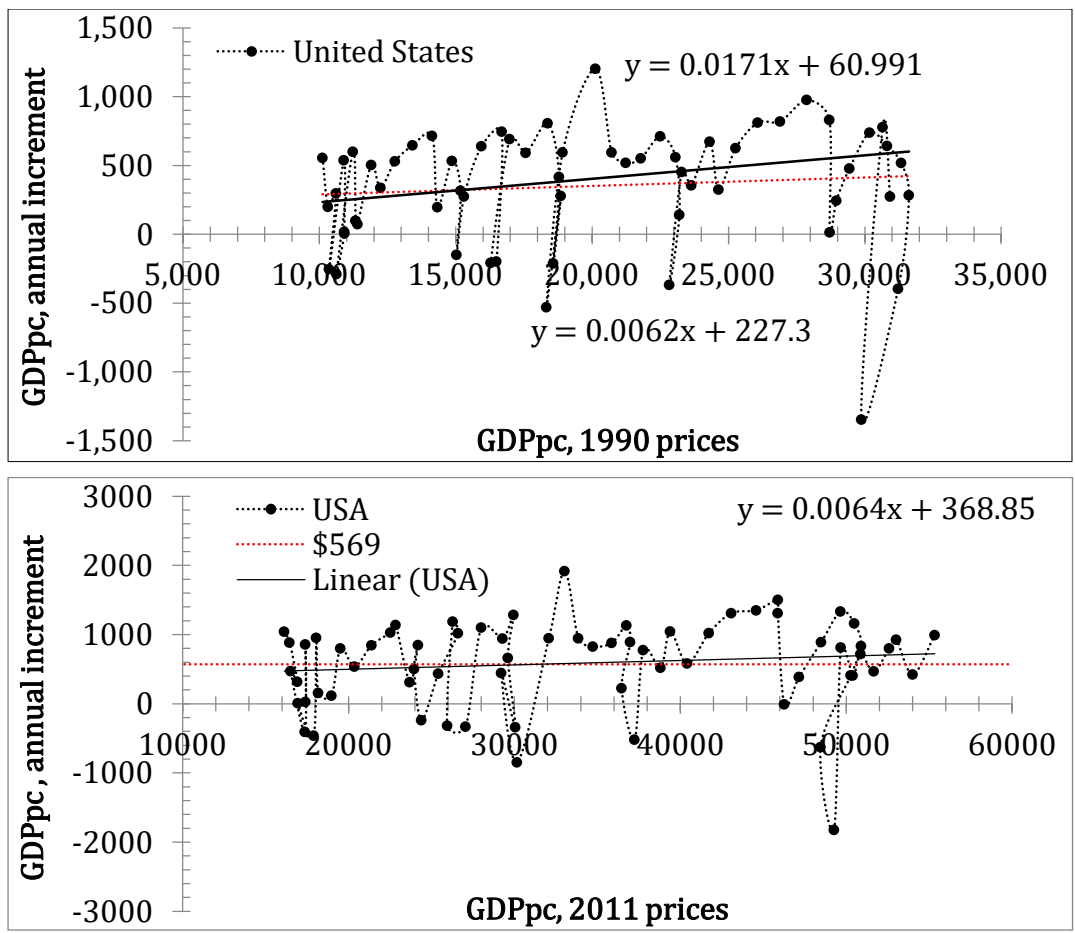

Figure 13. Same as in Figure 12 for the USA. The slope for the period between 1951 and 2018 (0.0064 $/$) is close but lower than that observed for the period between 1950 and 2011 (0.0171 $/$) because of the major fall in economic growth between 2008 and 2010. The mean value between 1951 and 2018 is $569, i.e. slightly higher than that in Austria.

We are going to demonstrate that the inertial growth model accurately predicts the economic growth is developed countries. Moreover, it gives an unbiased and theoretically justified view on the current rate of relative growth in GDPpc in various countries depending on the GDPpc level. One should not directly compare the relative rate of economic growth in China (GDPpc=$13102 in 2018) and in the USA (GDPpc=$55335 in 2018). One should compare the annual increments in the GDPpc and corresponding rates predicted by the model for inertial economic growth. In that sense, the USA growth rate is higher than that observed in China. However, population in China is 4 times larger than in the USA and this gives an impression of a faster total economic growth in the former.

## 5. The model for the period between 1960 and 2018

In this Section, we present the results of inertial growth analysis based on the Maddison Project Database data for an extended set of developed economies. Figure 14 shows three plots for Australia. The upper panel depicts the annual GDPpc increment between 1961 and 2018 with the average value (dashed red line) for the studied period of $618 (2011 US$ prices) with a standard deviation of $448 (see Table 1). The middle plot presents the same annual increment as a function of GDPpc level with the same mean value. In the lower panel, the relative growth rate of the GDPpc is a function of the GDPpc, where the relative rate is the ratio of the GDPpc increment and the level in the beginning of the one year period, i.e. the growth rate in 1961 is [GDPpc(1961)-GDPpc(1960)]/GDPpc(1960). The average increment line in the middle panel is transformed into the line $618/GDPpc($t$). The rate of economic growth is a decreasing (reciprocal) function of the GDP per capita. In the long run, the growth rate is approaching zero.

Australia demonstrates an excellent (much above the average) economic growth between 1992 and 2007. The global recession reduced the rate of growth but did not harm the overall growth much. The 2020 catastrophic fall induced by the COVID-19 pandemic may also push the GDPpc growth rate in Australia below the zero line, and thus, the slope 0.115 $/$ in the middle panel may drop closer to the mean GDPpc line. The Australian GDPpc level in 1960 was $14,013. The initial GDPpc value is an important parameter used in this study. Our model suggests that the growth rate in a fixed period depends on the start value and the mean annual increment.

Figure 15 presents the case of Austria, but now for the period between 1960 and 2018. The mean annual GDPpc increment line ($562±$472) is practically the same as the regression line for the increment with a slightly negative slope (-1.47 $/year) indicating that the annual increment has been decreasing since 1960. This observation is close to the findings in the previous Section, but the years between 1950 and 1959 were less successful for Austria and the slope for the longer period since 1951 is positive with the mean increment of only $547. The GDPpc level in 1960 was $10,391, *i.e.* is much lower than in Australia. The observation curve in the middle panel oscillates around the mean line but demonstrates extended periods with higher growth rates, *e.g.* 1994 to 2011, and short but deep troughs. When the growth rate is negative, the dependence of the annual increment on GDPpc has loops as shown in the middle panel in Figure 15. These loops indicate that the potential rate of growth increases with decreasing GDPpc. This is one of the key elements of the rebound observed after large recessions.

Belgium is presented in Figure 16. The mean annual GDPpc increment ($494 ± $439) is lower than that observed in Australia and Austria for the same period. The GDPpc level in 1960 was $11081, i.e. close to that in Austria. In Belgium, there were 6 years with a negative growth rate, and the corresponding falls were slightly deeper than in Austria. This is one of the reasons behind the lower mean annual increment. The negative slope of -4.5 $/year indicates that the average annual increment is decreasing with time and the economic growth decays faster than predicted by an inverse function of GDPpc.

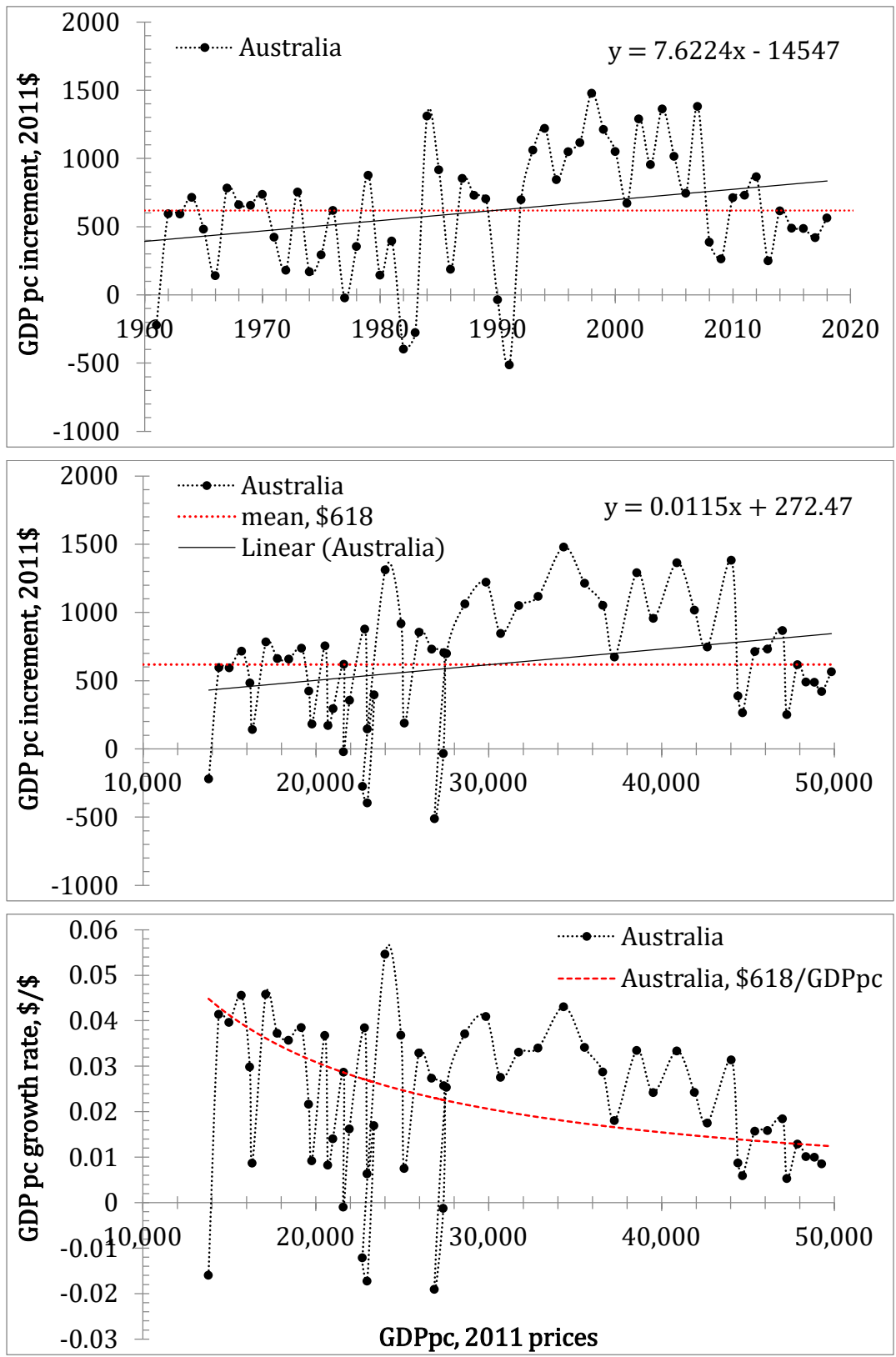

Figure 14. Upper panel: the annual GDPpc increment between 1961 and 2018 with the average value for the studied period of $618 (2011 prices). Middle panel: the same annual increment as a function of GDPpc level. Lower panel: the relative growth rate of the GDPpc as a function of the GDPpc.

A similar pattern is observed in Canada (Figure 17), but the mean annual increment is higher $533±$571. The GDPpc level measured in 1960 was $13,952, i.e. almost the same as in Australia. For Canada, the slope is also slightly negative -1.81 $/year, but this is likely the effect of the Great Recession. The performance was mostly below the average level since 2012. The red dashed line in the lower panel ($533/GDPpc) is a good approximation of data and illustrates the fall in the growth rate with increasing GDPpc. The same is valid for Austria and Belgium.

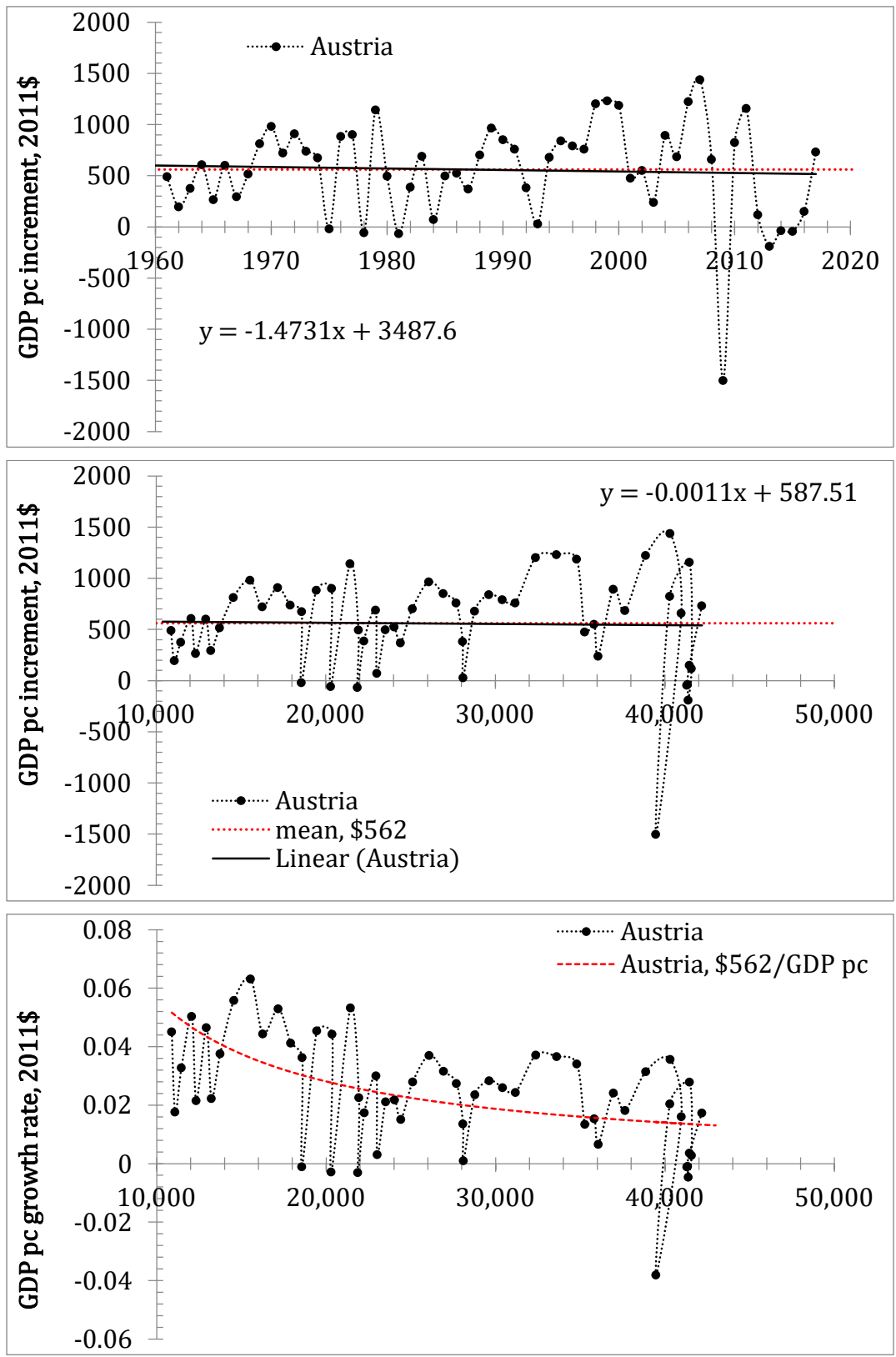

Figure 15. Same as in Figure 14 for Austria.

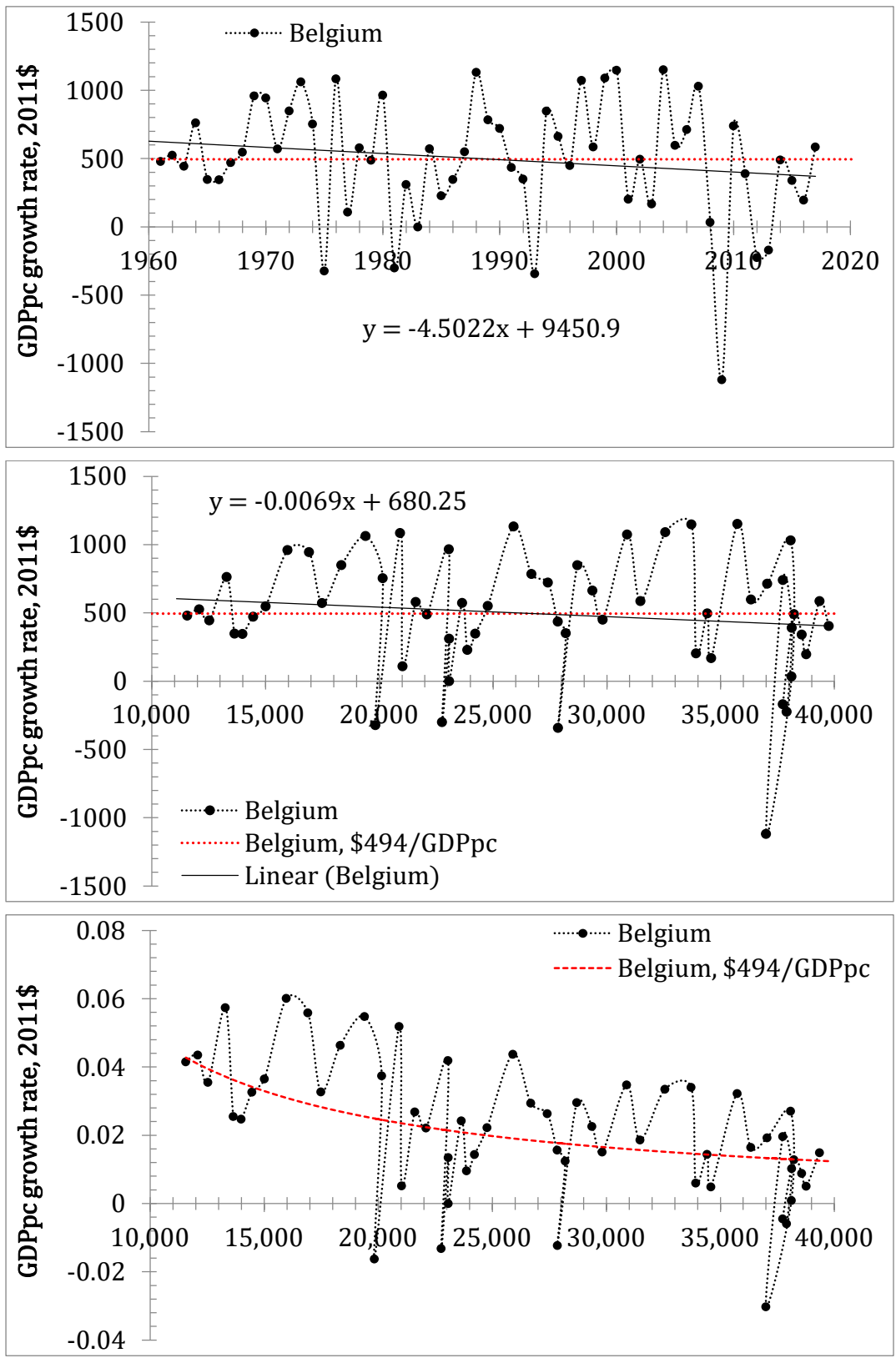

Figure 16. Same as in Figure 14 for Belgium.

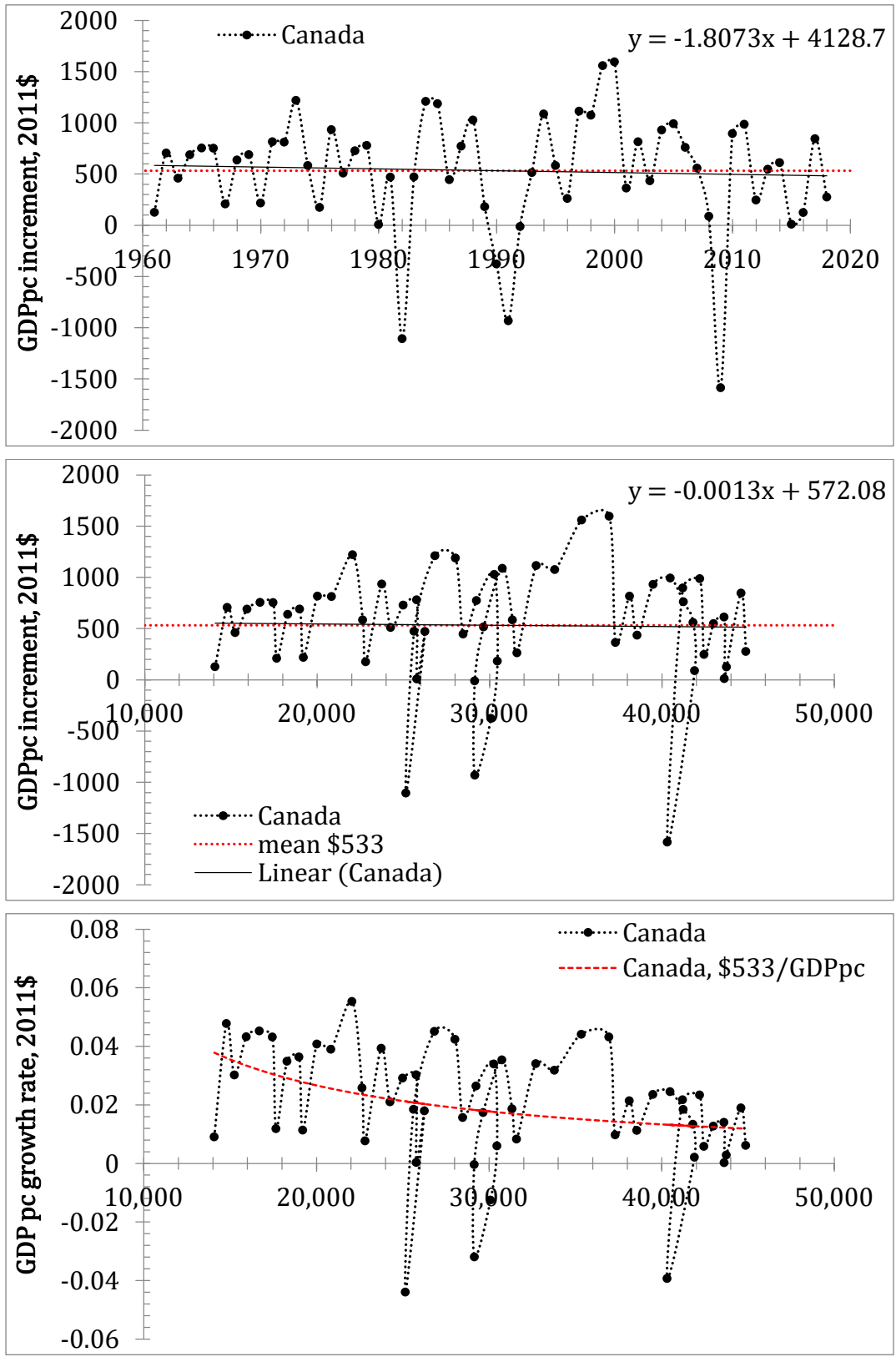

Figure 17. Same as in Figure 14 for Canada.

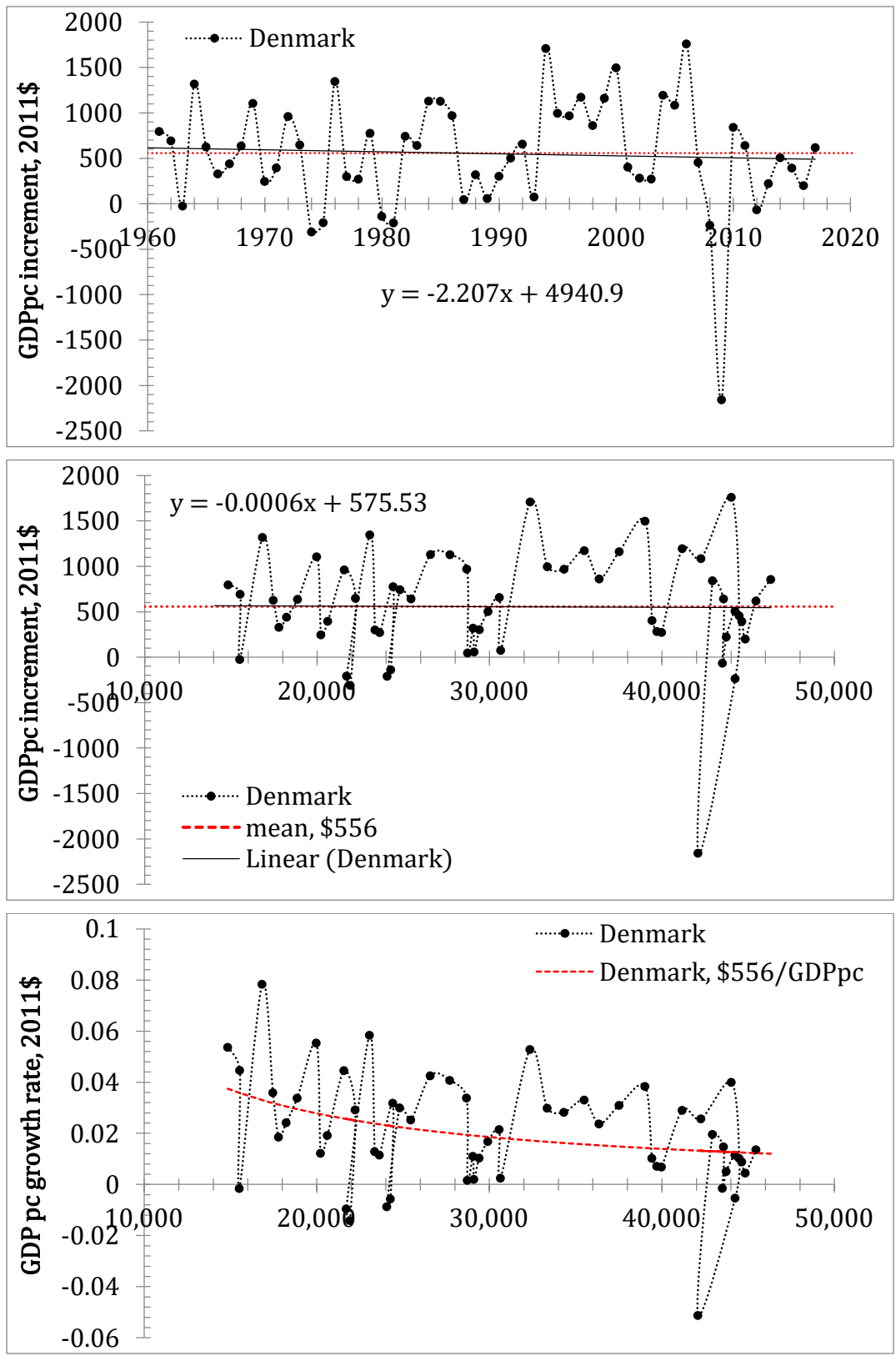

Figure 18. Same as in Figure 14 for Denmark.

The GDPpc level in Denmark (Figure 18) in 1960 was $14,046 and the annual increment between 1960 and 2018 is $556±$610. This is a successful country with high start value and healthy annual growth. In their 2006 and 2012 versions, Austria, Belgium, Canada, and Denmark all had positive slopes in the time regression lines. In 2018, the slope is negative. We assumed the possibility of this evolution as the result of economic forces returning the growth rate to the constant annual GDPpc increment. The negative deviations are likely to be compensated in the

future by a more intensive growth. There are non-economic forces, however, which are able to suppress economic development in the long run.

One of several countries with longer periods of slower economic growth is France. In Figure 19 we see that the trend is negative (slope is -6.65 $/year). This negative trend was also reported in 2006 and 2012. One can suggest that the underperformance in France likely started in the beginning of the 21$^{st}$ century. The only major reason for the extended period of slow growth might be related to the EU extension and the leading role of Germany in financial and industrial aspects of the EU overall development. There were no other specific economic or non-economic events or processes. The average annual GDPpc increment in France between 1960 and 2018 is $461±$388 (the lowest standard deviation among the countries studied in this Section). Between 2001 and 2018, the average increment is only $284±$483. This is a discouraging result – only 52% of the average annual growth between 1961 and 2000.

The OECD estimates for France are slightly larger than those reported by the MPD. However, the proportions between the periods before and after 2000 given by the OECD are worse: $253±$459 for the segment between 2001 and 2018. This is only 45% of the average annual increment between 1961 and 2000: $565±$326. The Total Economy Database provides real GDP per capita estimated with 2019 as a reference year, and thus, the average values are larger: $678±383 and $319±$576 before and after 2000, respectively. The ratio is 0.47. With the COVID-19 fall, the first quarter of the 21$^{st}$ century is an economic disaster for France.

The next case is about the extremely poor economic performance of Italy. Figure 20 illustrates the shift to slower growth around the year of 2000 with the negative readings in 2008, 2009, and 2012 to 2014. Figure 1 in Section 2 also shows that the GDPpc level in 2018 is lower than the peak value observed in 2007. The length and extent of this failure has not been observed in the data for Italy after 1960. The average annual increment is $430±$546, and since 2001: $92±$740. The average annual increment of $92 is only 15% of that observed in Italy between 1961 and 2000. The scattering is much larger than in France after 2000. The OECD and TED estimates (see Figure 7) for the period between 2001 and 2018 are both negative -$43±$727 and -$61±$926, respectively. The OECD does not report the GDPpc before 1970 and it is not possible to compare the estimates for the period between 1961 and 2000. The regression line in the upper panel in Figure 20 has a negative slope and the annual increment decreases by $10.1 per year on average. The dependence of the GDPpc growth rate (lower panel) is characterized by loops related by negative increments, and the cycle of deep falls and quick rebounds generate high scattering as measured in the most recent period.

Japan is well known for the extraordinary fast growth before the 1990s and dramatic slowdown since 1991. Figure 21 illustrates this behavior and the negative slope (-11.2 $/year) of the regression line manifests the 1991 transition. Japan is still in the low performance regime and the average of $557 between 1961 and 2018 will decrease in the future. The decrease in the average value will also affect the inertial growth line in the lower panel – it has to shift down.

The Kingdom of the Netherlands is a representative of the mid-size West-European set of economies. Austria and Denmark are two typical examples. The Netherlands, however, still have a positive slope of the regression line shown in Figure 22 and very high average annual increment: $591±$582. In the 21$^{st}$ century, the average increment is $592±$792, i.e. the data scattering is higher than for the period before 2000. As we suggested in 2006, any deviation should return to the constant annual increment, and the Netherlands will likely have some years of slow growth in the future. In 2006, the positive slope was much larger and the Great Recession moved the regression line close to the mean annual increment line.

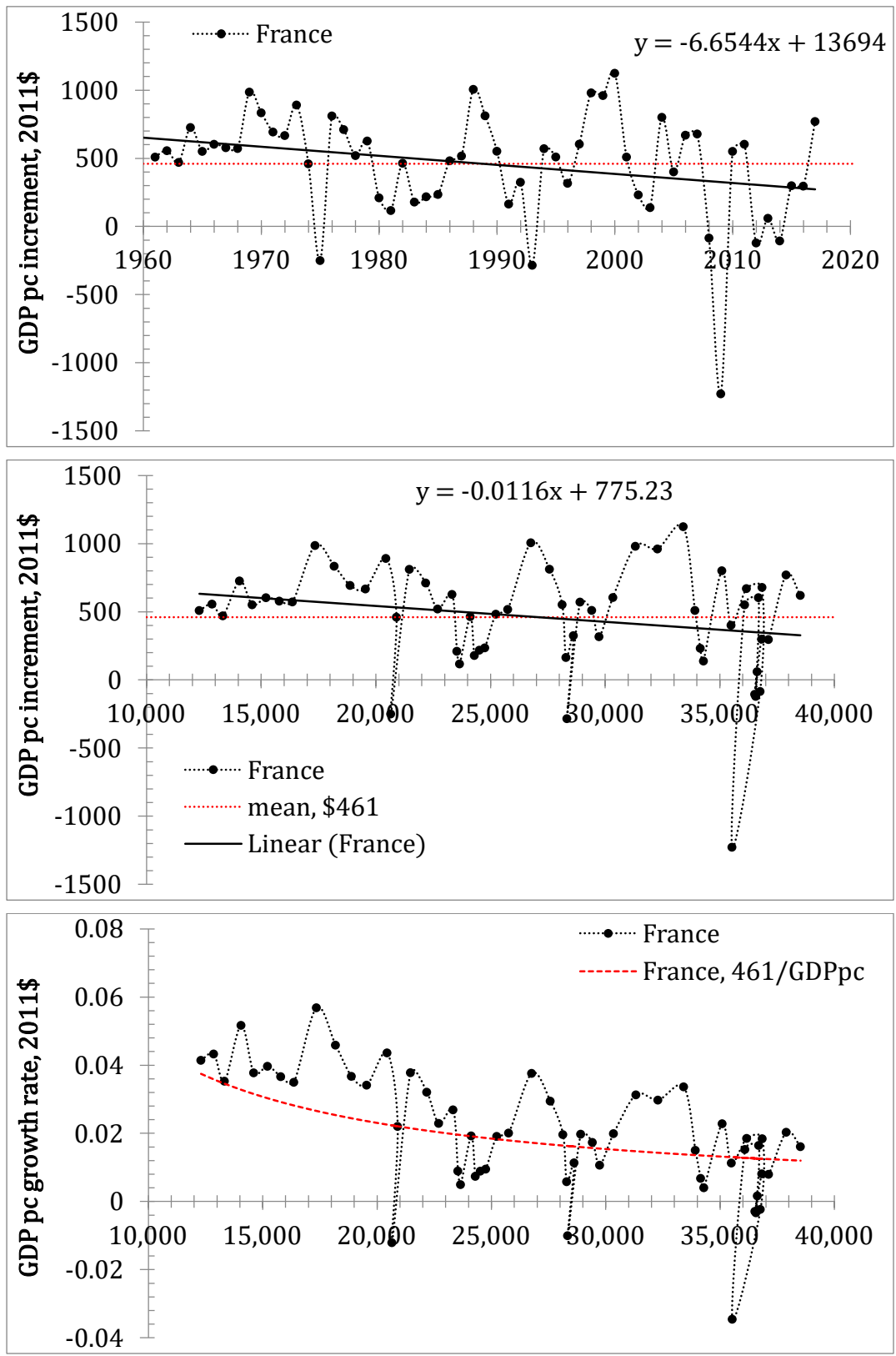

Figure 19. France. The upper panel: the annual GDPpc increment between 1961 and 2018 with the average value for the studied period of $461 (2011 prices). The middle panel: the same annual increment as a function of GDPpc level. The lower panel: the relative growth rate of the GDPpc as a function of the GDPpc.

Spain is a large European economy (#5) and it has the features similar to Italy and France: the average increment $456±$567 for the whole period since 1961, and a much lower value since 2001: $250±$827, i.e. 55% of that before 2000. The slope is negative (-4.1 $/year). There is a

significant difference, however. The initial value in 1960 was $5,037. France had $11,792 and Italy - $9,430. The initial value is defining for the growth rate, and the approximately same annual increment in France and Spain doubles the growth rate in the latter. In terms of rate, Spain has been growing much faster than France and Italy since 1960. The Spainish economy grew between 1960 and 2018 by a factor of 6.25 (real GDP per capita), while Italy grew by a factor of 3.64, and France by 3.27. Nevertheless, the GDPpc gap with France grew from $6,755 in 1960 to $7,019 in 2018.

Sweden is a successful country in terms of real GDPpc. In Figure 24, the annual increment time series oscillates around the mean value $546±$697. The largest negative spike (-$2,074) was observed in 2009, and the largest positive increment ($2518) was measured in 2010, with the total amplitude of change of $4,592 within two years. The regression line has a slightly positive slope (3.56 $/year). The period after 2000 was even more successful for Sweden ($630±$1,034) than the years between 1960 and 2000 ($509±$490). Sweden is likely a beneficiary of the EU.

Figure 25 presents Switzerland which has the largest annual GDPpc increment since 1960 ($776±$729) among the countries in this Section. The 21st century was more successful for Switzerland ($1007±$982) than the years between 1960 and 2000 ($672±$567). Therefore, the time series has a positive trend of 9.82 $/year, but even with this trend the inertial growth model is likely right and the COVID-19 pandemic will bring the regression line closer to the mean line.

When testing a hypothesis, one mandatory step is to test the extreme cases. The next four countries include 3 largest developed economies (USA, Germany, and the UK) and a small (1.1% of the US) and geographically isolated economy – New Zealand. Germany is a challenge for statistical analysis – there was a significant non-economic event – the reunification in 1991. Merging data from two pieces with different levels of economic development is an art rather than science.

The United Kingdom in Figure 26 is characterized by a low slope (0.5 $/year) of the regression line. The mean income ($419±$486) line almost coincide with the regression line validating the assumption of inertial growth. As in other developed countries, there are periods of excellent performance (e.g., 1993 to 2007) and a few recessions. The Great Recession gave two consecutive years with negative annual increments – 2008 (-$552) and 2009 (-$1,862). The growth rate theoretical curve in the lower panel evidences that the rate of GDPpc growth has been decreasing since 1960. The average increment in the 21st century is $340±$639, and the years between 1961 and 2000 are characterized by the mean annual increment $454±$400. The scattering before 2000 is much lower than after 2000 as related to the large amplitude spikes during the Great Recession. The ratio of the mean values is 0.75 (0.8 for TED) is the result of the recession rather than subordinate position in the EU. It is instructive to analyze economic consequences of Brexit in the next decades.

The USA (Figure 27) is the biggest developed economy and its statistical results are likely the most robust. The average annual increment in the USA between 1961 and 2018 is $643±$650, i.e. the largest among the studied developed countries except Switzerland. The corresponding standard deviation is very close to the mean value. The performance in the 21st century ($525±$755) is approximately 75% of that between 1960 and 2000 ($696±$599). The TED estimates this ratio as 99%. The regression line in the upper panel in Figure 27 has a weak slope (1.0 $/year) and is almost indistinguishable from the mean value line (dashed red). The evolution of the US economy was better described by a linear rather than exponential function of time since 1960. The exponential growth in real GDP during the same period is the result of the exponential population growth. The exogenous economic growth related to population has strict limits and this component has lower and lower influence with time: in the 21st century the US

population growth rate was 0.8% per year on average with lower rates in the 2010s. Between 1960 and 2000, the population growth was 1.1% on average.

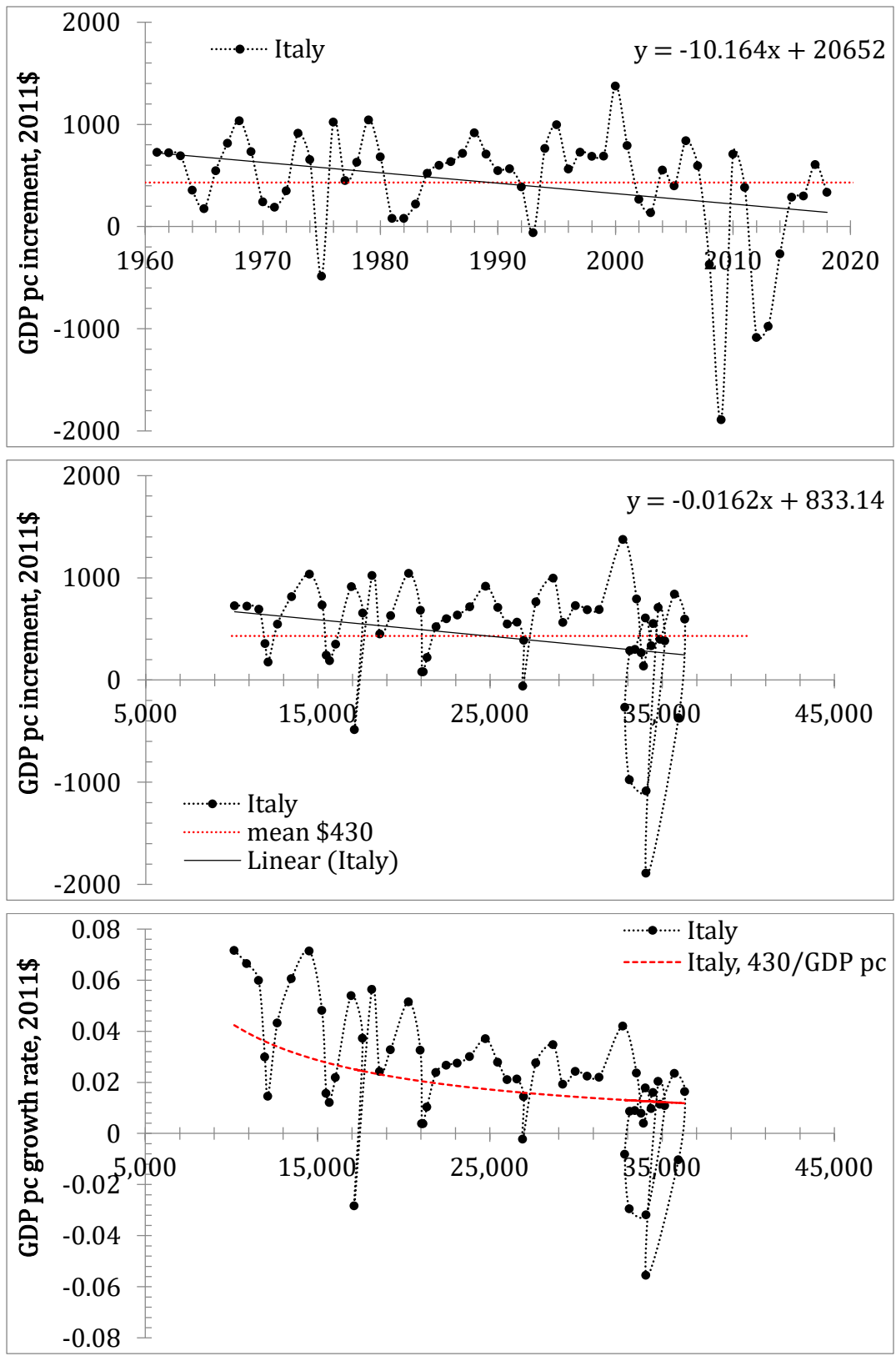

Figure 20. Same as in Figure 14 for Italy.

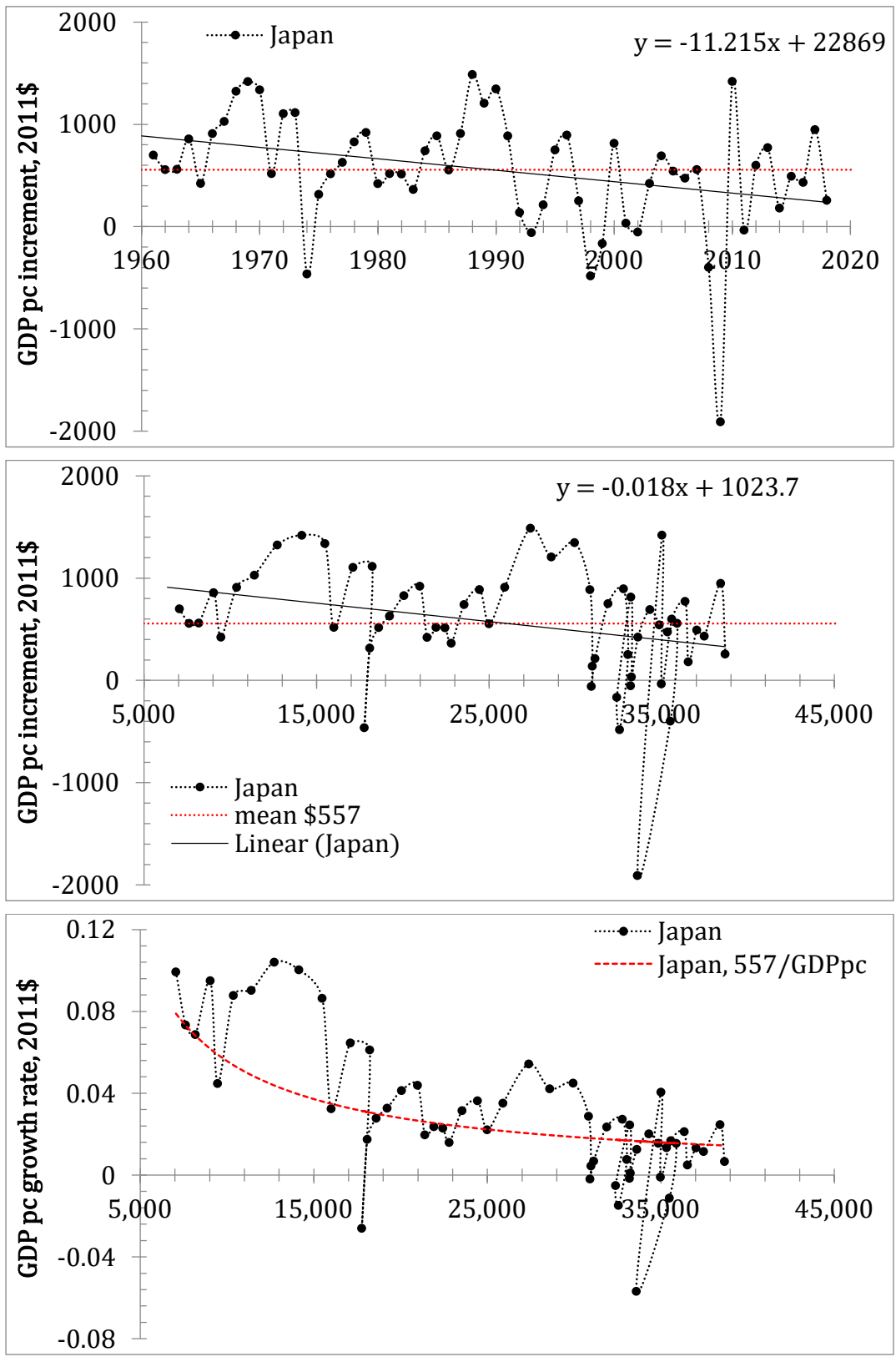

Figure 21. Same as in Figure 14 for Japan.

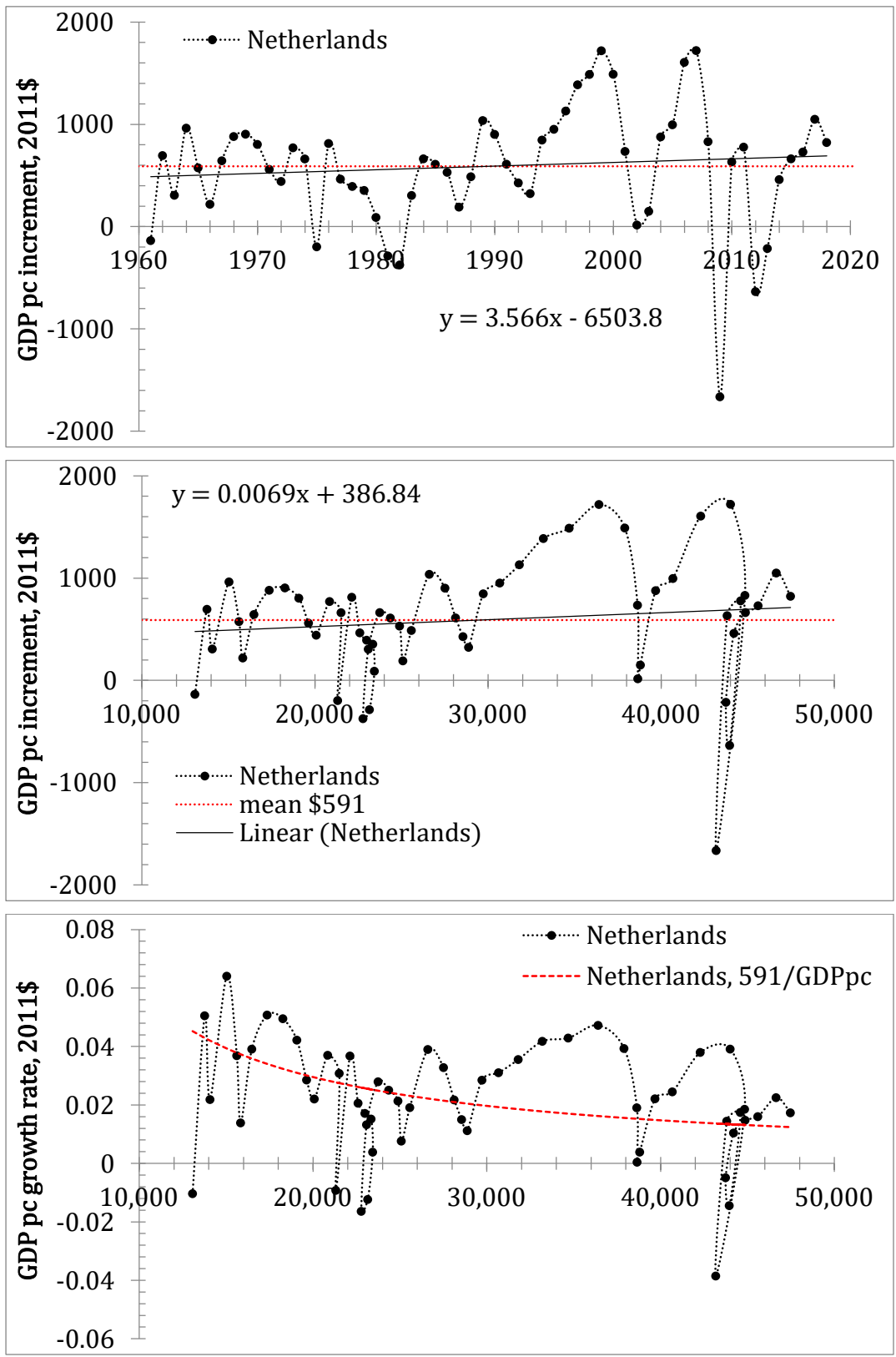

Figure 22. Same as in Figure 14 for Netherlands.

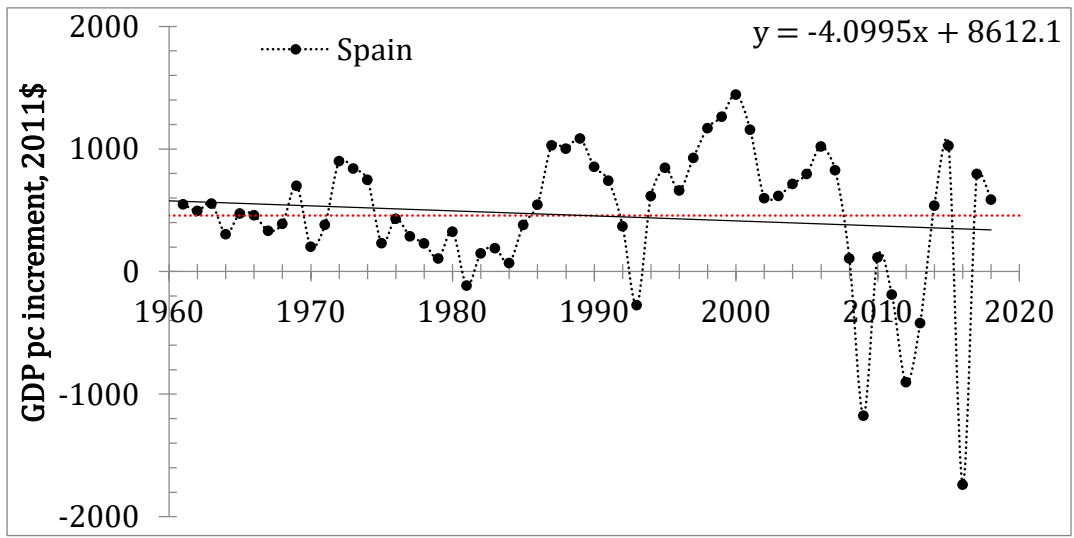

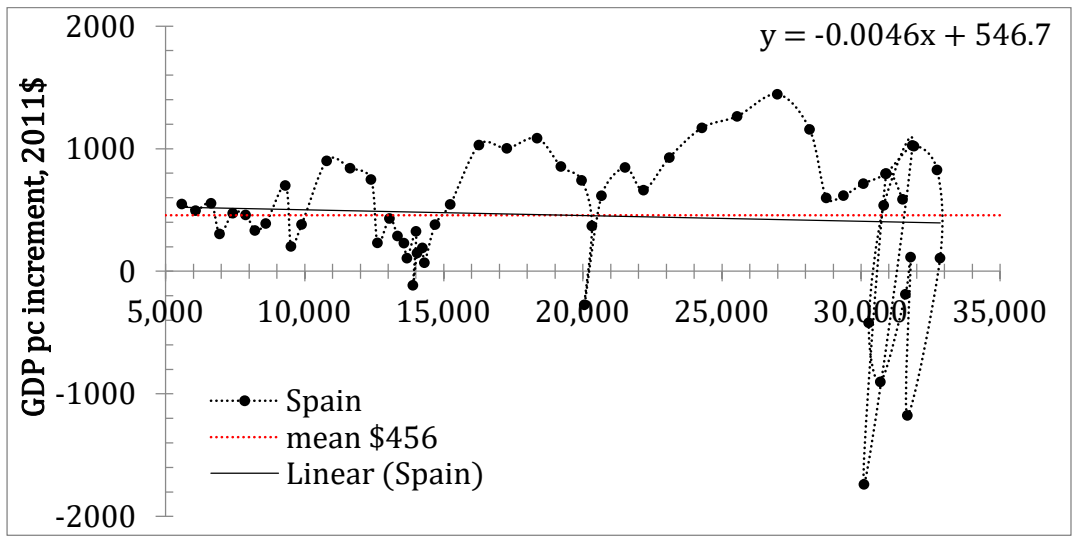

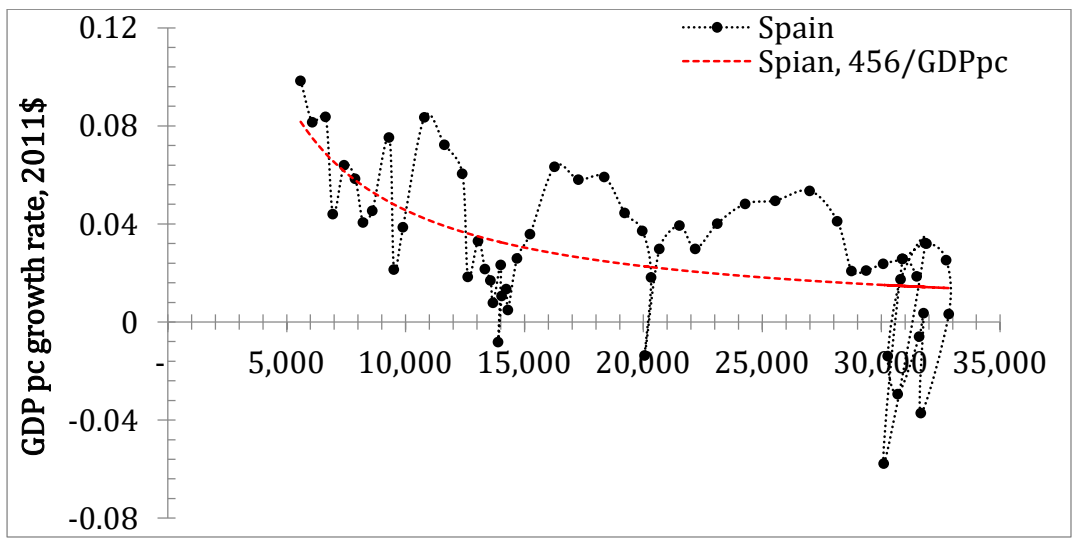

Figure 23. Same as in Figure 14 for Spain.

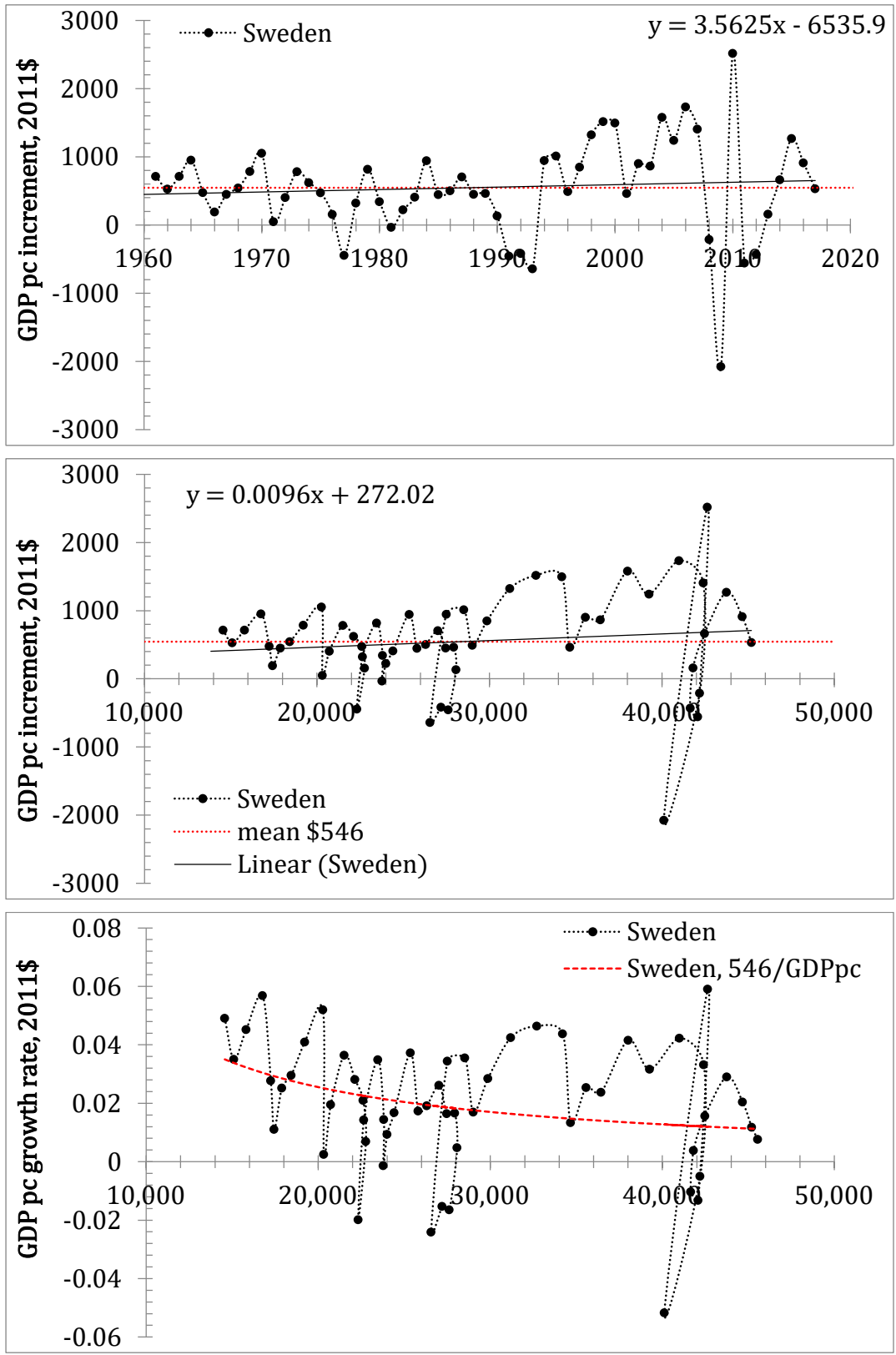

Figure 24. Same as in Figure 14 for Sweden.

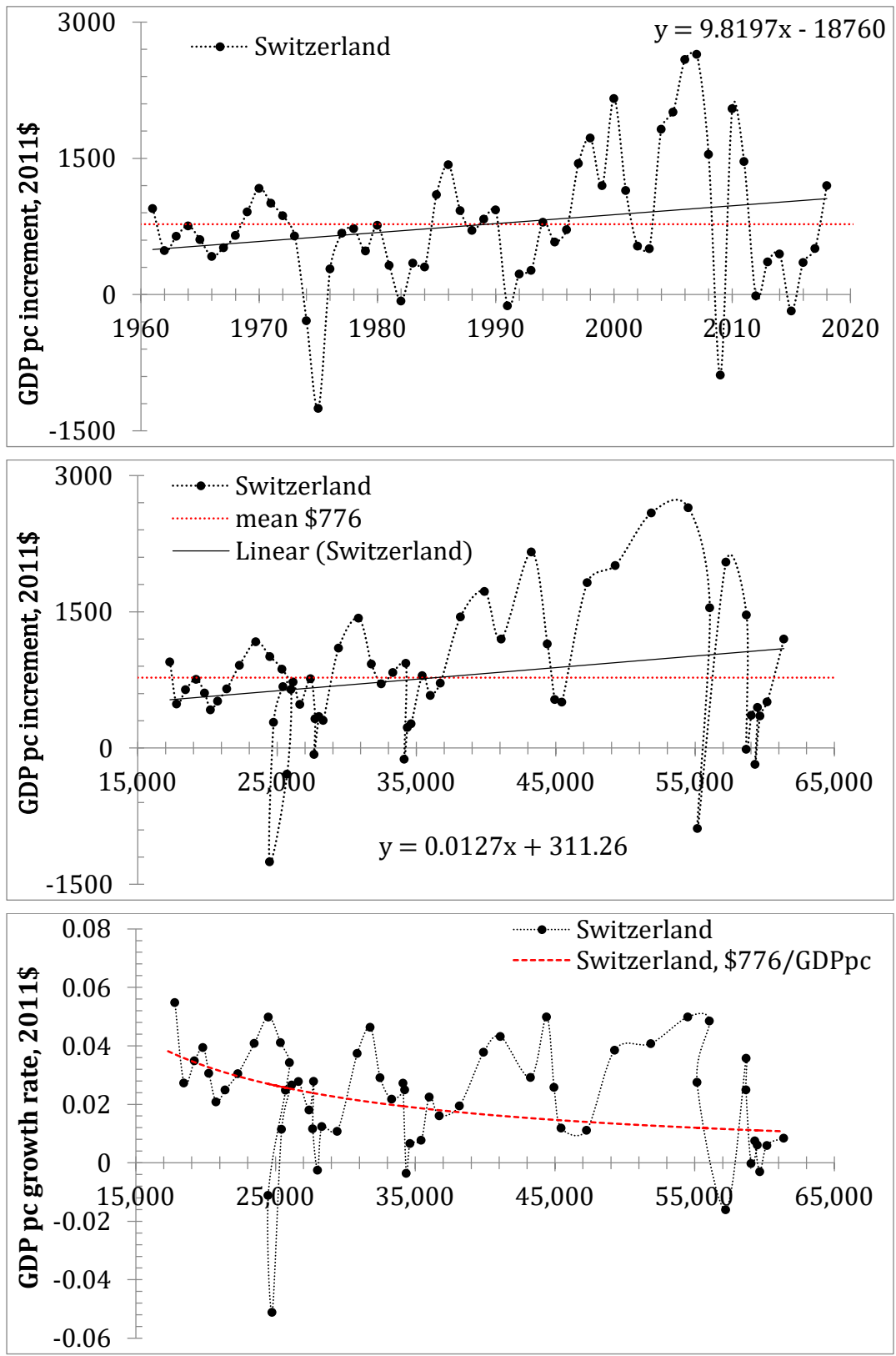

Figure 25. Same as in Figure 14 for Switzerland

Germany (Figure 28) is an example of excellent economic performance in the 21[st] century with the mean annual increment of the GDP per capita of $712±$914. (The Great Recession was expressed by one negative year with a GDPpc increment of -$1,752.) With this growth, the regression line in the upper panel has a significant positive slope of 6.42 $/year. The period before 2000 is characterized by a lower mean income $527±$430. The improvement in economic performance since 2000 relative to the pre-2000 period is 135%. This is the highest growth among the countries in this Section. The mean income for the whole period is \$584±\$615. After the Euro zone was organized with the ECB in Frankfurt, Germany has been demonstrating enhanced economic performance.

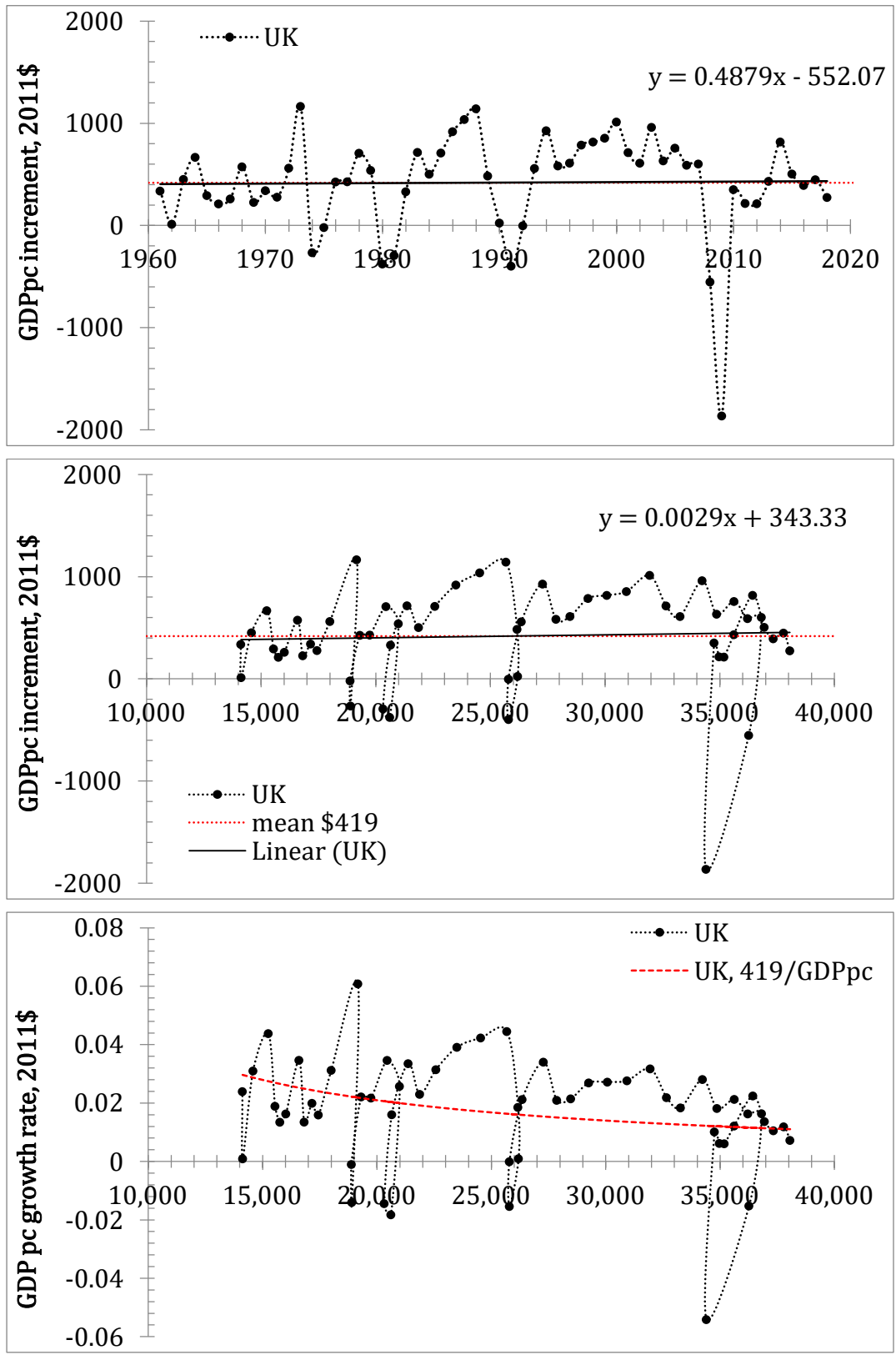

Figure 26. Same as in Figure 14 for the UK.

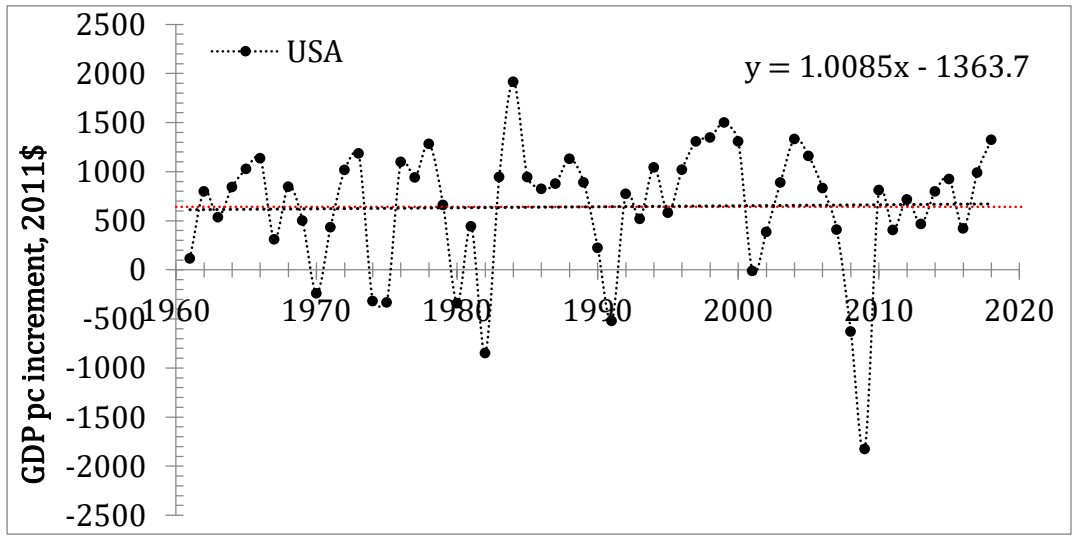

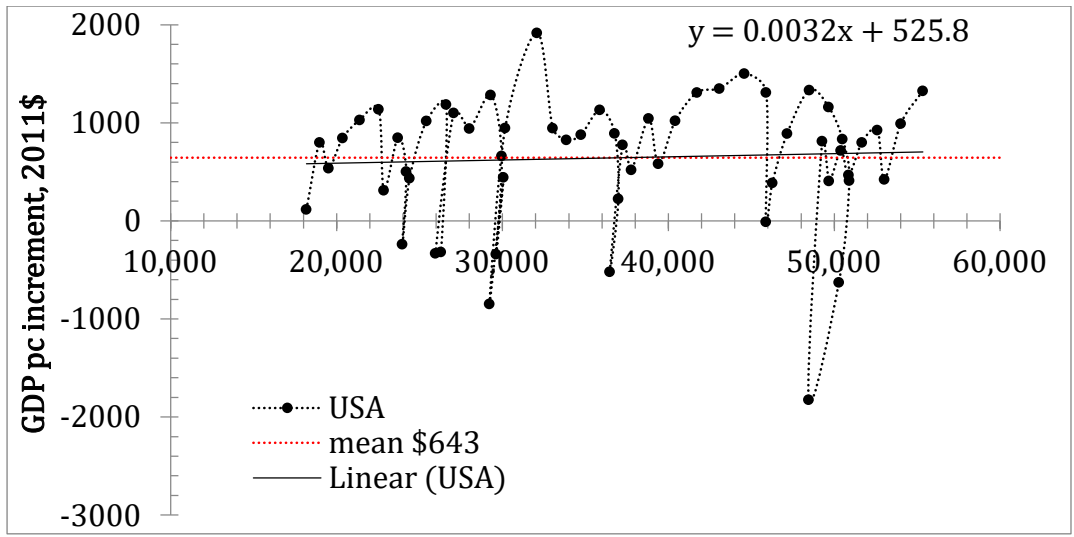

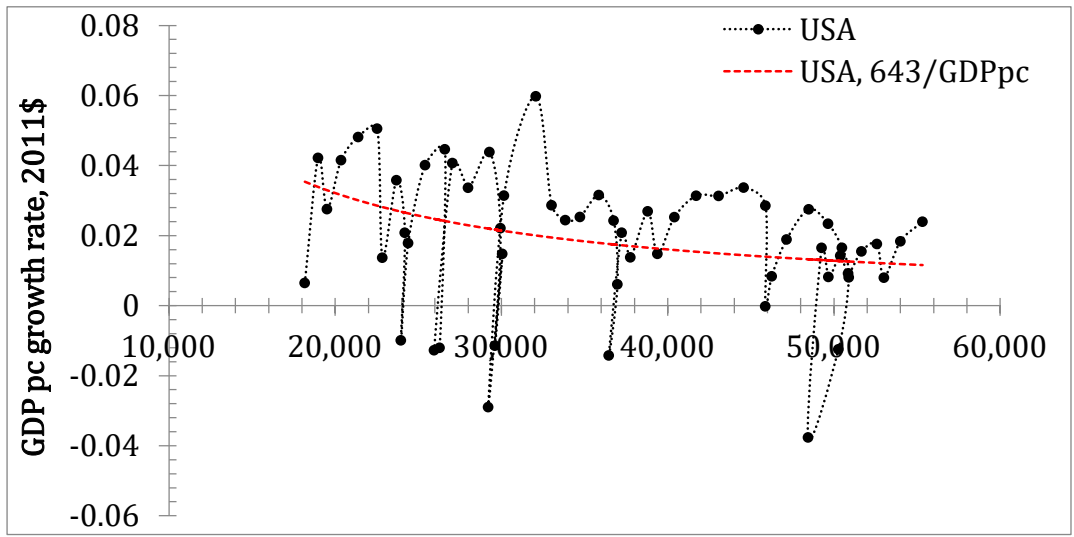

Figure 27. Same as in Figure 14 for the USA.

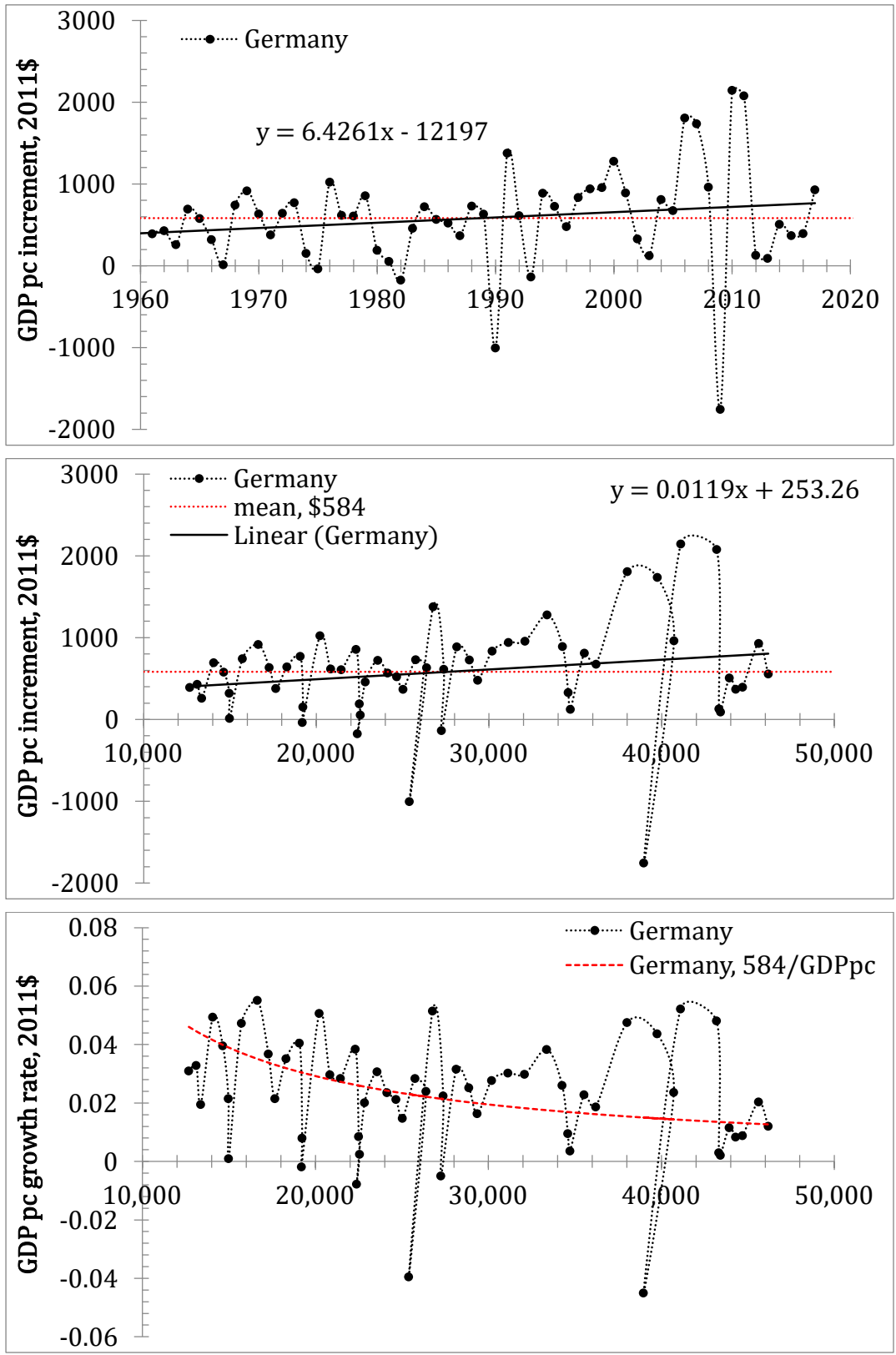

Figure 28. Same as in Figure 14 for Germany. The period before 1990, i.e. before the reunification, is characterized by much lower annual GDPpc increment. Statistics might be slightly biased.

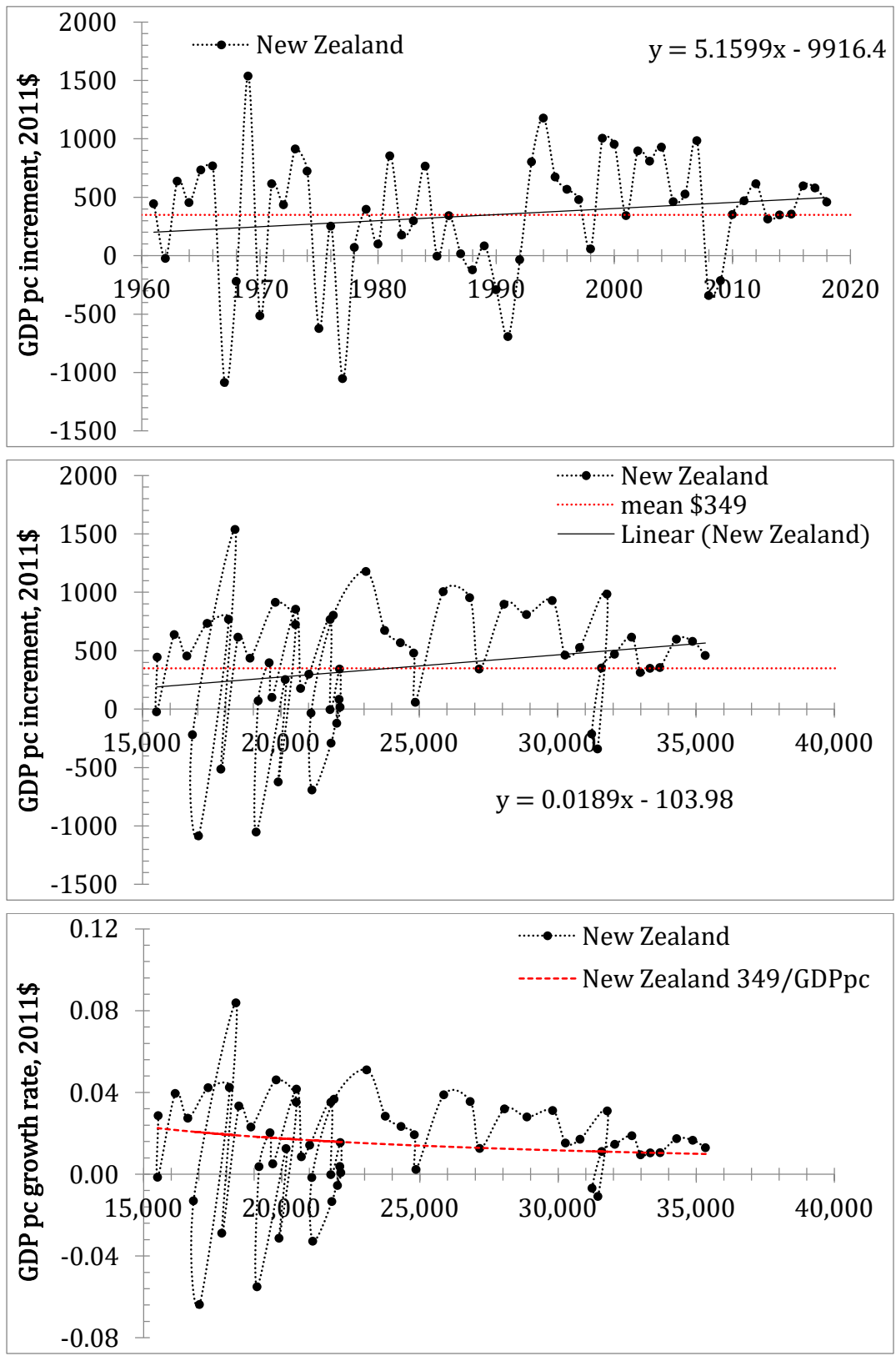

Figure 29. Same as in Figure 14 for the New Zealand.

New Zealand (Figure 29) is an example opposite to Germany – the growth between 1960 and 2018 is characterized by the mean $349±$523. The period between 1960 and 2000 was accompanied by several recessions – eight negative annual increments. The 21st century is slightly more successful with two poor years during the Great Recession. The slope of the regression line is positive 5.2 $/year. The initial GDP per capita in 1960 was $15,087, i.e. close to the top among the developed countries with USA and Switzerland having higher initial values (Table 1). The low mean annual increment with the high initial value makes the rate of growth very low for New Zealand as the lower panel in Figure 29 show.

Overall, the concept of constant annual GDPpc increment is supported by the MPD data for the developed countries. In the long run, the biggest (and complete) developed economies demonstrate a decaying rate of economic growth as expressed by our model of inertial economic growth of the real GDP per capita. At the same time, these countries are characterized by quite different economic performance as represented by the mean annual increment. For example, the average annual GDPpc increment in the USA is $643 per year (2011 US$) between 1960 and 2018, i.e. every US citizen virtually gets every year by $643 dollars more than in the previous year. In France, the average increment is only $461 and, in the long run, the US citizens become richer than the French citizens. This income gap increases with time but the growth rate of the GDP per capita may be larger in France because the base (GDPpc level) is lower than in the USA. The difference between actual increment and growth rate may deceive French people, who may be confused by the rate as the only indicator of real growth.

The European Central Bank may cause an uneven reaction of the previously independent states with various economic measures against inflation and slow growth rate. Germany significantly increased its annual increment in real GDP per capita since the mid-1990s. Spain, Italy and France suffer painful decrease in the annual increment in the 21$^{st}$ century. There is no direct evidence that the EU participation is the only cause of the drop in economic performance but the difference between the periods before and after 2000 makes the Economic and Monetary Union the main suspect. The United Kingdom also experienced a tangible slowdown (75% relative to the before-2000 period) relative to Germany (135%), but this effect might also be related to the Great Recession: the fall in the UK is similar to that in the USA.

## 6. Economies with the highest growth rate

The MPD gives many examples of poor economic performance, which can be explained by various reasons and causes. These reasons are usually very convincing. In this Section, we present the results of real GDP per capita analysis for nine small and mid-size economies demonstrating extremely high economic performance: Norway, Ireland, South Korea, Taiwan, Hong Kong, Singapore, Saudi Arabia, Bahrain and Qatar. The reasons of such excellent performance are less self-explanatory. We are not going to discuss these reasons here and just present data and the overall misfit with our model. These countries are small or have very specific economic activity and do not influence the global economy. One can consider them as exotic and insignificant outliers.

Norway is presented in Figure 30. The years before 1980 were not too successful, but show progressive increase in the real GDP per capita. The 1980s were turbulent and ended by a jump from -$167 in 1988 to $3,423 in 1997. The most successful period was observed between 1997 and 2007, which mainly define the extremely high mean annual increment between 1960 and 2018 - $1,260±$1074, i.e. by a factor of 2 larger than in the USA. The Great Recession had a minor effect with only one poor year -2009. Interestingly, the worst year in the history since 1960 was 2013 with the increment of -$176. The 2010s were closer to the 1960s in terms of economic growth and the future evolution is likely linked to the oil price, which is the main driver of the Norwegian economic miracle. Other industries and services were not able to compensate the oil price fall after 2008. The regression line in Figure 30 has a large positive slope and deviates from the mean increment line. Such a discrepancy is hardly expected in large economies where a deep fall in one activity is compensated by slower growth in many others.

Norway (GDP~$350 billion in 2019) is approximately of the same size as New Zealand (~$250 billion), but the latter has no economic advantages like oil or gas.

In Figure 31, the case of Ireland is presented with the mean annual increment of $998±$1329. The principal difference with Norway is the deep fall in 2008 and 2009 as related to the Great Recession and the return to the extremely high growth in 2014. This means that Ireland is tightly related to the growth in the largest economies as providing some specific services. In that sense, Ireland is not a price setter and its future depends on the economic development. The growth trend is positive and it is rather defined by the poor 1960s and a successful period between 1995 and 2006 as well as the most recent period.

South Korea (Figure 32) started its economic conquest in the 1980s and has been showing the growth rate above the average ($627±$537). The average increment is not larger than in the USA, but the initial conditions in 1960 ($1,548) and the level in 2018 ($37,928) make South Korea one of the most successful countries in the studied period. The shift to a faster growth in the 1980s divides the whole period into two segments below and above the mean increment line. The regression line slope is positive and expresses the success in the most recent segment. One cannot exclude that the future growth will be closer to the mean increment line.

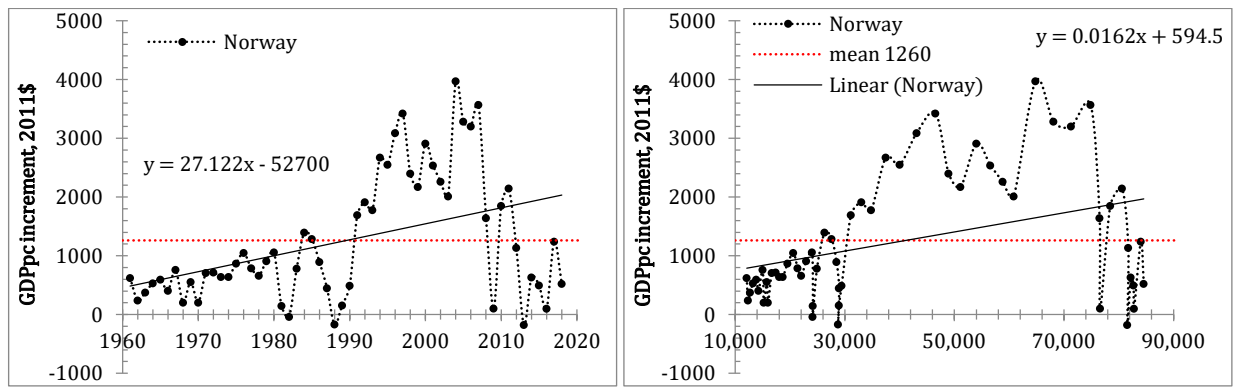

Figure 30. Norway. Left panel: the annual GDPpc increment between 1961 and 2018 with the average value for the studied period of $1260 (2011 US$). Right panel: the same annual increment as a function of GDPpc level.

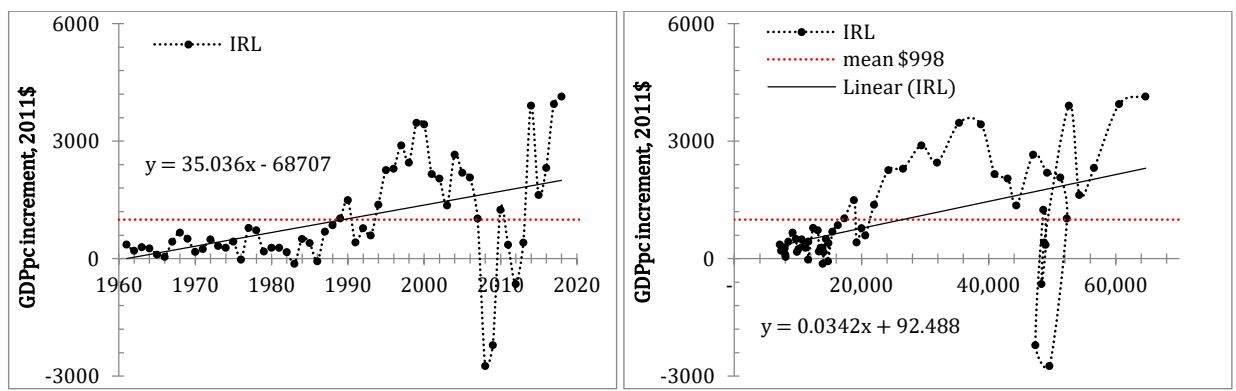

Figure 31. Same as in Figure 30 for Ireland.

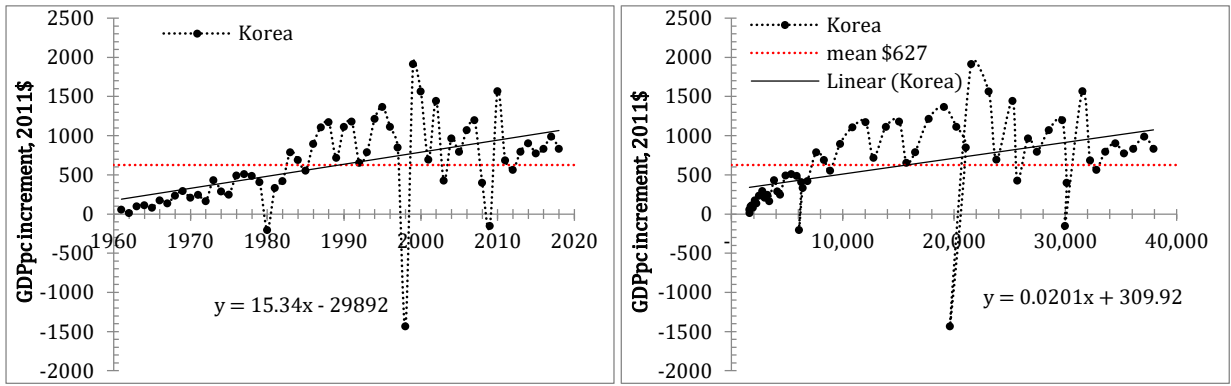

Figure 32. Same as in Figure 30 for South Korea.

The evolution of the real GDPpc in Taiwan is similar to that in Singapore as Figures 33 and 34 show. The main difference is a lower mean increment in Taiwan is $733±$655 than in Singapore has $1120±$1309. As a result, in 2018 the real GDP per capita in Taiwan was $44,664 and in Singapore - $68,402. Hong Kong in Figure 35 is an example of a mid-size developed country like Sweden with a positive slope in the regression line 8.4 $/year and a mean increment $789± $883.

Saudi Arabia and Qatar in Figures 36 and 37 are two extremely successful economies both driven by oil production, and thus, demonstrating extreme fluctuations in the annual GDPpc increment. As a bigger economy, Saudi Arabia shows a lower positive slope of the regression line 17.5 $/year with 150 $/year in Qatar. At the same time, the shapes of the increment curves are rather similar and tightly related to the oil market.

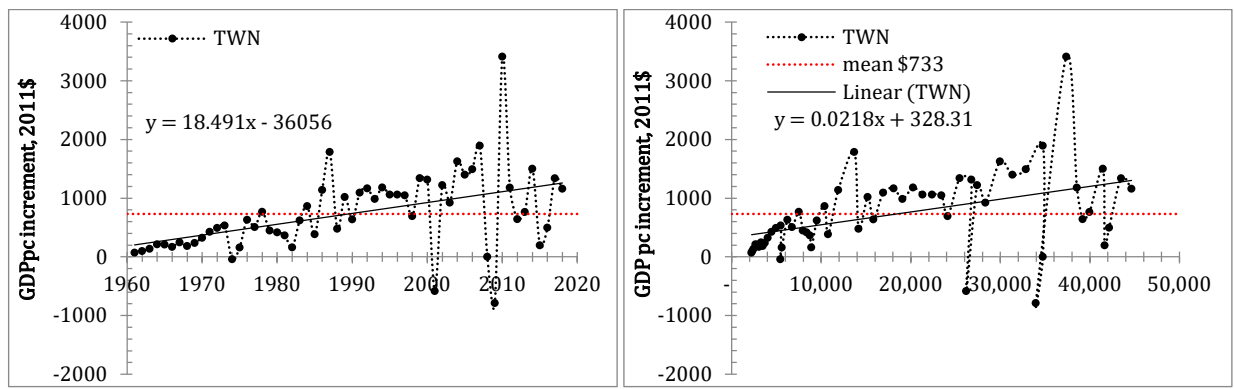

Figure 33. Same as in Figure 30 for Taiwan.

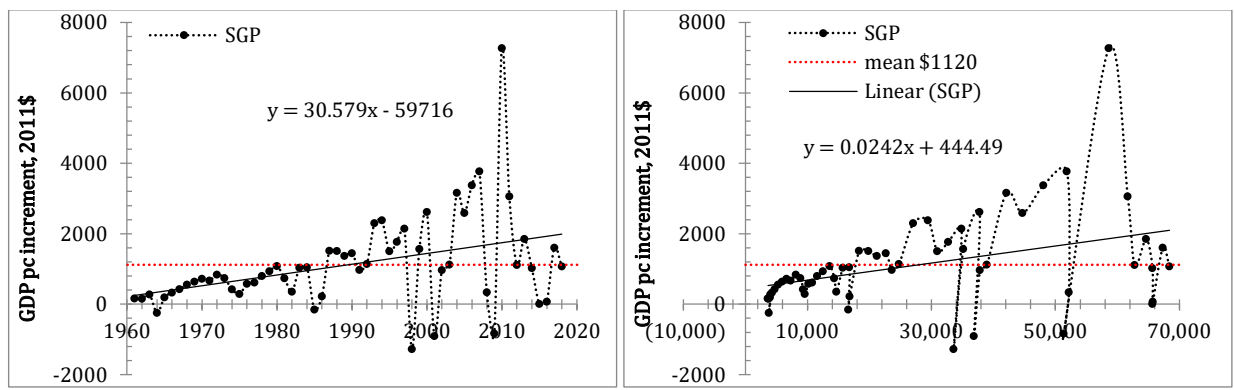

Figure 34. Same as in Figure 30 for Singapore

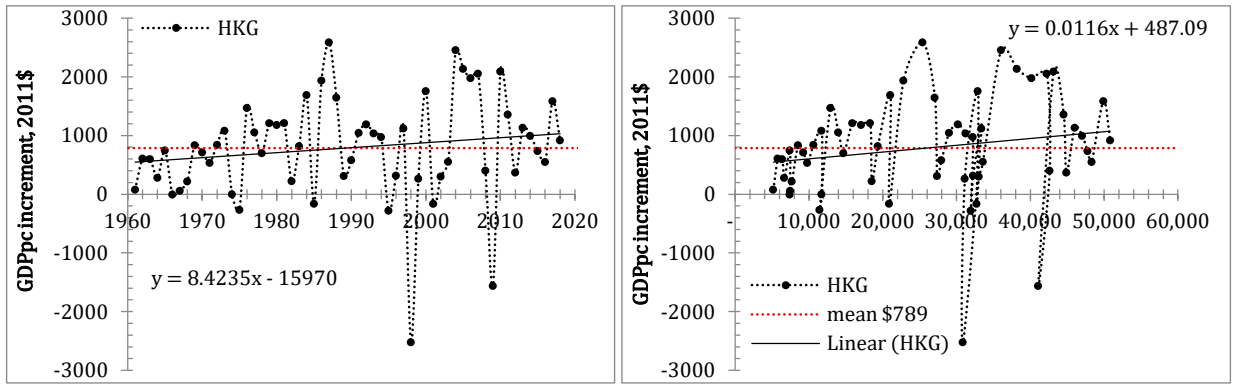

Figure 35. Same as in Figure 30 for Hong Kong.

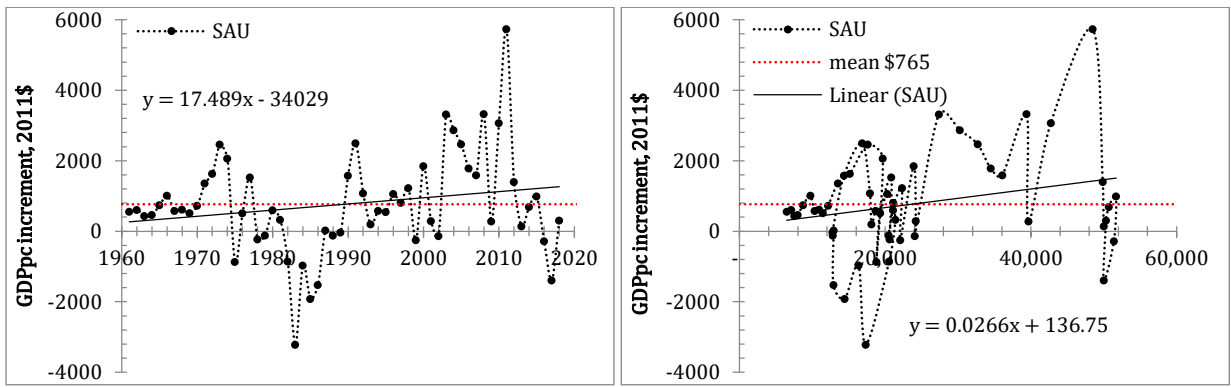

Figure 36. Same as in Figure 30 for Saudi Arabia.

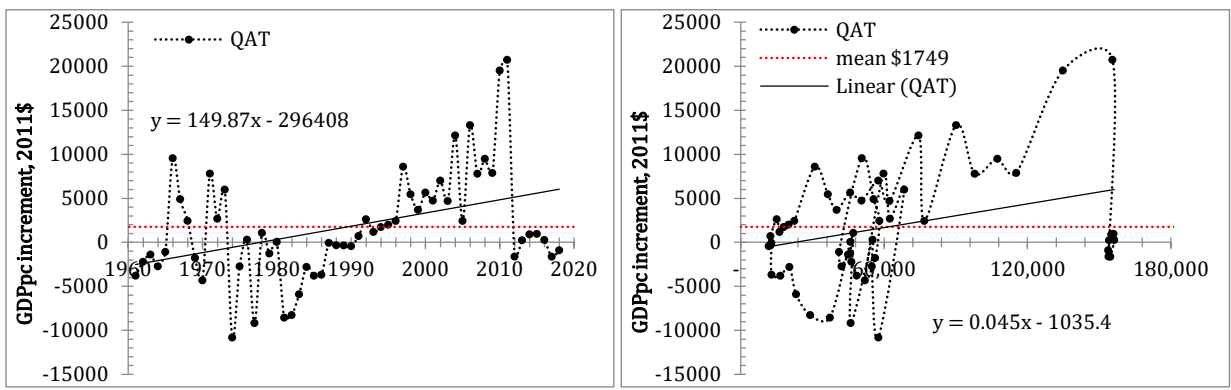

Figure 37. Same as in Figure 30 for Qatar.

These best performers would not be able to perform so well by themselves. They are important parts matching the specific demand from the full diversity created by the largest economies. For example, with the change in demand priorities, *e.g.* transition to green energy, the faith of oil countries may change dramatically. Financial services are also changing fast and banks now are moving to digital format. With globalization of services and dissolving sovereignty of national states the global business will take form these states only needed components and ignore the other activities and related population.

## 7. BRICS

BRICS was organized in 2006 by Brasilia, Russia, India and China, *i.e.* by four economies from the top 10. South Africa joined in 2009. The idea behind the club was mutual assistance to provide the fastest economic growth. Here, we study the real GDP per capita in these five countries as the principal indicator of economic growth. Brasilia (Figure 38) demonstrates mostly inertial growth with a constant annual increment of $183±$404 per year since 1960. No dramatic fall or surge over the studied period was observed, except the drop of -$1953 in 2016. Participation in BRICS did not give Brasilia any significant economic push with the average annual increment of $210 since 2006 which is compatible with the long-term mean increment.

In Figure 39, the evolution of the annual GDPpc increment in the Russian Federation is presented as a function of time (left panel) and GDPpc (right panel). The annual GDPpc increment between 1960 and 2018 is $330±$682 (2011 US$). This value is lower than in developed countries (New Zealand has $349 for the same period). The period after 1999 was the most successful in the Russian history since 1960 and the average annual growth was $806±$815 between 1999 and 2018 and $648±$908 between 2006 and 2018, *i.e.* the Great Recession included. The mean increment in the 21$^{st}$ century is one of the largest worldwide. The fall in 2015 and 2016 was just a short term deviation and the Russian economy returned to the healthy growth rate (right panel in Figure 39). Two distinct periods in the Russian history since 1960 are related to the change in political and economic concept in 1991. This change, despite the deep fall in the mid-1990s, seems to be a positive one in the long run and we expect that the future development will follow the inertial growth scenario with constant annual increment in the range from $500 to $700. This is possible in cooperation with two RIC countries also experiencing extremely fast growth in the annual increment of GDPpc since the 1990s.

Figure 40 presents the case of India. This is a country with poor population (low GDP per capita) and with a long history of non-equivalent exchange with developed countries (*e.g.*, the UK). On average, the GDP per capita has been increasing by $97±$108 per year since 1960. The annual increment increased from $45 in 1960 to $357 in 2018 in the longest stretch of intensive economic growth. We also expect India to reach a stationary growth regime with the annual GDPpc growth above $500, as the regression line slope of 14 $/year since 2000 suggest. This will move it to the second position in real GDP level in 10 to 15 years.

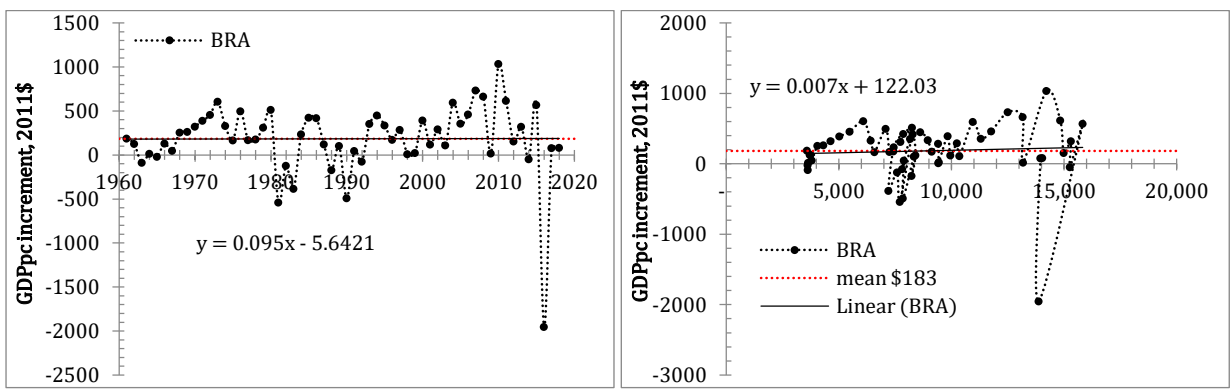

Figure 38. Brasilia. Left panel: the annual GDPpc increment between 1961 and 2018 with the average value for the studied period of $183±$404 (2011 US$). Right panel: the same annual increment as a function of GDPpc level.

The first position in the list of the largest economies (real GDP PPP) belongs to China. The evolution of real GDPpc since 1960 presented in Figure 41. The average increment is $208 (±$219) per year. The years after 2000, were extremely successful for the Chinese economy with the average increment of $498±$204 since 2006, and $454±$189 since 2000. These values are already in the club of developed countries (especially if to remember poor performance of the largest European economies except Germany) and we will not be surprised to find the Chinese economy in the list of best performers worldwide. It is worth noting here that the relative growth rate is the ratio of the increment and the GDPpc level. The latter is much lower for China and the rate of growth has to be much higher than in developed countries with the same GDPpc increment.

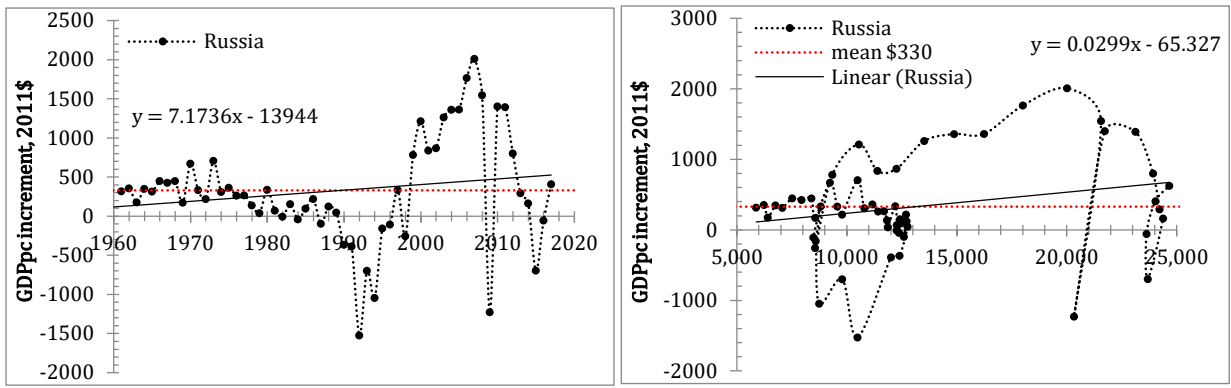

Figure 39. Russia. Left panel: the annual GDPpc increment between 1961 and 2018 with the average value for the studied period of $330 (2011 prices). Right panel: the same annual increment as a function of GDPpc level.

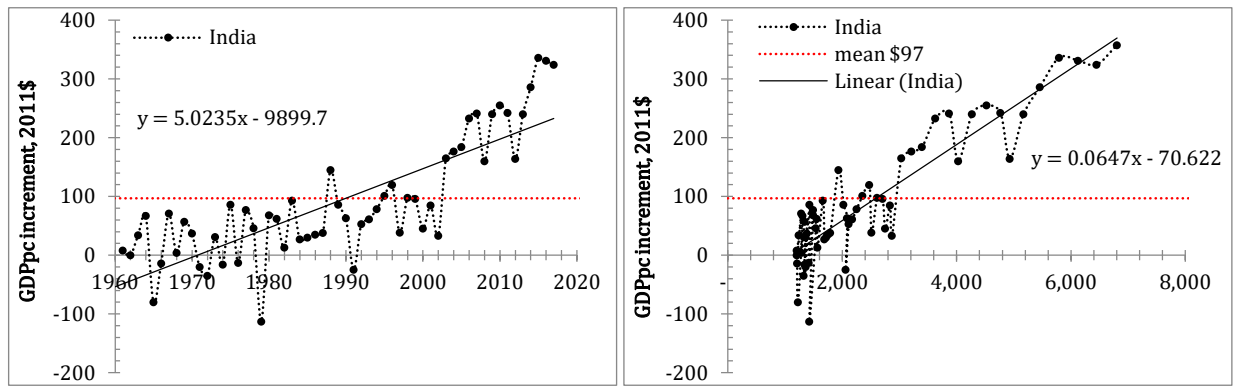

Figure 40. Same as in Figure 38 for India.

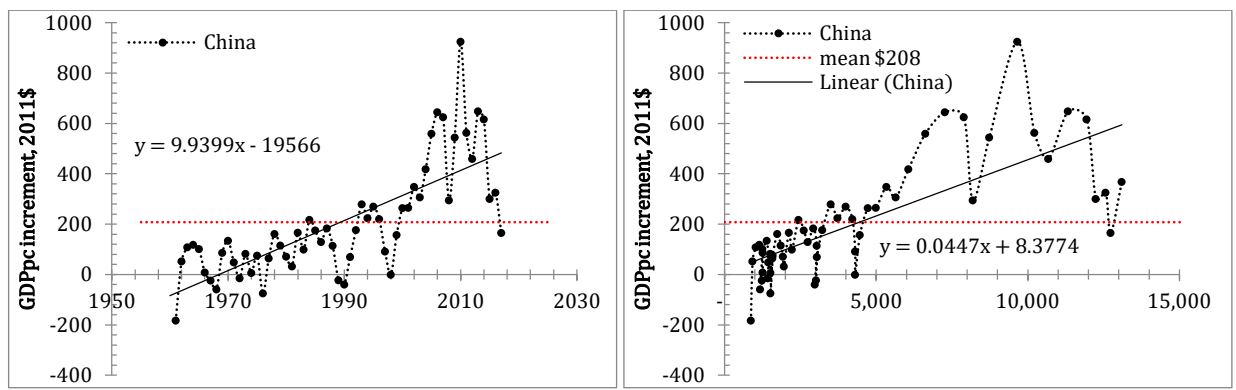

Figure 41. Same as in Figure 38 for China.

Figure 42 presents South Africa. This is the smallest BRICS economy, but it also reveals fast growth since the mid-1990s. The average increment since 1960 is $126±$208 and $255±$262 since 2006. There is a period of underperformance since 2014, which stopped the linear growth segment observed since 2000.

The BRICS countries are all different in many ways – from language to history. They decided to cooperate within the BRICS club because of many different reasons except one – promote the fastest economic growth using all possible methods of cooperation not affecting own interest (i.e. they try to avoid the non-equivalent exchange). It is hard to say which part of the observed economic growth belongs to this cooperation, but the club is still working. We will follow the news about BRICS cooperation and the rate of economic growth.

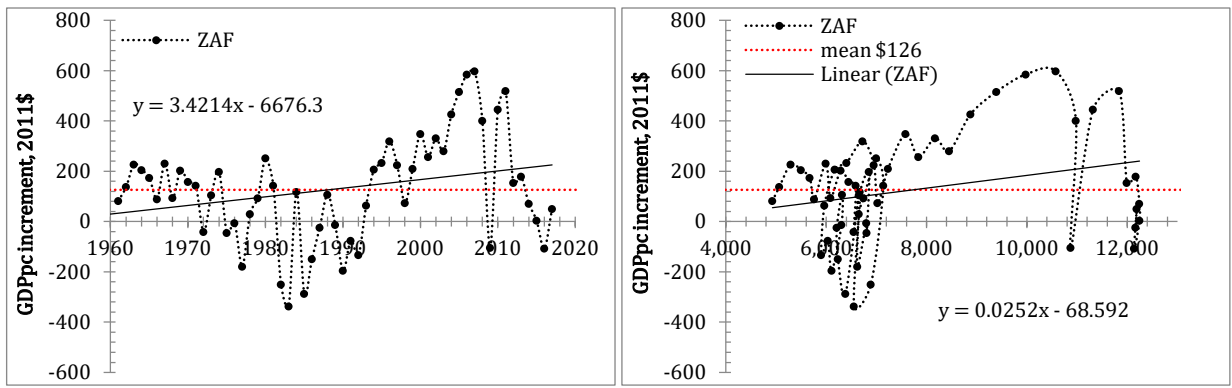

Figure 42. Same as in Figure 38 for South Africa.

## 8. Eastern Europe

In this Section, we present four East European countries: Bulgaria, Hungary, Poland, and Romania. Their behavior and overall performance is similar to those demonstrated by Russia with a break in the GDPpc growth in the 1990s. Romania is likely an outlier with the total GDPpc increase by a factor of 12, which is an example of outstanding performance. It is likely that the initial GDPpc value of $1,605 in 1960 is heavily underestimated considering that neighboring Bulgaria had $4,642 at the same time. We ignore smaller countries as insignificant for the statistics in this study. Czech Republic and Slovakia had a major break and 1991 and their statistics is biased.

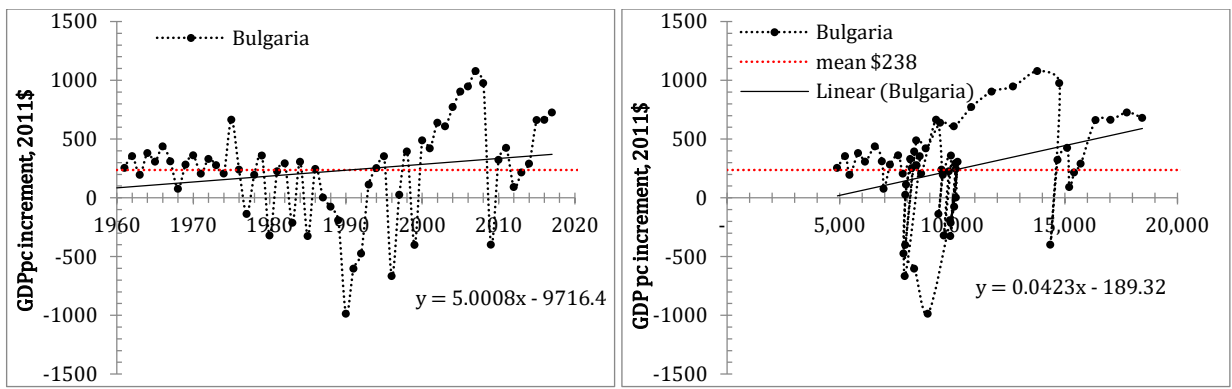

Figure 43. Bulgaria. Left panel: the annual GDPpc increment between 1961 and 2018 with the average value for the studied period of $238 (2011 prices). Right panel: the same annual increment as a function of GDPpc level.

The evolution of the GDPpc is presented in Figures 43 through 46. The mean increments since 1960 in 3 countries is below the lowest in Section 5, and Poland has a larger value, $389 per year, than New Zealand. Since 2000, all countries demonstrate much faster growth: Poland $810±$344 per year, Romania $711±$773 per year, Hungary $700±$524 per year, and Bulgaria $554±$354 per year. The EU support was crucial for the accelerated growth, and the poor development in France, Italy and Spain during the same years may manifest the unbalanced redistribution of financial subventions.

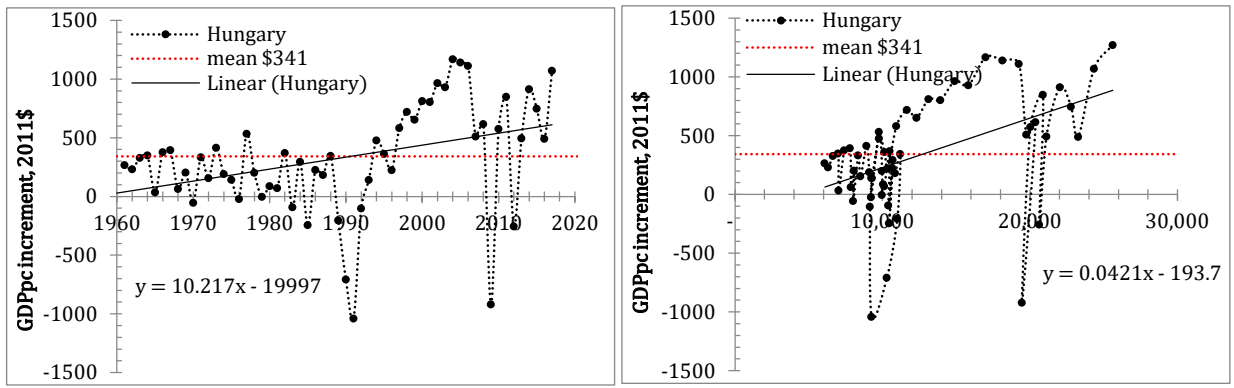

Figure 44. Same as in Figure 43 for Hungary.

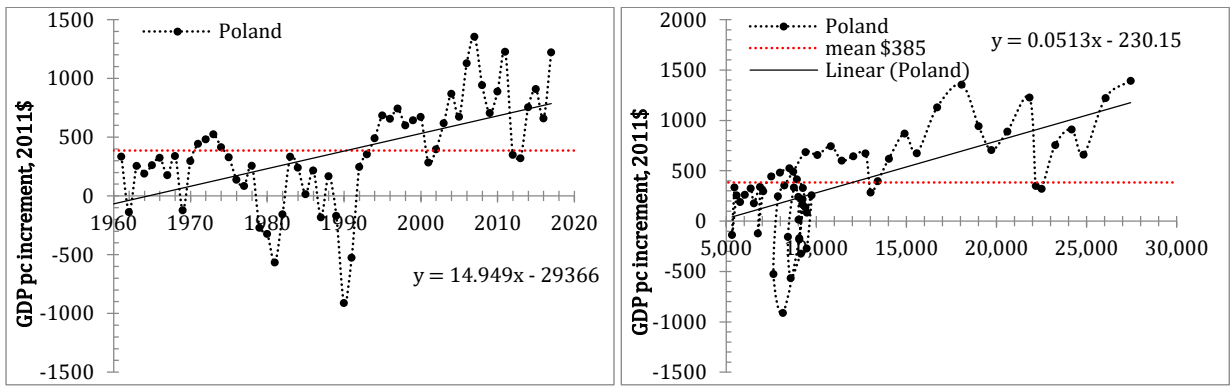

Figure 45. Same as in Figure 43 for Poland.

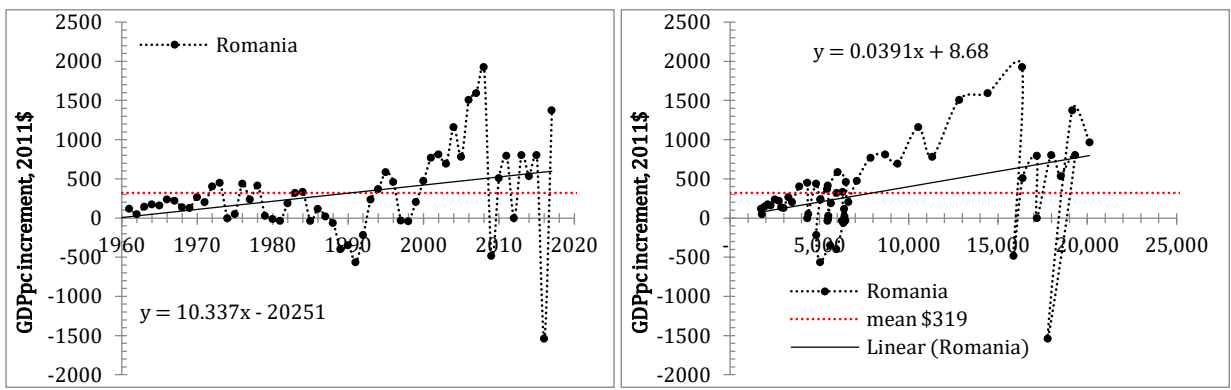

Figure 46. Same as in Figure 43 for Romania.

## 9. Arab Spring

We avoid economic discussion in this Section and just present data. The influence of non-economic factors on economic development in several Maghreb and Middle East countries is synchronized and almost identical - failure. Non-economic factors are rarely positive – almost two decades of healthy economic growth in the presented countries was suddenly stopped (except Morocco). This is especially negative outcome because the gaps in economic development are almost impossible to compensate and the deficit in real GDP per capita will affect the population in these countries in the future.

In 2011, Arab Spring started and spread over many countries. Almost all of the involved economies suffered significant drop in the growth of real GDP per capita. Algeria is presented in

Figure 47. In 2011, the GDPpc increased by $616 after ten years of the above $400 annual increment. In 2012 and 2013, the increment was just $188 and $115. In 2017 and 2018, the increment was negative. Therefore, the Arab Spring did not accelerate real economic growth in Algeria.

Egypt (Figure 48) was on an accelerating route since the earlier 1990s and reached the annual increment of $621 in 2007 and $560 in 2010. This is much better than in many developed countries during the same period. In 2011, the annual increment dropped to $18 and the low performance was observed till 2013. Currently, Egypt tries to return to the fast growth track, but the loss (accurately, lost gain) of a few thousand US dollars per person is not possible to recover.

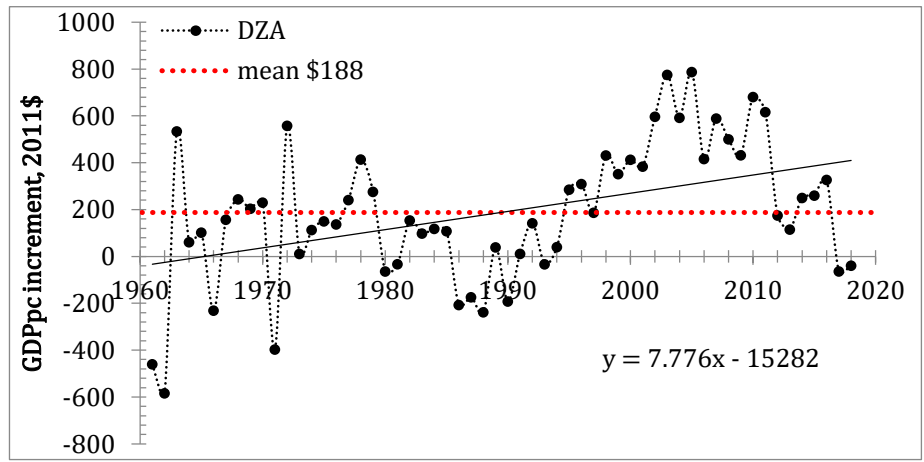

Figure 47. Algeria. Annual GDPpc increment between 1961 and 2018 with the average value for the studied period of $188 (2011 prices).

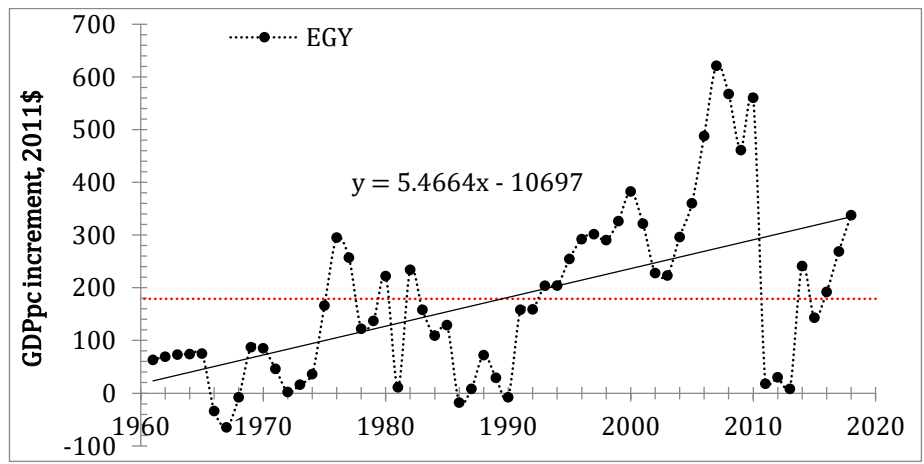

Figure 48. Same as in Figure 47 for Egypt.

The mean annual increment in Iran (Figure 49) was $234 for the period since 1960. The years after 2000 are characterized by intensive growth with the peak value of $1,281 in 2007. The Great Recession moved the annual increment to the mean level, but the following years before 2011 were successful again. The Arab Spring moved the growth in the red zone with -$1,279 in 2012.

Iraq (Figure 50) has a lower annual increment of $146 since 1960. There were several wars inducing negative growth rates. Since 2004, the growth was very fast but ended in 2012-2013. All periods with negative growth are forced by non-economic factors. In Figure 51, Libya is presented. The years between 2003 and 2011 were extremely successful with the annual

increment around $2,300. In 2010, Libya was the richest African country with $29,157 per head. In South Africa, the GDPpc was $11,319 in 2010. The fairy tale ended in 2011 with the fall of around $16,200 per head. The rebound of $16,870 in 2012 is likely a statistical bug but it was compensated by the fall of $16,440 in 2013. In 2018, the GDP per capita was $15,013. Morocco in Figure 52 does not show any success of failure examples with very low mean increment of $109 between 1960 and 2018. The GDP per head was $8,451 in 2018.

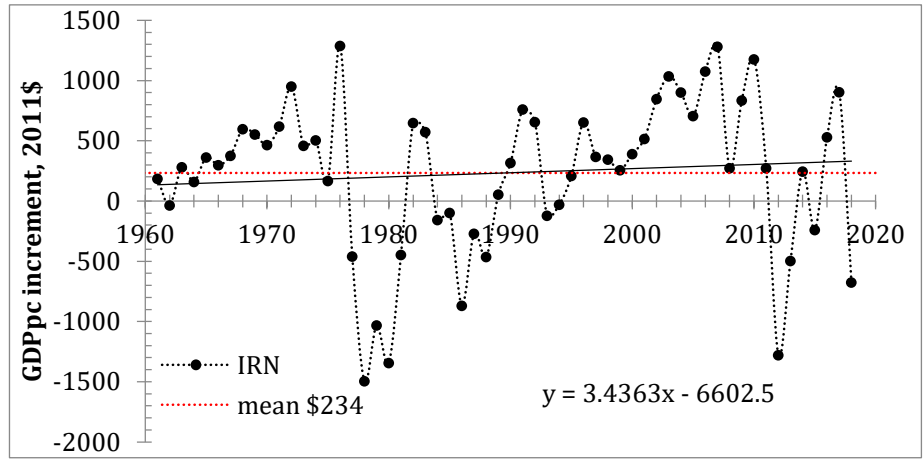

Figure 49. Same as in Figure 47 for Iran.

In Figure 53 Tunisia is presented with the mean annual increment of $159. This is an example of slow economic growth. The years after 2002, were more successful with the annual growth up to $513 in 2007. In 2011, the GDP per capita fell by $285 and the following years were mostly below the mean line.

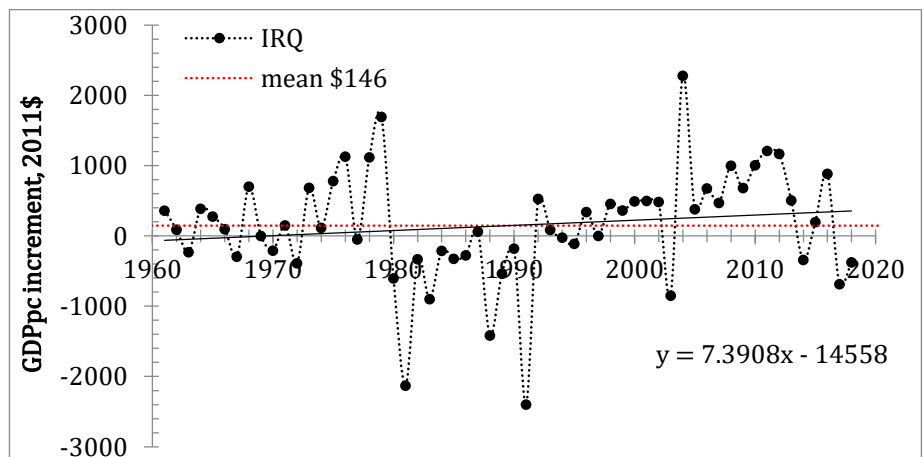

Figure 50. Same as in Figure 47 for Iraq.

Bahrain in Figure 54 is an oil producing country with very high mean increment of $609±$798 per year. In 2011, it had the peak increase followed by a fall to the mean level. After 2013, the annual GDPpc increment was on a falling trajectory with negative increments in 2017 and 2018. Oman (Figure 55) experiences severe problems with real economic growth since 2011. In 2010, the GDPpc was $46,182 and only $36,438 in 2018. The mean GDPpc is $609±$1294, i.e. the same as in Bahrain but with higher scattering.

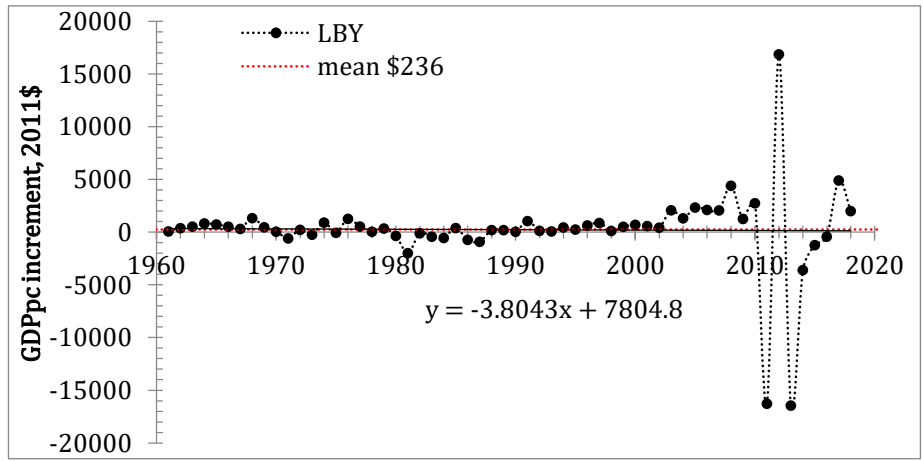

Figure 51. Same as in Figure 47 for Libya.

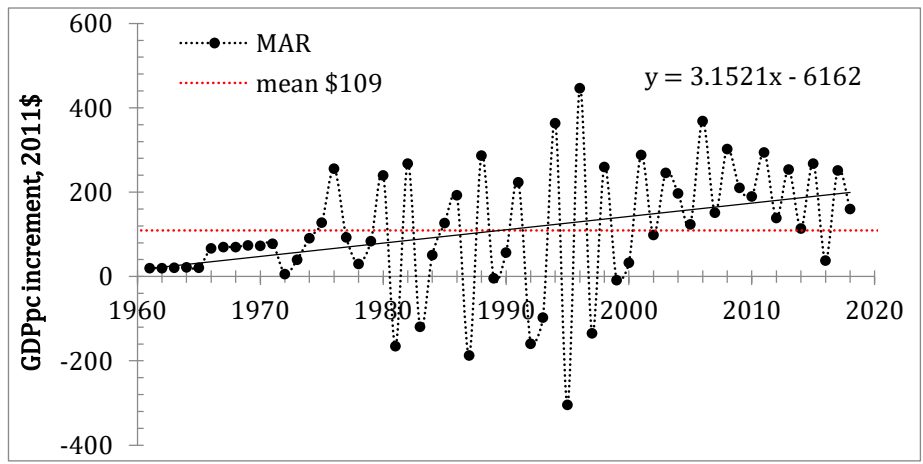

Figure 52. Same as in Figure 47 for Morocco.

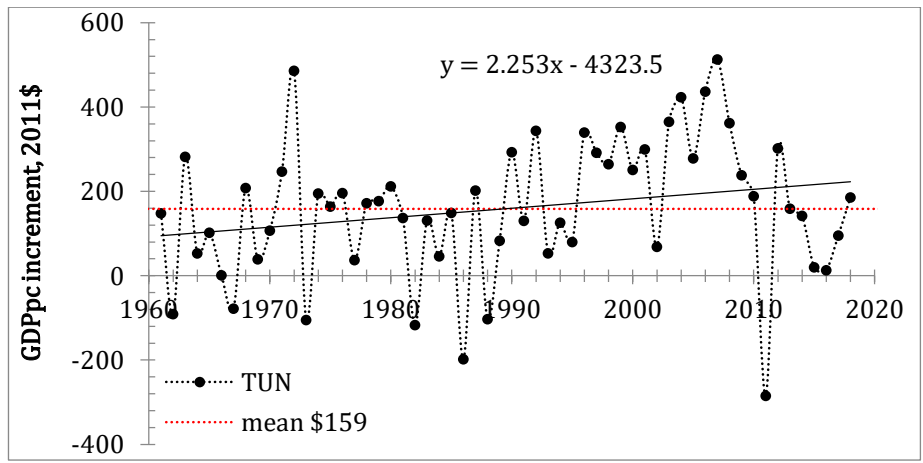

Figure 53. Same as in Figure 47 for Tunisia.

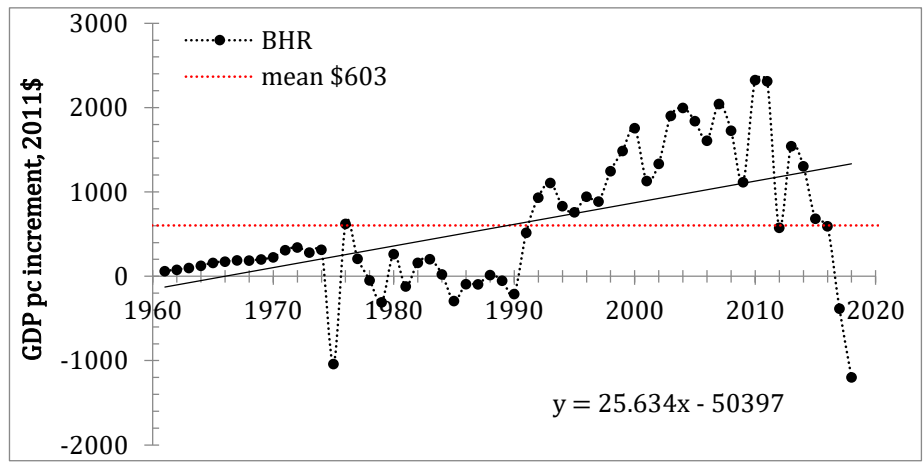

Figure 54. Same as in Figure 47 for Bahrain.

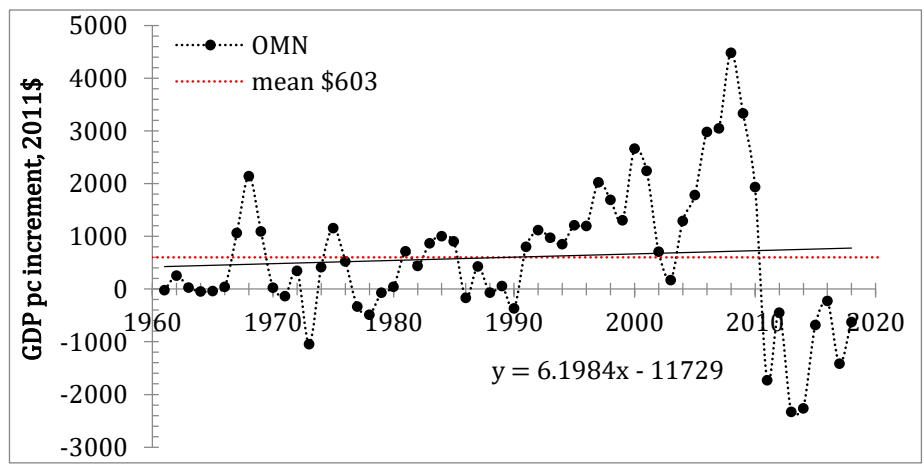

Figure 55. Same as in Figure 47 for Oman.

## 10. Cross country comparison

We have presented a large number of the real GDPpc country histories between 1960 and 2018 as listed in the Maddison Project Database. The full range of initial conditions, GDPpc(1960), and total growth, GDPpc(2018)/GDPpc(1960), creates an impression of chaotic behaviour and may compromise our understanding of the actual growth processes in various countries. In this Section, we present a general framework allowing consistent analysis of the MPD data and assessment of relative economic performance for a given country.

The principal GDPpc inertial growth equation is the following: $g(t) = dlnG(t)/dt = A/G(t)$. Instead of integrating this equation between 1960 and 2018 and finding the total growth GDPpc(2018)/GDPpc(1960), we directly calculate this ratio. Figure 56 shows several theoretical curves of total GDPpc growth in 58 years, GDPpc(58)/GDPpc(0), for four different constant annual GDPpc increments, $A$. For example, the curve with $A$=\$300 gives the total growth by a factor of 2, when the initial GDPpc(0)=\$17,200, and the total growth of 4 for GDPpc(0)=\$5,800. Because of the GDPpc level difference in 1960, the total growth, and thus, the annual GDPpc growth, in rich countries is lower than in poor countries with the same annual GDPpc increment.

A good example of the initial basis effect on growth rate was given in 2020 when the GDP in the USA fell by 5% in the first quarter, additional 31.4% in Q2, and increased by 33.4% in Q3. At first glance, the rise in Q3 should compensate the fall in Q2. However, the 31.4% drop of real

GDP gives a new base for the following rise. For the sake of simplicity, let's assume the level in the beginning of Q2 as 1. Then the GDP at the end of Q2 is 0.686 (68.6%). The rise of 33.4% in Q3 with a new basis of 0.686 gives the level of 0.915 (0.686*1.334) to Q1. Therefore, the rise in Q3 does not return the economy to the initial level. One should not pay attention to only the relative growth rate, but to compare the GDPc level and the (average) annual increment.

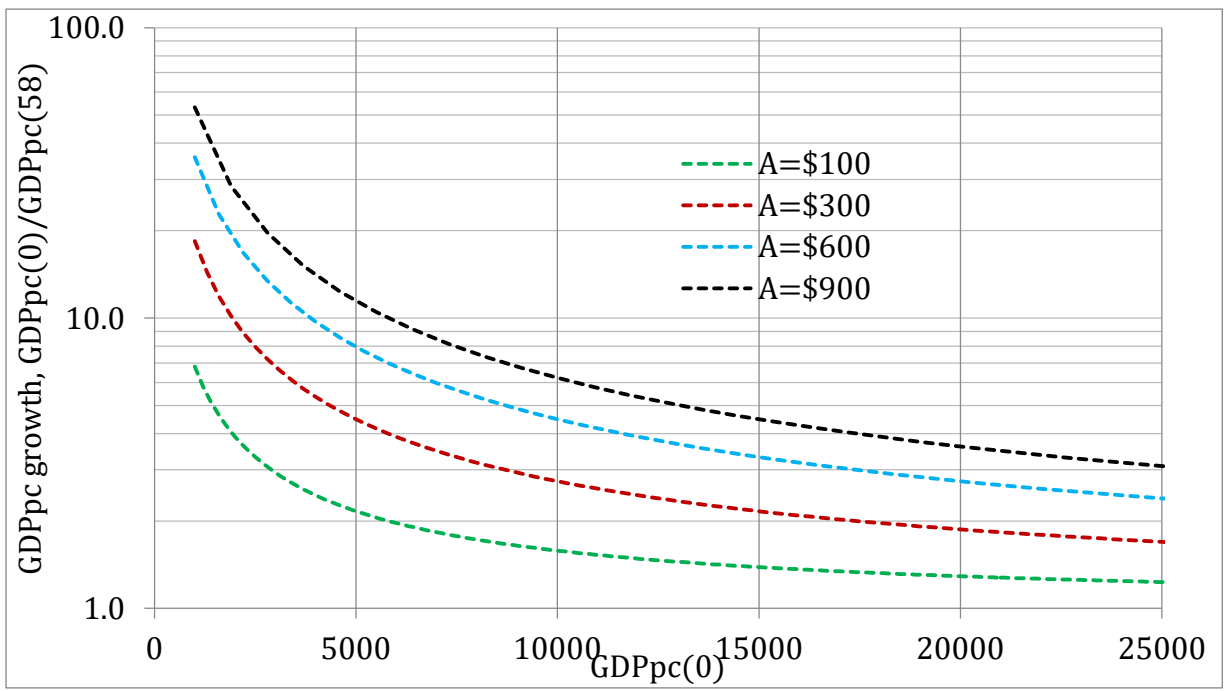

Figure 56. Total growth as a function of initial GDPpc value, GDPpc(0), and the annual increment, *A*.

In Figure 57, we merged the theoretical GDPpc growth curves for the 58-year period between 1960 and 2018 presented in Figure 56, and the estimates of the total GDPpc growth for the same period obtained from the MPD. Now the interpretation of economic performance is straightforward. In 1960, the USA had the largest GDPpc=$18,057 (ARE and Qatar are excluded) and the total GDPpc grew by a factor of 3.04. The latter value seems to be low compared to the fast growing economies like China and Malaysia (MYS). However, China and Malaysia grow from the very low level and their GDPpc annual increments are between $200 and $300, when the USA had ~$640 per year. Therefore, China and Malaysia grew at a lower rate than needed to decrease the absolute GDPpc gap with the USA. In the long run, they would not catch the GDPpc level in the USA.

Many West European countries, Japan, Canada, and Australia are all close to the *A*=$600 line. Their performance since 1960 was overall stationary with a few short recessions. Norway has lower total growth than Ireland, but taking into account the initial GDPpc, one can say that Norway had faster economic growth in terms of annual increment. The UK, France and Italy were rather the relative losers among the most developed countries, likely because of non-equivalent exchange with Germany (DEU) within the EU. New Zealand performed not well since 1960 despite very high start level in 1960. Switzerland outperformed all developed economies except exotic cases of Ireland and Norway. At this point, it is important for the underperforming developed economies to understand that their lag behind the best performers will be increasing in the future. The causes of different growth rates observed within the EU might be internal and external.

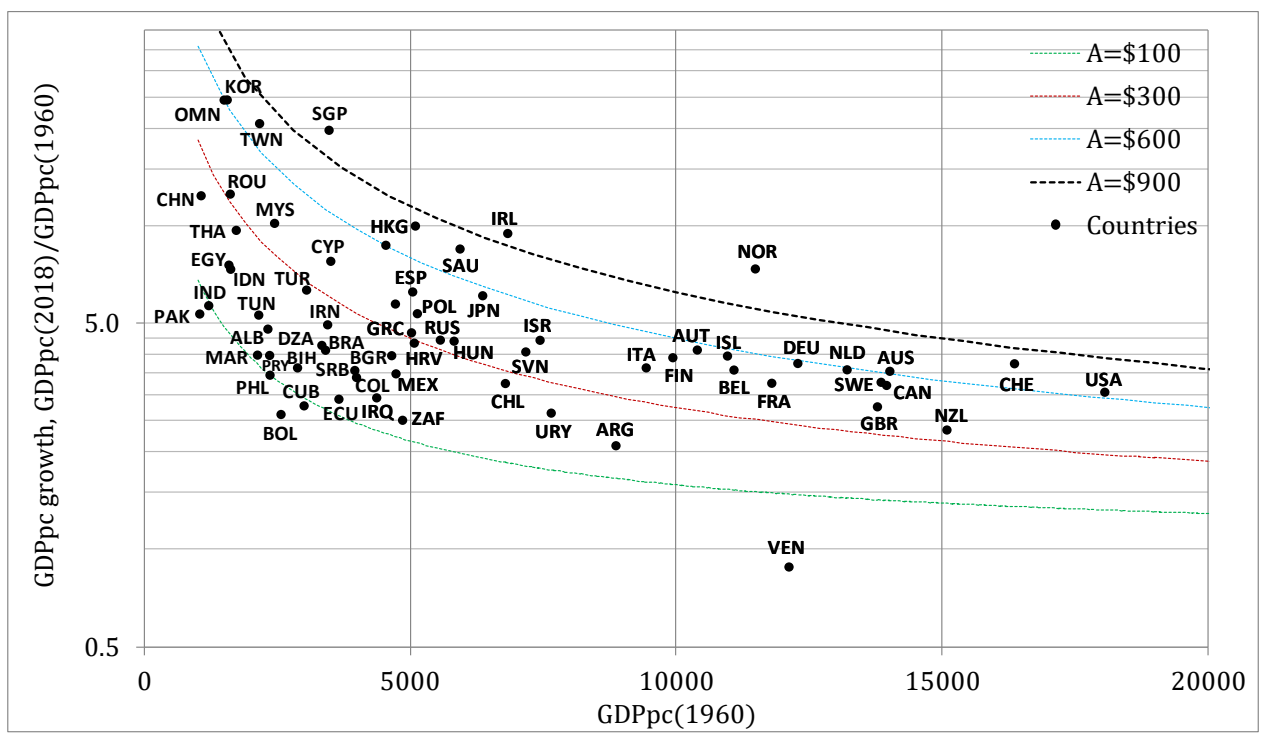

Figure 57. Theoretical curves of the total GDPpc growth between 1960 and 2018, and the estimates for various countries. Country tags are borrowed from the MPD and correspond to international abbreviations.

Four Latin American countries (Argentine, Uruguay, Chile and Venezuela) have high initial GDPpc values. Their annual increments reside in the range between $A=\$100$ and $A=\$300$, which is practically the same for many other countries with lower start values: Brasilia, Colombia, Mexico, Paraguay. Bolivia and Cuba have poor start conditions and the annual increment near $100. The worst growth history belongs to Venezuela which lost 12% of the initial GDPpc in 1960. Therefore, the only difference between the Latin American countries shown in Figure 57 is the start GDPpc level.

East European countries are close to $A=\$300$ line, but the break in the 1990s demonstrates that their growth rate can be higher in the future. Russia has all chances to move close to $A=\$600$ line in the near future judging by the GDPpc history since 2000. Despite the incredible growth period, China is still in the club of not excellent performers with good chances to move to $A=\$500$ level, i.e. join the bunch of developed courtiers. India is just in the beginning of growth history and the linear growth in the annual GDPpc increment opens bright perspectives.

The estimates in Sections 5 through 10 are obtained for the longer period between 1960 and 2018 in order to validate the theory of inertial growth with constant annual increment. The most developed countries with stable and stationary economic development, i.e. in the absence of larger non-economic forces like war or natural disasters, are well described by the inertial growth model. Smaller developed economies may experience periods of faster growth. The economic forces behind such a stellar performance as demonstrated by Norway and Ireland cannot be transferred to other developed countries like New Zealand, Greece and Portugal. This means that their performance does not reject the inertial growth approach and we likely need more time for the assessment in view that any advantage may disappear any time.

Other countries presented in this study experienced the influence of various large-scale non-economic forces since 1960. The post-Soviet and post-socialist countries all had a dramatic transition from socialist to capitalist economic development, and this transition was always accompanied by deep fall in the real GDP per capita. This transition period is obvious in the corresponding GDPpc trajectories and divide the period of lower and higher real economic growth as reported by the MPD. China and India had poor start conditions and their progress since the mid-1990s is not finished yet in a stationary regime. We can expect further acceleration in the GDPpc growth. The number of countries with non-stationary growth is large and almost all these countries demonstrate higher growth in the 21[st] century. At the same time, many developed countries are on a decelerating trajectory. Therefore, the 21[st] century may manifest a global redistribution of economic power and some loss in the competitiveness of the West as such. We analyze the real GDPpc in the 21[st] century in Section 12.

## 11. Alternative measures of real GDP per capita

We have presented dozens of real GDP per capita graphs using the Maddison Project Database. We trust professionalism of the MPD team and their GDPpc estimates. They use raw data, however, which might be biased overall and specifically for our study. First of all, when we say "per capita" or "per head" what do we mean specifically - the total population, working age population or the population in labour force? The main question is: Who is driving real economy? When modelling real economic growth in the mainstream economics, the GDP per capita is using the working-age population (not the values published in the MPD). There are several reasons behind this approach. Two of them are: 1) children do not work and do not produce goods and services; 2) Children do not have income; at least their incomes are not determined in the annual Current Population Surveys and decennial censuses. The comprehensive analysis of the other GDPpc definition is a task for the future as well as the comparison between different data source, e.g. the Maddison Project Database and the Total Economy Database of the Conference Board. In this study, we just present some examples illustrating the necessity of these studies.

The importance of the difference between total and working-age population is illustrated in Figure 58 with the USA and Mexico as examples. Mexico had a larger portion of children between 1960 and 1980 and then the USA took the lead. It is not excluded that the growth in the portion of Hispanic and Latino Americans due to migration to the USA assisted the elevated birth rate and children proportion. Correction to the difference between the total and working-age population is an interesting option to be exercised in any case. One could expect that the overall data consistency has to be improved. On the other hand, there are countries with official and non-official migration. The share of such people in the workforce may vary from country to country. For example, Qatar uses more than 2 millions of foreign workers.

The "per capita" definition is crucial for our study, but the real GDP definition is also important. The real GDP cannot be measured as such. Instead, nominal GDP and GDP deflator are used to estimate the real GDP. The GDP deflator is a complex variable, which include direct measurements of real price change and an estimate of the price change for new goods and services as well as the goods and services changing own quality and utility. Moreover, there are new items included in the nominal GDP estimates (e.g. imputed rent). All these aspects in the estimation of real (and nominal) GDP are well described and reported by national agencies (e.g., BEA). As a result, one can often see a striking comment that the GDP estimates are not compatible over time.

This is an extremely difficult problem to reconstruct the real GDP time series, and then harmonize them between various countries. We are not going to solve such a difficult problem and leave this work to the Maddison Project team. There is one important problem, however, which we are able to address in this study - the change in calculations of the CPI and GDP deflator (dGPD). Figure 59 presents two curves in the USA - dGDP and CPI time series normalized to their values in 1929, i.e. these curve present the cumulative price growth since 1929. One can suppose that the CPI and dGDP are close economic variables. The data show quite different picture - the curves start to diverge in 1979 (new definition to GDP was introduced). Since the dGDP and CPI were the same before 1979 and then started to diverge due to the definition change it is instructive to apply the same definition of inflation to the whole series and obtain a consistent time series. Otherwise, the real GDP in the USA is overestimated since 1979 since the dGDP is underestimated relative to the CPI used before 1979. For the GDPpc estimates in this study, the dramatic change in the definition of inflation might change the long term trends and the model of inertial growth.

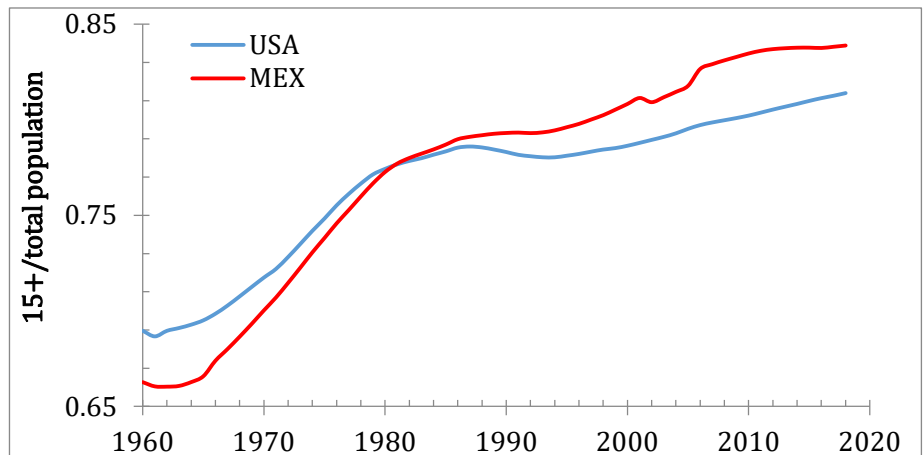

Figure 58. The evolution of the working-age and total population ratio in the USA and Mexico since 1960.

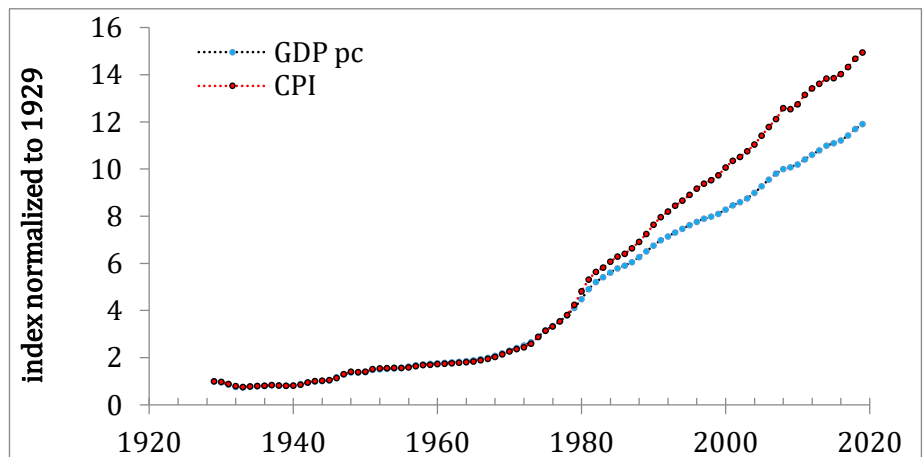

Figure 59. The CPI and dGDP cumulative inflation in the USA since 1929. Two curves start to diverge in 1979.

The effect of the inflation change in the USA is illustrated in Figure 60. We have corrected the portion of the GDPpc after 1979 to the ratio of the dGDP and CPI. This effectively allows using the CPI time series as inflation in the real GDP estimates. One can also use the same dGDP definition before 1979, as shown in Figure 61. This is a simple procedure since the ratio of CPI and dGDP is 1.22 since 1979. We take the same ratio and extend the corrected dGDP time series into the past.

In Figure 60, one can see that the deviation from the linear regression line observed after 1979 completely disappear. In other words, the GDP per capita in the USA is likely biased in favour of better real economic growth, which is actually just a statistical trick. Unfortunately, this nasty trick is used by other developed countries.

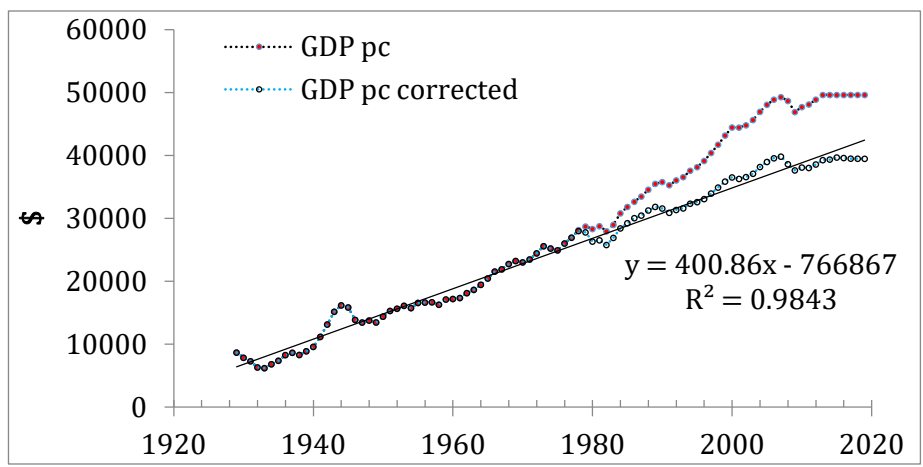

Figure 60. Real GDP per capita in the USA corrected for the difference between the CPI and dGDP.

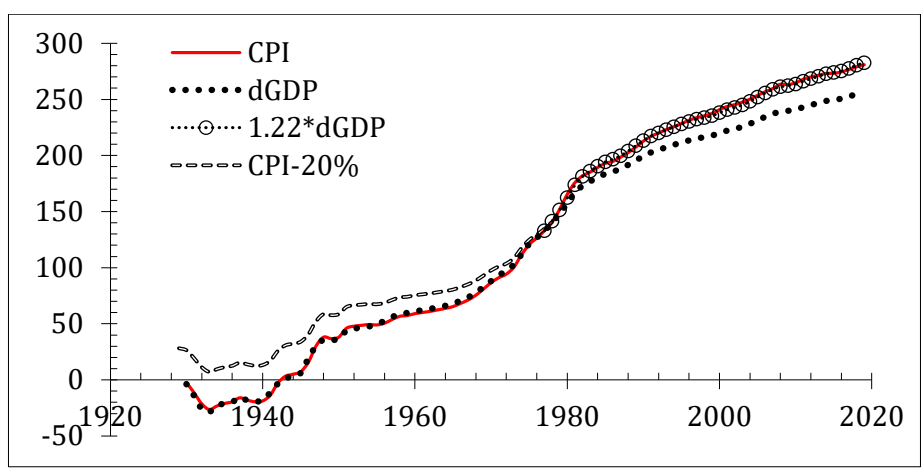

Figure 61. The extension of the dGDP definition into the past. The dGDP is taken as CPI-20% before 1979.

## 12. Economic performance in the 21st century

Brexit forces us to think about the future of the EU. We have presented a series Figures analyzing economic growth in selected countries using one invariant - constant annual increment in the real GDP per capita. This invariant defines the inertial part of real economic growth as observed in all large developed economies since the 1950s (no accurate data is available before). Statistical analysis demonstrates that this invariant is actually a constant for a given country but varies between the countries. Therefore, it might be used to evaluate relative performance of a given country.

Figure 62 compares several EU countries and splits the period from 1960 to 2018 into three sub-periods: 1960-1979, 1980-1999, and 2000-2018. Statistically, variations in the annual increment in the GDPpc between these periods are important for the estimates of the linear regression, i.e. the long-term behavior. For example, France and Italy demonstrate significant decrease in the average increment from around $600 in the first period to $330 in France and $160 in Italy between 2000 and 2018. We have formulated this decay in terms of the gradual lost of economic competitiveness (efficiency) compared to Germany within the EU. This situation is likely fixed

and neither Italy nor France is able to get back to the pre-EU competitiveness level. The case of the UK, which is also demonstrating lower performance than Germany and the Kingdom of the Netherlands, gives a reasonable solution – to leave the EU and fight for the efficiency out of the EU framework which they actually do not control. Portugal and Spain should probably join such a move. If the EU is a multi-speed union then Italy and Portugal are driving in first gear.

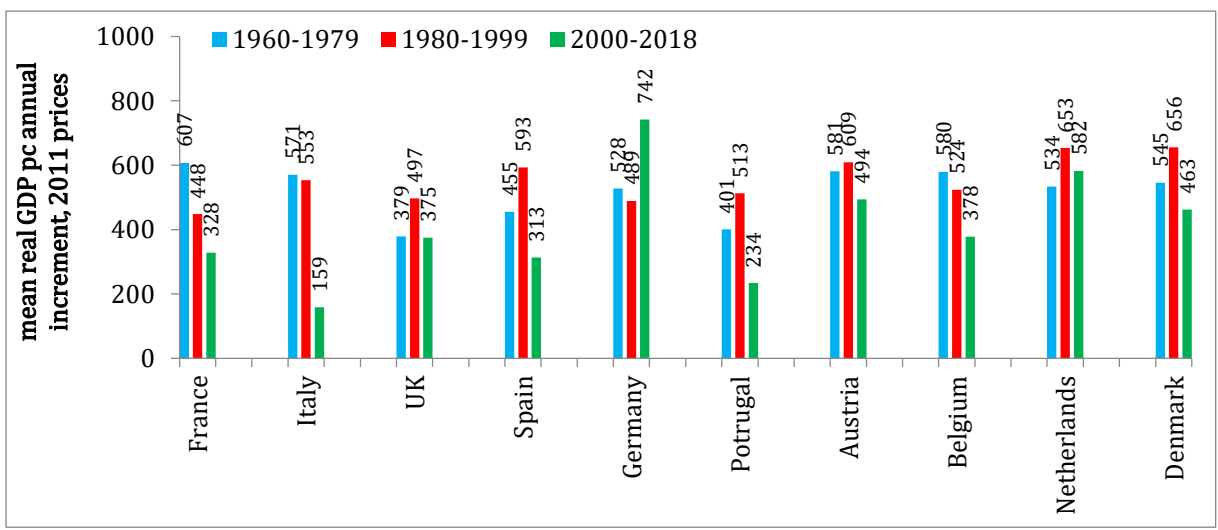

Figure 62. Real GDP per capita (2011 prices) in selected EU countries. Data are obtained from the Maddison Project database.

The real GDP per capita is likely the best measure of economic progress. It can be an expression of various driving forces behind the growth, however. The largest (practically full spectrum of industries and services as well as the related income inequality and social problems) economies are competing between each other in all political (*i.e.*, economic and military) aspects. The military power is very important to protect the non-equivalent exchange with smaller economies. The overall economic influence (the margin of non-equivalent exchange) of a given country can be described by three parameters – the increment of real GDP per capita, military power, and total population. The smallest countries can be extremely successful in a few specific activities, however. Bigger countries have to fight. Clearly, the military interaction in the key trade focus points (Suez canal, Panama canal, Malacca strait) grows in intensity and probably in the next decade may come closer to the global war: China squeezes the USA out of the strait of Malacca droplet by droplet, Russia has a military base in Syria and will get a new technical base for warships in Sudan (Red Sea); China and Russia both invest in Venezuela and provide military support. Yet, Russia and China do not control these focus points but can ruin them and stop the trade any moment. In all the strategic points, West loses control gained after the Former Soviet Union closed own military bases worldwide.

In Figure 63, we illustrate the most recent economic developments within the framework used to analyze the period since 1960. We have selected the 21[st] century as a representative sample describing the dynamic changes in the relative economic power change. For example, the USA suffers a fall in the mean annual increment below $600. At the same time, Russia is characterizes by a dramatic increase to the level near $800. China is also moving up and the next target is $600. The UK, France, Japan, Spain and a few smaller economies created a cluster near the $300 line. Italy fell to the $100 level together with Greece and Portugal. Only Germany improved own performance moving to the level above $700. From the East European countries, Poland demonstrated the largest gain and moved closer to $800. Romania and Hungary were more successful than other post-socialist countries (e.g. Bulgaria, Croatia, Serbia, and Slovenia). Considering the fact that the GDPpc level in China, Russia and East European countries are

lower than in the most developed countries and the annual GDPpc increment is higher, their growth rates are much higher than in the Western economies.

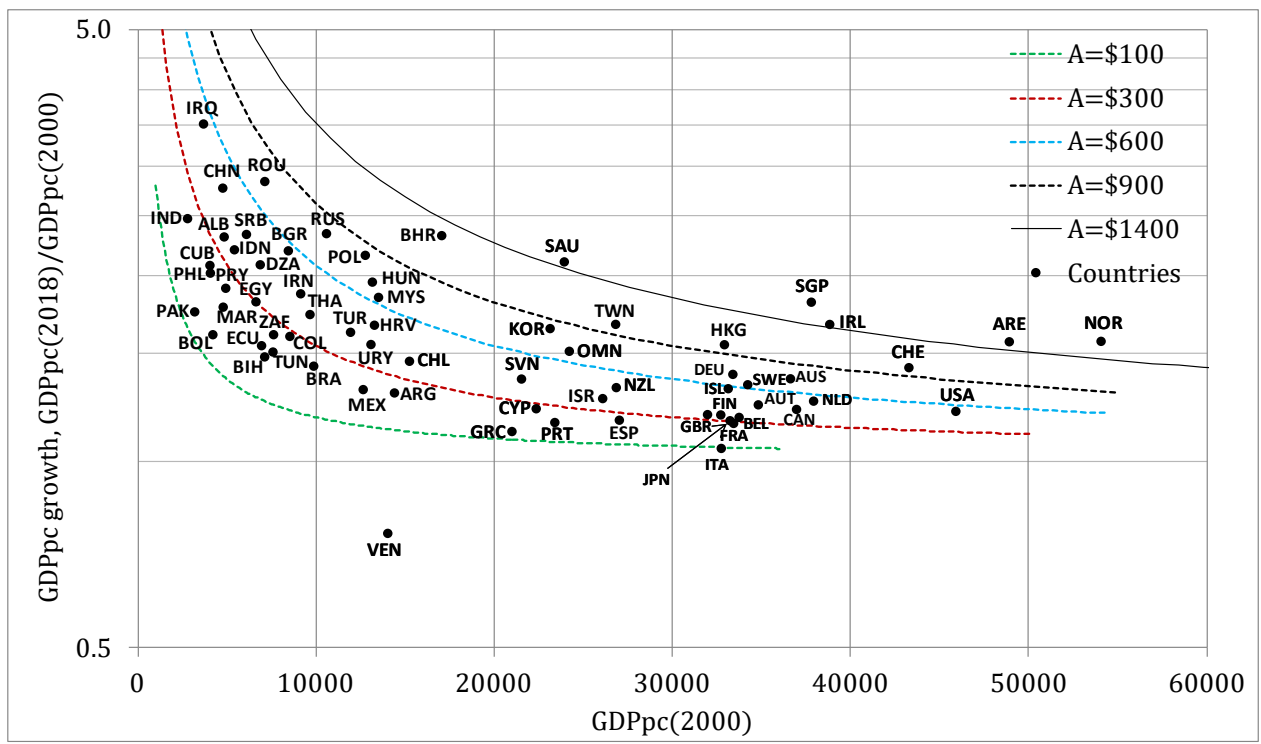

Figure 63. Theoretical curves of the total GDPpc growth between 2000 and 2018, and the estimates for various countries. Country tags are borrowed from the MPD and correspond to international abbreviations.

In Appendix 1, we present a table listing the real GDP per capita annual increments in the 21[st] century and total population (2018), for all countries presented in the Maddison Project Database. Position #5 is occupied by Saudi Arabia. Its wealth is based on oil and military protection by the USA. Position #10 belongs to Switzerland. It is difficult to call this country "small" in terms of economy, but it has a few very specific businesses likely subject to increasing influence from abroad. The Russian Federation is number 20: the unbeaten military power with nuclear weapon and the largest territory is well supported by the sixth place in the real GDP (PPP) list. The growth rate in real GDP per capita was extremely healthy ($784 per year on average) after the recovery from the dramatic fall associated with the transition to capitalism from socialism. Germany occupies position 27 and is likely the major beneficiary of the EU. The success of East European countries is related to the EU subvention, which is close to end, however. This is the reason of the two-speed Europe. The UK (73), France (83), Italy (92), Spain (89) are almost the worst performers in the EU – lost their power relative to Germany. The USA is #. China is #47, but its economic and military power is based on the largest population despite lower real GDP per capita.

## 13. Discussion

The results of the thorough analysis presented in this paper validate the original assumption that the growth rate in the developed counties has been falling according to the inertial growth relationship (4). The observed decrease in the rate of growth contradicts the expectation of a constant growth rate (*i.e.*, exponential GDP growth) as suggested by the mainstream economic approach. Regular actors of the global economy and financial markets ground their strategies on the assumption of the transient zero-mean fluctuations around the constant growth rate. The gap between the real and expected growth is a potential source of economic, financial, and social problems.

Among many other parameters, companies, firms, and enterprises base their development plans on the mid- and long-term expected economic growth as the expression of moving balance between potential demand and targeted supply. The decaying rate of economic growth has never been a part of this approach with the mainstream opinion of the long-term exponential growth. When the cumulative gap reaches some critical value the economy makes self-adjustment and returns to the real trajectory of economic growth. This could be expressed as an economic recession. In the past, such gaps were growing at a higher rate because the rate of economic growth was higher and its fall was much faster according to relationship (4). With the decreasing rate of growth, recessions have to occur less often since the deceleration of the growth rate at the current levels of the real GDP per capita in developed countries (see Section 5) is almost negligible.

In the world of decaying rate of economic growth, and thus, the long-term revenue decrease, financial institutions have to look for the places with higher growth rates where investments provide higher returns (likely with some elevated risk). The profit generating industries and services are forced to move to these higher-growth-rate places. With time, the growth rate decreases (growing GDPpc lowers the rate) event in these places and the revenue as well. The possibilities to retain the historical revenue level are shrinking. It is not excluded that the new methods to return the financial profit to the desired level are associated with global economic and social redesign expressed in the forced creation of such zones of higher revenue.

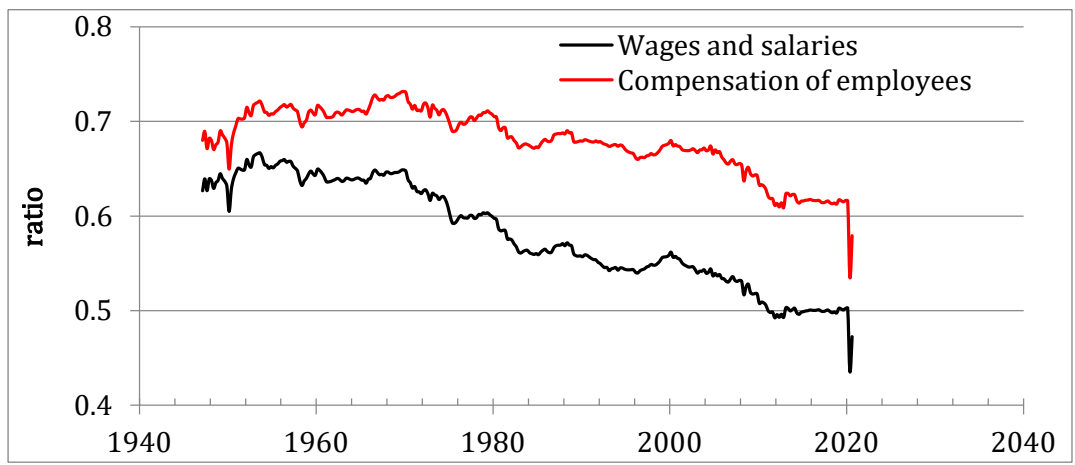

Figure 64. The shares of "Compensation of employees" and "Wages and salaries" in the "Personal income".

In developed countries, the abandoned employees of the removed industries and services lose their labor-price-setting power, and thus, the share of personal income related to the job. In the USA, the share of "compensation to employees" in the total personal income has been falling from 0.732 (absolute peak in the time series reported by the Bureau of Economic Analysis) in 1969 to 0.609 in 2013 (the COVID-19 fall to 0.535 is not considered). The share of "wages and salaries" in the personal income decreased from 0.649 in 1969 to 0.492 in 2011, i.e. wages and salaries have been falling much faster than the compensation of employees. Figure 64 presents both curves as reported in Table 2.1 "Personal income and its disposition" available from the BEA. The most dramatic fall in both economic variables was observed between 2006 and 2013. Such discouraging falls may result in political turmoil.

The future economic giants are China and India with the developed countries doomed to degradation and extinction in terms of relative economic power. China and India not only improve the total GDP as the economies with the largest population but also experience a quantitative jump to the level of stationary and sustainable growth in real GDP per capita. This is a qualitative change – they become developed countries with complete economies and corresponding price-setting power. It is not possible to estimate their potential, i.e. the annual GDP per capita increment, in the next few decades but it can reach the current level of the USA as the principal price setter. The trends are stable and promising.

Russia is almost in the self-consistent, sustainable, and stationary growth state with an annual GDPpc increment above $600. This value is measured in the 21[st] century and is one of the largest worldwide. The future depends on the potential of resistance to the increasing pressure in the field of economic and political cooperation with various actors. There are natural partners and opponents. Brasilia is likely a failed state in an economic sense. It demonstrates a stationary regime (i.e. no economic jump as in Russia, China, and India) since 1960 with the annual GDPpc increment of $183.

East European countries fully depend on the EU. Their economic performance can be successful only within the EU markets and system. Germany is the EU driver as the country with the highest economic potential and the largest annual GDPpc growth rate. The negative side of this leadership is the progressive decay in the rate of growth in France and Italy. They pay by underperformance for the rise in East Europe. The UK's future is not clear because the configuration of economic cooperation and competition with all possible legal and not-so-legal measures is changing fast.

Finally, in the world of the rapid growth of the future economic behemoths and stagnation of the most developed countries conflicts are inevitable. Unfair trade restrictions, political pressure, media attacks, propaganda, military aggression, and other dimensions of these conflicts may only rise in amplitude and extent. These conflicts involve new countries in the global clash, which also includes the clash of civilizations as an additional dimension. This is only because the growth in real GDP per capita is a linear function of time. In the exponential economic world, the lead of developed countries would be eternal as they had better start conditions and the exponent provides the increasing economic gap. In the linear economic world, the lead in GDPpc is constant, the chasing countries grow faster and the gap is shrinking in relative terms.

### References

Hodrick, R and E. Prescott (1980), "Postwar U.S. business cycles: an empirical investigation", Discussion Paper, Northwestern University

Kitov, Ivan, 2006. "Real GDP per capita in developed countries," MPRA Paper 2738, University Library of Munich, Germany.

Kitov, Ivan, Kitov, Oleg, 2012. "Real GDP per capita since 1870", Papers 1205.5671, arXiv.org.

Kitov, Ivan, Kitov, Oleg, 2021. The link between unemployment and real economic growth in developed countries. (under preparation)

Maddison Project Database, 2020. Maddison Project Database 2020 | Releases | Groningen Growth and Development Centre | University of Groningen (rug.nl)

Total Economy Database, 2020. Total Economy Database™ - Key Findings | The Conference Board (conference-board.org)

OECD Database, 2020. OECD Statistics

Appendix 1.

| Mean GDP pc increment since 2000 | Country | population, x1000 |
|---|---|---|
| 5999 | Qatar | 2606 |
| 1702 | Singapore | 5996 |
| 1697 | Norway | 5313 |
| 1649 | Kuwait | 4229 |
| 1528 | United Arab Emirates | 9619 |
| 1467 | Saudi Arabia | 33344 |
| 1438 | Ireland | 4859 |
| 1249 | Bahrain | 1584 |
| 1230 | Turkmenistan | 5410 |
| 1007 | Switzerland | 8668 |
| 997 | China, Hong Kong SAR | 7350 |
| 993 | Taiwan, Province of China | 23324 |
| 920 | Lithuania | 2799 |
| 889 | Kazakhstan | 18323 |
| 867 | Equatorial Guinea | 797 |
| 842 | Seychelles | 95 |
| 823 | Republic of Korea | 51635 |
| 818 | Poland | 38371 |
| 790 | Czechoslovakia | 16015 |
| 784 | Russian Federation | 146476 |
| 761 | Czech Republic | 10605 |
| 748 | Montenegro | 614 |
| 735 | Australia | 24855 |
| 732 | Slovakia | 5409 |
| 724 | Romania | 19191 |
| 722 | Latvia | 1927 |
| 712 | Germany | 84152 |
| 708 | Panama | 3801 |
| 694 | Hungary | 9774 |
| 690 | Azerbaijan | 9648 |
| 683 | Oman | 4725 |
| 644 | Malta | 487 |
| 632 | Malaysia | 31143 |
| 630 | Sweden | 10175 |
| 589 | Estonia | 1292 |
| 580 | Mongolia | 3103 |
| 574 | Iceland | 353 |
| 557 | Bulgaria | 7167 |
| 534 | Belarus | 9528 |
| 532 | Netherlands | 17232 |
| 525 | United States | 327835 |
| 521 | Trinidad and Tobago | 1215 |
| 510 | Iraq | 40012 |
| 507 | Dominican Republic | 10385 |
| 487 | Croatia | 3861 |
| 473 | New Zealand | 4840 |
| 465 | China | 1385439 |
| 455 | Austria | 8892 |
| 449 | Former Yugoslavia | 21424 |
| 448 | Serbia | 7078 |
| 440 | Canada | 37206 |
| 439 | Iran (Islamic Republic of) | 82239 |
| 433 | Botswana | 2263 |
| 430 | Slovenia | 2067 |
| 427 | Puerto Rico | 3295 |
| 411 | Algeria | 41490 |
| 409 | Turkey | 85935 |
| 409 | Luxembourg | 610 |
| 405 | Denmark | 5794 |
| 396 | Uruguay | 3396 |
| 394 | Georgia | 4922 |
| 390 | Thailand | 67515 |
| 388 | Gabon | 2083 |
| 383 | Israel | 8460 |
| 383 | Chile | 18676 |
| 364 | Uzbekistan | 30023 |
| 359 | Indonesia | 259494 |
| 351 | Armenia | 2907 |
| 350 | Albania | 3063 |
| 348 | Peru | 31382 |
| 345 | Finland | 5516 |
| 340 | United Kingdom | 66746 |
| 335 | Belgium | 11429 |
| 326 | Mauritius | 1364 |
| 323 | Sri Lanka | 21732 |
| 320 | Angola | 22638 |
| 304 | Japan | 125848 |
| 298 | Costa Rica | 4994 |
| 298 | Egypt | 102184 |
| 291 | Libya | 6688 |
| 286 | Lebanon | 7708 |
| 284 | France | 67029 |
| 280 | Colombia | 48128 |
| 277 | TFYR of Macedonia | 2119 |
| 270 | Cyprus | 1022 |
| 258 | Ukraine | 43952 |
| 255 | South Africa | 55341 |
| 250 | Spain | 47380 |
| 247 | Paraguay | 7099 |
| 240 | Cuba | 11116 |
| 233 | Brazil | 211308 |
| 233 | Lao People's DR | 7234 |
| 233 | Argentina | 44695 |
| 228 | Philippines | 112106 |
| 225 | India | 1298136 |
| 225 | Viet Nam | 97076 |
| 222 | Myanmar | 54400 |
| 216 | Jordan | 9116 |
| 216 | Mexico | 121656 |
| 211 | Tunisia | 11478 |
| 209 | Republic of Moldova | 3453 |
| 207 | Ecuador | 16499 |
| 205 | Morocco | 35496 |
| 204 | Portugal | 10221 |
| 188 | Bosnia and Herzegovina | 3947 |
| 185 | Dominica | 74 |
| 182 | Saint Lucia | 166 |
| 180 | Cabo Verde | 568 |
| 177 | El Salvador | 6368 |
| 176 | Swaziland | 1087 |
| 172 | Nigeria | 202534 |
| 171 | Namibia | 2554 |
| 151 | Tajikistan | 8604 |
| 145 | Bangladesh | 167475 |
| 140 | Bolivia (Plurinational State of) | 11306 |
| 138 | Greece | 10662 |
| 131 | Pakistan | 219964 |
| 120 | Ghana | 27858 |
| 117 | Zambia | 17255 |
| 116 | Nicaragua | 6085 |
| 109 | Cambodia | 16409 |
| 107 | Kyrgyzstan | 5718 |
| 96 | U.R. of Tanzania: Mainland | 56367 |
| 95 | Honduras | 9002 |
| 92 | Italy | 60447 |
| 86 | Congo | 5165 |
| 86 | Guatemala | 17232 |
| 85 | Mauritania | 3840 |
| 84 | Sao Tome and Principe | 204 |
| 81 | Kenya | 51372 |
| 80 | Afghanistan | 34941 |
| 79 | Sudan (Former) | 41774 |
| 79 | Djibouti | 884 |
| 76 | Côte d'Ivoire | 27623 |
| 74 | Chad | 15253 |
| 59 | Ethiopia | 106468 |
| 58 | Nepal | 29718 |
| 51 | Cameroon | 26374 |
| 50 | Rwanda | 11916 |
| 48 | Sierra Leone | 6312 |

| 47 | Uganda | 42331 |
|---|---|---|
| 42 | Lesotho | 1962 |
| 41 | Jamaica | 2672 |
| 35 | Senegal | 16356 |
| 31 | Guinea | 11855 |
| 29 | Comoros | 821 |
| 29 | Mali | 16507 |
| 25 | D.R. of the Congo | 95239 |
| 25 | Gambia | 2093 |
| 17 | Burkina Faso | 19700 |
| 17 | Benin | 12018 |
| 17 | Togo | 8176 |
| 15 | Niger | 21234 |
| 15 | Guinea-Bissau | 1833 |
| 12 | Haiti | 10788 |
| 10 | Madagascar | 25684 |
| 8 | Malawi | 20217 |
| -3 | D.P.R. of Korea | 25381 |
| -4 | Mozambique | 25998 |
| -5 | Burundi | 11229 |
| -9 | Liberia | 4810 |
| -9 | State of Palestine | 4615 |
| -19 | Central African Republic | 5745 |
| -33 | Zimbabwe | 14097 |
| -34 | Barbados | 293 |
| -107 | Yemen | 30217 |
| -182 | Venezuela (Bolivarian Republic of) | 28890 |
| -240 | Syrian Arab Republic | 16931 |